\documentclass[a4paper,fleqn,usenatbib,useAMS]{mnras}

\usepackage[T1]{fontenc}
\usepackage{ae,aecompl}

\usepackage{graphicx}	
\usepackage{amsmath}	
\usepackage{amssymb}	

\newcommand{\msun}{$M_{\odot}$}

\title[Magnetic fields in O-type multiple systems]{Are magnetic fields universal in O-type multiple systems?}

\stepcounter{footnote}

\author[Hubrig et al.]{
S.~Hubrig$^{1}$\thanks{Corresponding author: shubrig@aip.de},
S.~P.~J\"arvinen$^{1}$,
I.~Ilyin$^{1}$,
M.~Sch\"oller$^{2}$,
R.~Jayaraman$^{3}$
\\
$^{1}${Leibniz-Institut f\"ur Astrophysik Potsdam (AIP), An der Sternwarte~16, 14482~Potsdam, Germany} \\
$^{2}${European Southern Observatory, Karl-Schwarzschild-Str.~2, 85748 Garching, Germany} \\
$^{3}${MIT Kavli Institute and Department of Physics, 77 Massachusetts Avenue, Cambridge, MA 02139, USA}\\
}

\date{Accepted XXX. Received YYY; in original form ZZZ}

\pubyear{2023}

\begin{document}
\label{firstpage}
\pagerange{\pageref{firstpage}--\pageref{lastpage}}
\maketitle

\begin{abstract}
Although significant progress has been achieved in recent surveys of the magnetism in massive stars,
the origin of the detected magnetic fields remains to be the least understood topic in their studies.
We present an analysis of 61 high-resolution spectropolarimetric 
observations of 36 systems with O-type primaries,
among them ten known particle-accelerating colliding-wind binaries 
exhibiting synchrotron radio emission. 
Our sample consists of multiple systems with components at different evolutionary stages with wide and
tight orbits and different types of interactions. 
For the treatment of the complex composite spectra of the multiple systems, we used 
a special procedure involving different line masks populated for each element separately.
Out of the 36 systems, 22 exhibit in their LSD Stokes~$V$ profiles
definitely detected Zeeman features, among them seven systems with colliding winds.
For fourteen systems the detected Zeeman
features are most likely associated with O-type components whereas for three systems we suggest an association with 
an early B-type component.
For the remaining five systems the source of the field is unclear.
Marginal evidence for the detection of a Zeeman feature is reported for eleven systems
and non-detection for three systems.
The large number of systems with definitely detected Zeeman features 
presents a mystery, but probably indicates that multiplicity plays a definite role in the generation of 
magnetic fields in massive stars.
The newly found magnetic systems are supreme candidates for spectropolarimetric monitoring
over their orbital and rotation periods
to obtain trustworthy statistics on the magnetic field geometry and the distribution of field strength.
\end{abstract}




\section{Introduction}
\label{sec:intro}

Most recent observational surveys indicate that $\sim$7~per cent of O-type stars
host strong, large-scale organised fields 
\citep{Grunhut2017,Schoeller2017}.
The presence of a magnetic 
field in these stars can significantly modify their mass-loss and alter
their angular momentum, subsequently affecting their evolution
(e.g.\ \citealt{udDoula2008}; \citealt{TakahashiLanger2021}).
On the other hand, the principal question on the origin of the detected magnetic fields 
still remains unanswered, presenting currently the least 
understood topic in studies of massive stars.
It has been argued that magnetic 
fields could be fossil, dynamo generated, or may be generated by strong 
binary interaction, i.e.\ in stellar mergers, or during a mass transfer or 
common envelope evolutionary phase 
(e.g.\ \citealt{Tout2008}; \citealt{Ferrario2009}; \citealt{Schneider2016}). 
The currently most popular scenarios consider strong binary interaction and 
merging events.  Mass transfer or stellar merging may rejuvenate the mass gaining
star, while the induced differential rotation is thought
to be the key ingredient to generate a magnetic field
(e.g.\ \citealt{Wickramasinghe2014}).

In the context of this scenario, O stars are of special interest as the fraction of close
binaries for these objects is so high that many of them interact with a companion during their 
main sequence evolution through mass transfer or merging
(e.g.\ \citealt{Sana2012}).
As an example, by 
combining spectroscopy, interferometry, and photometry for the triple system 
HD\,150136,
\citet{Mahy2018}
showed that the masses estimated through their analysis (42.8\,$M_\odot$, 
29.5\,$M_\odot$, and 15.5\,$M_\odot$) are much smaller than those expected 
from evolutionary models (55.4\,$M_\odot$, 30.6\,$M_\odot$, and 
22.4\,$M_\odot$).
These results imply that evolutionary tracks computed for 
single stars cannot be used in studies of a majority of the O-type stars 
that interact with the other components in binary and multiple systems and 
that this interaction dominates the evolution of massive stars 
(e.g.\ \citealt{Holgado2020}).
No search for the presence of a magnetic field in HD\,150136
using high-resolution spectropolarimetry has been carried out yet.
With respect to already known magnetic massive stars, \citet{Frost2021}
combined ten years of spectroscopic and interferometric data for the 
magnetic Of?p binary HD\,148937 with an orbital period of 29\,yr and 
reported that the magnetic primary, although more massive, appears younger, 
suggesting that a merge or mass transfer took place in this system.

\citet{Offner2022} reported that 
the formation of multiple star systems with  separations of about 0.1\,pc
takes place during the earliest phases of star formation and that
the majority of such systems form and evolve to their final configuration during the time period spanned by the collapse of 
dense cores through the end of mass accretion. To study the impact of magnetic fields in such a process,
\citet{Wurster2019} used non-ideal MHD simulations including ambipolar diffusion, Ohmic dissipation,
and the Hall effect. The authors concluded that 
non-ideal magnetic processes do not inhibit the formation of multiple systems and that such systems form  
independent of the initial field strength.

Surprisingly, despite of the fact that multiplicity is widespread among massive stars, 
previous spectropolarimetric surveys of massive stars indicated that the 
incidence rate of magnetic component(s) in close binaries with 
$P_{\rm orb}<20$\,d is very low, less than 2~per cent
(e.g.\ \citealt{Alecian2015}).  
Of the dozen O stars currently confirmed to possess magnetic fields, about 
half are known to belong to wide binary systems where components can be 
treated in the spectropolarimetric analysis as single stars where
the structure and evolution depend only on intrinsic stellar properties. However, for stars in dense clusters, or in binary
and multiple systems, the evolution can be influenced by interaction with neighbouring stars.
As of today, only one short-period binary system, Plaskett's star,
is known to contain a hot, massive, magnetic star
\citep{Grunhut2013}.
Obviously, these results 
are difficult to comprehend, pointing out the need for a re-analysis of available spectropolarimetric observations
using a special procedure for the treatment of complex composite spectra of double and 
multiple systems. 

In particular the employment of high-resolution high signal-to-noise 
($S/N$) observations is necessary to properly interpret the features appearing in the Stokes~$V$ spectra.
Advantageously, the ESO archive contains a number of high-resolution ($R\approx110\,000$) HARPS\-pol
(High Accuracy Radial velocity Planet Searcher 
polarimeter; \citealt{Harps}) observations acquired for
O type stars belonging to binary and multiple systems. Almost all of them were observed within the framework of the 
ESO Large programme ``The B fields in OB stars'' (BOB). Spectropolarimetric observations of a number of
multiple systems with O-type primaries were acquired with ESPaDOnS 
(the Echelle SpectroPolarimetric Device for Observations of
Stars) with $R\approx65\,000$ within the framework of
the CFHT (the Canada-France-Hawaii Telescope) Large program ``Magnetism in Massive Stars'' (MiMeS),
available in the CFHT science archive.
For our re-analysis of spectropolarimetric data,
we initially concentrated on HAPRS\-pol data,
which have a higher spectral resolution than ESPaDOnS observations.
However, as in the process of this analysis a number of multiple
systems indicated the presence of Zeeman features in their Stokes~$V$ spectra,
we decided to also analyze ESPaDOnS observations of already known multiple systems with O-type primaries.
In total, we were able to gather reduced spectropolarimetric data for 36 multiple systems with O-type primaries.
Half of the targets were observed only on single epochs,
thirteen targets were observed twice, and 
five targets were observed three times.   

Importantly, our target list includes ten
particle-accelerating colliding-wind binaries (PACWBs) exhibiting synchrotron radio emission.
The catalog of PACWBs
based on radio observations was presented by \citet{DeBeckerRaucq2013}. It included exclusively 
massive non-degenerate stars in binary or higher multiplicity systems with wind-wind interaction regions.
The radiation mechanism of such systems requires 
two main ingredients: a population of relativistic electrons and a magnetic field, which can also 
play an important role in the process of the electron acceleration. 
Although no magnetic field detection was previously achieved
in the few systems studied by \citet{Neiner2015}, 
we believe that a magnetic survey of a more representative sample of PACWBs is necessary to investigate the role of 
magnetic fields in the generation of their synchrotron radio emission. 
Further, in addition to PACWBs, our sample contains two X-ray colliding wind binaries
(X-rayCMBs), one binary suggested to present colliding winds, and one high-mass X-ray binary (HMXB).
Apart from colliding wind binaries, we included in our sample a number of ordinary  main-sequence systems with noninteracting 
components, systems called in the literature ``twins'' of Plaskett's Star that are defined as short-period massive binaries 
that interact or have interacted in the recent past, two systems with blue
supergiant components, and two systems with Wolf-Rayet stars. 

The structure of this paper is as follows. 
In Sect.~2 we give details on the available archival spectropolarimetric observations and describe the methodology of our analysis. 
The results of the magnetic field measurements for each target are presented in Sect.~3.
In Sect.~4 we discuss the
implication of the reported field detections on future studies of massive stars.

\section{Observations and analysis}
\label{sect:obs}

\begin{table*}
\caption{
The sample of the investigated multiple systems.
The first column gives the HD number of the star
and the second column the visual magnitude.
This is followed by the spectral classification and the corresponding reference in Cols.~3 and 4
and the multiplicity indicator and corresponding reference in Cols.~5 and 6.
Here, VB denotes a visual binary, SB1/2/3 spectroscopic binaries with single, double, and triple line systems,
SBE a spectroscopic ecplising binary, and '?' an uncertainty of the multiplicity classification.
Cols.~7 and 8 list notes concerning the nature of the objects and the corresponding reference.
\label{tab:obsspt}
} 
\centering
\begin{tabular}{rllclclc}
\hline
\multicolumn{1}{c}{HD} &
\multicolumn{1}{c}{$V$} &
\multicolumn{1}{c}{Spectral} &
\multicolumn{1}{c}{Reference} &
\multicolumn{1}{c}{Multiplicity} &
\multicolumn{1}{c}{Reference} &
\multicolumn{1}{c}{Notes} &
\multicolumn{1}{c}{Reference} \\
\multicolumn{1}{c}{number} &
\multicolumn{1}{c}{} &
\multicolumn{1}{c}{Classification} &
\multicolumn{1}{c}{} &
\multicolumn{1}{c}{} &
\multicolumn{1}{c}{} &
\multicolumn{1}{c}{} &
\multicolumn{1}{c}{} \\
\hline
  1337  & 6.1 & O9.2II+O8V((f))             & (1)  & SB2       & (13) & X-rayCWB   & (36) \\
 17505  & 7.1 & O6.5III((f))n+O8V           & (2)  & SB3       & (7)  & & \\
 35921  & 6.9 & O9.2II(n)+O9.2III(n)+BO.2IV & (3)  & quadruple & (3)  & & \\
 36486  & 2.4 & O9.5\,II + B1V              & (4)  & triple    & (4)  & PACWB      & (25) \\ 
 36512  & 4.6 & O9.7V                       & (5)  & SB2       & (14) &  & \\ 
 36879  & 7.6 & O7V(n)((f))z                & (5)  & SB2?       & (14) &  & \\ 
 37366  & 7.6 & O9.5IV+B0-1V                & (6)  & SB2       & (6)  & &\\
 37468  & 3.8 & O9.5V+B0.2V(n)              & (3)  & triple    & (3)  & PACWB      & (25) \\
 47839  & 4.7 & O7V+B1.5/2V                 & (5)  & SB2       & (15) & PACWB      & (25) \\
 48099  & 6.4 & O5V((f))z+O9:V              & (2)  & SB2       & (16) & X-rayCWB   & (37) \\
 57061  & 4.4 & O9II                        & (7)  & SB1+SBE   & (17) & &\\ 
 75759  & 6.0 & O9V+B0V                     & (2)  & SB2       & (18) & &\\ 
 92206C & 7.5 & O6.5V((f))+O6.5V((f))       & (2)  & SB3       & (19) & &\\ 
 93129A & 7.9 & AaAb O2If*+O3.5V((f))       & (7)  & VB        & (20) & PACWB      & (25) \\ 
 93161A & 7.9 & O8V+O9V                     & (2)  & SB2       & (21) & &\\ 
101205  & 6.5 & O7 II: (n)                  & (1)  & quadruple & (22) & &\\ 
147165  & 2.9 & O9.5(V)+B7(V)               & (8)  & quadruple & (23) & &\\ 
149404  & 5.5 & O7.5 I(f)+ON9.7             & (7)  & SB2       & (24) & CWB        & (24) \\ 
151804  & 5.3 & O8Iaf                       & (7)  & SB2?      & (25) & PACWB      & (25) \\ 
152218A & 7.6 & O9.5IV(n)+B0V               & (2)  & SB2       & (26) & & \\ 
152236  & 4.8 & B1Ia-0ek                    & (9)  & SB2?      & (14) & BSG, cLBV  & (38) \\
152246  & 7.3 & O9IV                        & (7)  & triple    & (27) & & \\ 
152248  & 6.1 & O7Iabf+O7Ib(f)              & (2)  & SB2       & (28) & PACWB      & (39) \\
152408  & 5.8 & O8Iape                      & (7)  & VB?       & (29) & PACWB, BSG, WR\,79a & (25) \\ 
152590  & 9.3 & O7.5Vz                      & (7)  & SB2E      & (7)  & & \\ 
153919  & 6.5 & O6Iafcp                     & (7)  & SB1E      & (30) & HMXB       & (31) \\
155775  & 8.6 & O9.7III                     & (7)  & SB2       & (31) &  & \\ 
164794  & 5.9 & O3V((f+))+O5V((f))          & (10) & SB2       & (32) & PACWB      & (25) \\ 
166546  & 7.2 & O9.5IV                      & (7)  & SB3?      & (14) & & \\ 
167659  & 7.4 & O7II-III(f)                 & (7)  & SB2       & (14) & & \\ 
167971  & 7.5 & O7.5III+O9.5III+O9.5I       & (11) & SB3       & (33) & PACWB, BSG & (25) \\ 
190918  & 6.8 & WN5+O9I                     & (2)  & SB2       & (34) & PACWB, WR  & (25) \\ 
191201  & 7.3 & O9.5III+B0IV+O9.7III        & (3)  & SB3       & (35) & &\\ 
204827  & 7.9 & O9.5IV                      & (7)  & SB2       & (14) & &\\ 
209481  & 5.5 & O9III+ON9.7                 & (12) & SB2       & (31) & Algol      & (40) \\
305524  & 9.3 & O6.5Vn((f))z                & (7)  & SB2?      & (14) & & \\ 
\hline
\end{tabular}

Notes:
(1) -- \citet{Holgado2020};
(2) -- \citet{MaizApellaniz2004};
(3) -- \citet{MaizApellaniz2019};
(4) -- \citet{Harvin2002};
(5) -- \citet{Holgado2022};
(6) -- \citet{Boyajian2007};
(7) -- \citet{Sota2014};
(8) -- \citet{BeaversCook1980};
(9) -- \citet{Buscombe1969};
(10) -- \citet{Rauw2012};
(11) -- \citet{DeBecker2018};
(12) -- \citet{Mahy2011};
(13) -- \citet{Linder2008};
(14) -- this work;
(15) -- \citet{Burssens2020};
(16) -- \citet{Garmany1980};
(17) -- \citet{MaizApellanizBarba2020};
(18) -- \citet{Thackeray1966};
(19) -- \citet{Mayer2017};
(20) -- \citet{Walborn1973};
(21) -- \citet{Naze2005};
(22) -- \citet{Sana2014};
(23) -- \citet{Pigulski1992};
(24) -- \citet{Thaller1998};
(25) -- \citet{DeBeckerRaucq2013};
(26) -- \citet{Rosu2022};
(27) -- \citet{Nasseri2014};
(28) -- \citet{Hill1974};
(29) -- \citet{Mason1998};
(30) -- \citet{Falanga2015};
(31) -- \citet{Lefevre2009};
(32) -- \citet{Fabry2021};
(33) -- \citet{Ibanoglu2013};
(34) -- \citet{Richardson2021};
(35) -- \citet{Burkholder1997};
(36) -- \citet{PortegiesZwart2002};
(37) -- \citet{Berdyugin2016};
(38) -- \citet{Mahy2022};
(39) -- \citet{Sana2001};
(40) -- \citet{Wang2022}.
\end{table*}

The list of targets, their visual magnitudes, spectral classifications, and multiplicity information
with related literature sources are presented in Table~\ref{tab:obsspt}.
Further, we added notes concerning the nature of the targets.

HARPS\-pol and  ESPaDOnS spectropolarimetric observations
usually consist of four subexposures recorded at different positions of
the quarter-wave retarder plate. The HARPS\-pol spectra cover the spectral range 3780--6910\,\AA{},
with a small gap between 5259 and 5337\,\AA{}, whereas the ESPaDOnS spectra extend from 3700 to 10480\,\AA{}. 
The normalization of the spectra to the continuum level was described in detail by \citet{Hubrig2013}.
The assessment of the longitudinal magnetic field measurements 
is presented in our previous papers (e.g.\ \citealt{Hubrig2018}; \citealt{Jarvinen2020}). 
Similar to our previous studies, to increase the 
signal-to-noise ratio ($S/N$) by a multiline approach, we employed the least-squares deconvolution (LSD) technique. 
The details of this technique, 
as well as how the LSD Stokes~$I$, Stokes~$V$, and diagnostic null spectra are calculated, were 
presented by \citet{Donati1997}.

\begin{figure}
\centering 
\includegraphics[width=0.460\textwidth]{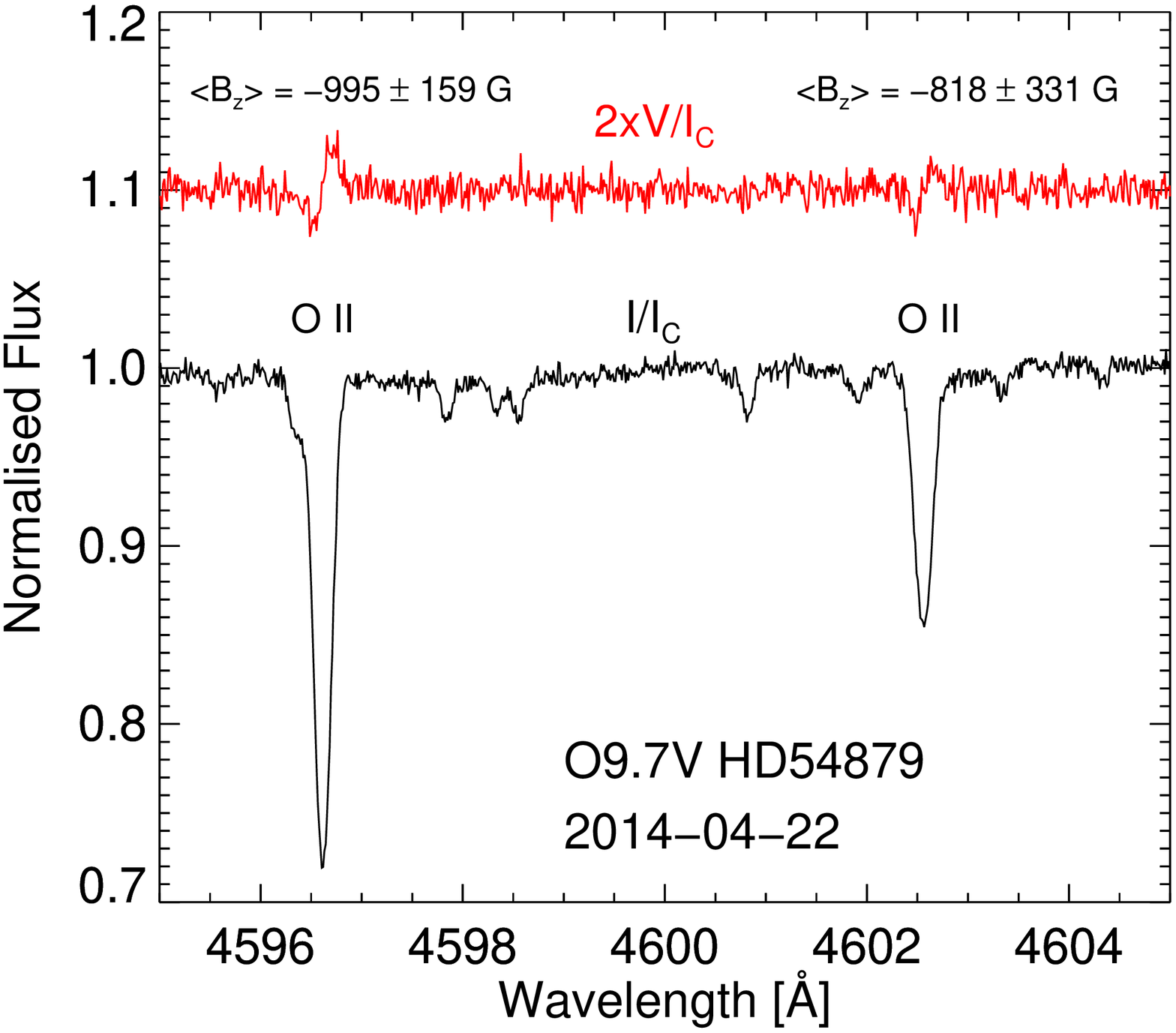}
\caption{
Stokes~$I$ (black) and $V$ (red) spectra of the magnetic O9.7V star
HD\,54879 observed with HARPS\-pol in 2014.
The two \ion{O}{ii} lines at $\lambda\lambda$4596 and 4602 have equal Land\'e factors of 0.9
and show different intensities in the spectrum.
The effect of the different strengths of the lines used for the magnetic field measurements
on the amplitude of the corresponding Zeeman features
and therefore on the measurement accuracy is clearly visible.
}
   \label{fig:hd54}
\end{figure}

 As already mentioned above, a careful analysis using high quality spectropolarimetric observations
and the utilisation of a special procedure are crucial to detect 
and properly interpret the Zeeman features observed in the polarimetric spectra of multiple systems.
In such systems, 
the shapes of blended spectral lines look different depending 
on the visibility of each component at different orbital phases.
Additionally, the amplitudes of the Zeeman features are much lower in multiple systems in comparison to
the size of these features in single stars.
Thus, for the LSD analysis of multiple systems, it is preferable not to populate the line mask 
with weak spectral lines.
In Fig.~\ref{fig:hd54} we show the effect of the different strengths in the lines
used in the magnetic field measurements on the 
amplitude of the corresponding Zeeman features and therefore on the measurement accuracy
in the magnetic star HD\,54879.
The amplitude of the Zeeman feature corresponding to the weaker \ion{O}{ii} line at 4602\,\AA{} is significantly smaller than 
for the stronger \ion{O}{ii} line at 4596\,\AA{} having both a Land\'e factor of 0.9.
When extracting the magnetic field strength of the two lines individually,
the accuracy of the mean longitudinal magnetic field measurement $\left< B_{\rm z} \right>$
using the weaker line has decreased by more than a factor of two.
Moreover, in contrast to wide binary systems where the magnetic components can be 
treated in the spectropolarimetric analysis as single stars, Zeeman features are usually blended in composite spectra of tight
multiple systems. The presence of severe distortions in the shapes of the Zeeman features
hinders  magnetic field detections in multiple systems and only dedicated spectropolarimetric monitoring over the
orbital and rotation periods can permit to make solid conclusions on the magnetic field strength and geometry.

\begin{figure}
\centering 
\includegraphics[width=0.460\textwidth]{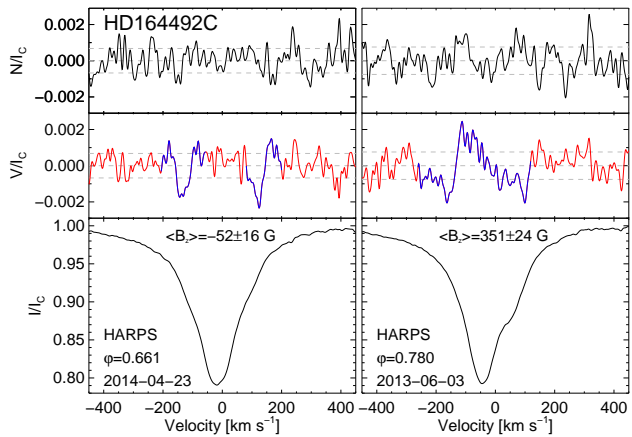}
\caption{
LSD Stokes~$I$, $V$, and diagnostic $N$ spectra of the massive magnetic
triple system HD\,164492C in the Trifid nebula calculated for HARPS\-pol spectra
obtained at two different observing epochs.
As all three components are blended,
we observed severe distortions of the Zeeman features in the Stokes~$V$ spectra.
The presence of the features indicates the magnetic nature of at least one of the components
\citep{Gonzalez2017}.
}
  \label{fig:trifid}
\end{figure}

The effect of the blending and severe distortion of Zeeman features in multiple systems was demonstrated in
the work of \citet{Gonzalez2017}, who studied the massive magnetic triple system HD\,164492C in the Trifid nebula.
This system consists of a close binary (component Ca) composed of B-type stars
and a more distant rapidly rotating early-B He-rich magnetic component Cb.  
As all three components in the HARPS\-pol spectra remain blended at all observing epochs, Zeeman features in the Stokes~$V$ spectra
exhibit complex configurations and atypical shapes compared to the characteristic Zeeman $S$-shape features usually detected in single 
magnetic massive stars. 
In Fig.~\ref{fig:trifid} we present our LSD analysis of two HARPS\-pol observations of this system at two different 
epochs using a line mask containing \ion{He}{i} and \ion{Si}{iii} lines. The clearly visible features in the Stokes~$V$ spectra indicate
the magnetic nature of at least one of the components, even if their shapes significantly deviate from the typical
Zeeman Stokes~$V$ profiles observed in single magnetic stars. 
Due to severe blending, the presence of a magnetic field
can remain undetected in some targets when only single or very few observations are available or when 
spectropolarimetric observations are obtained with low $S/N$.
We note that a considerable departure of the Stokes $V$ profiles from an $S$-shape can also be due to the  
presence  of  magnetic fields of mixed polarities over the visible hemisphere (see Fig.~7 in \citealt{Jarvinen2020}),
the rotational Doppler effect in fast rotating stars (e.g.\ \citealt{Mathys1995}), or the presence of gradients in velocity 
and magnetic field as a function of optical depth (e.g.\ \citealt{Jarvinen2021}, and references therein).

Based on the complications existing in the analysis of the presence of magnetic fields in multiple systems discussed above,
it appears reasonable to search in the Stokes~$V$ spectra of O-type systems for both Zeeman features with typical 
and atypical shapes, similar to the study of the triple system HD\,164492C.
To evaluate, whether the detected features are spurious or definite detections, 
we followed the generally adopted procedure to use the false alarm probability (${\rm FAP}$) based on the 
reduced $\chi^{2}$ test statistics \citep{Donati1992}:
the presence of the Zeeman feature
is considered as a definite detection DD
if ${\rm FAP} \leq 10^{-5}$,
as a marginal detection (MD) if  $10^{-5}<{\rm FAP}\leq 10^{-3}$,
and as a non-detection (ND) if ${\rm FAP}>10^{-3}$.

\begin{table*}
\caption{
The logbook of the observations and the results of the magnetic field measurements for the
stars in our sample.
The first column gives the HD number of the star followed by the name of the used spectropolarimeter, with H for
HARPS\-pol and E for ESPaDOnS.
The third column presents the MJD values at the middle of the exposure, while in the fourth
column we show the signal to noise ratio measured in the Stokes~$I$ spectra
in the spectral region around 5000\,\AA{}.
The line mask used, the false alarm probability (FAP) values, the detection flag --
where DD means definite detection, MD marginal detection, and ND no detection --
the measured LSD mean longitudinal magnetic field strength,
and a remark on the Stokes~$V$ feature where the field is diagnosed (if applicable),
are presented in Columns~5--9.
\label{tab:obsall}
}
\centering
\begin{tabular}{rllrccc r@{$\pm$}ll}
\hline
\multicolumn{1}{c}{HD} &
\multicolumn{1}{c}{Instr.} &
\multicolumn{1}{c}{MJD} &
\multicolumn{1}{c}{$S/N$} &
\multicolumn{1}{c}{Line} &
\multicolumn{1}{c}{FAP}&
\multicolumn{1}{c}{Det.} &
\multicolumn{2}{c}{$\left< B \right>_{\rm z}$} &
\multicolumn{1}{c}{Remark} \\
\multicolumn{1}{c}{number} &
\multicolumn{1}{c}{} &
\multicolumn{1}{c}{} &
\multicolumn{1}{c}{} &
\multicolumn{1}{c}{mask} &
\multicolumn{1}{c}{} &
\multicolumn{1}{c}{flag} &
\multicolumn{2}{c}{(G)} \\
\hline
1337    & E  & 55\,113.345 & 609 & \ion{He}{i/ii}, \ion{O}{iii}     & $<10^{-10}$       & DD & $-$274 & 52  \\
17505   & E  & 55\,169.207 & 390 & \ion{He}{i/ii}, \ion{O}{iii}     & $<10^{-10}$      & DD & \multicolumn{2}{c}{} \\
35921   & E  & 55\,878.373 & 463 & \ion{He}{i/ii}, \ion{O}{ii/iii}  & $<10^{-10}$       & DD & 538 & 172 \\
36486   & H  & 56\,650.270 & 337 & \ion{He}{i}         &  $>10^{-3}$            & ND & \multicolumn{2}{c}{} \\
        & H  & 56\,651.037 & 608 & \ion{He}{i}         & $4\times10^{-4}$  & MD & 358    & 17  \\
        & H  & 56\,655.296 & 286 & \ion{He}{i}         &  $>10^{-3}$                 & ND & \multicolumn{2}{c}{} \\
36512   & H  & 56\,650.085 & 159 & \ion{He}{i/ii}, \ion{C}{iii}, \ion{O}{ii/iii}, \ion{Si}{iii/iv} & $3\times10^{-4}$  & MD & 123    & 10 \\
        & H  & 56\,653.178 & 300 & \ion{He}{i/ii}, \ion{C}{iii}, \ion{N}{iii}, \ion{O}{iii}  &  $>10^{-3}$          & ND & \multicolumn{2}{c}{} \\
        & H  & 57\,092.026 & 504 & \ion{He}{i/ii}      & $3\times10^{-4}$  & MD & $-$19    & 2 \\
36879   & H  & 57\,317.371 & 184 & \ion{He}{i/ii}, \ion{C}{iv}, \ion{N}{iv}, \ion{O}{iii}, \ion{Si}{iv}  & $6\times10^{-10}$ & DD & 301    & 42 \\
        & E  & 55\,080.586 & 390 & \ion{He}{i/ii}, \ion{C}{iv}, \ion{N}{iv}, \ion{O}{iii}, \ion{Si}{iv}  &  $>10^{-3}$ & ND & \multicolumn{2}{c}{} \\
37366   & E  & 54\,698.621 & 522 & \ion{He}{i/ii}, \ion{C}{iii/iv}, \ion{Si}{iii/iv}  & $2\times10^{-4}$  & MD & \multicolumn{2}{c}{} \\
37468   & H  & 56\,651.221 & 434 & \ion{He}{i/ii}, \ion{Si}{iii/iv}  & $8\times10^{-5}$  & MD & \multicolumn{2}{c}{} \\
        & E  & 56\,971.585 & 424 & \ion{He}{i/ii}     & $9\times10^{-4}$  & MD & \multicolumn{2}{c}{} \\
47839   & E  & 55\,960.491 & 475 & \ion{He}{i/ii}, \ion{C}{iv}    & $<10^{-10}$       & DD & 282    & 58 \\
48099   & E  & 55\,961.441 & 451 & \ion{He}{i/ii}, \ion{C}{iv}, \ion{Si}{iv}   & $7\times10^{-6}$  & DD & \multicolumn{2}{c}{} \\
57061   & H  & 57\,092.152 & 761 & \ion{He}{i}, \ion{C}{iii}, \ion{O}{iii}       & $3\times10^{-10}$ & DD & $-$285  & 144 \\
75759   & H  & 56\,651.301 & 452 & \ion{He}{i}, \ion{C}{iii}, \ion{N}{iii}, \ion{Si}{iii/iv}     & $2\times10^{-4}$  & MD & $-92$   & 9 \\
        & H  & 57\,092.217 & 442 & \ion{He}{ii}, \ion{C}{iv}        & $3\times10^{-4}$  & MD & $-113$  & 10  \\
92206C  & H  & 56\,445.014 & 134 & \ion{He}{i}         & $<10^{-10}$       & DD & \multicolumn{2}{c}{} \\
        & H  & 56\,446.050 & 171 & \ion{He}{i/ii}      &   $>10^{-3}$                & ND & \multicolumn{2}{c}{} \\
93129A  & H  & 57\,556.963 & 348 & \ion{He}{ii}         & $<10^{-10}$       & DD &  \multicolumn{2}{c}{}\\
93161A  & H  & 57\,557.031 & 204 & \ion{He}{i/ii}, \ion{C}{iv}, \ion{O}{iii}    &    $>10^{-3}$               & ND & \multicolumn{2}{c}{} \\
101205  & H  & 56\,446.987 & 421 & \ion{He}{i/ii}, \ion{O}{iii}     & $9\times10^{-6}$  & DD & \multicolumn{2}{c}{}  \\
147165  & H  & 56\,769.190 & 285 & \ion{He}{i}         & $1\times10^{-5}$  & MD & \multicolumn{2}{c}{} \\
        & H  & 57\,176.969 & 406 & \ion{He}{i/ii}, \ion{C}{ii}, \ion{N}{ii}, \ion{Si}{iii/iv}  &   $>10^{-3}$ & ND & \multicolumn{2}{c}{} \\
149404  & H  & 57\,557.191 & 536 & \ion{He}{i/ii}, \ion{C}{iv}, \ion{Si}{iii/iv}   & $5\times10^{-4}$  & MD &  \multicolumn{2}{c}{} \\ 
        & E  & 56\,816.367 & 408 & \ion{He}{i/ii}, \ion{Si}{iii/iv}    & $7\times10^{-4}$  & MD & \multicolumn{2}{c}{} \\
151804  & H  & 55\,708.336 & 468 & \ion{He}{i/ii}      & $6\times10^{-4}$  & MD & \multicolumn{2}{c}{} \\
152218A & H  & 56\,447.091 & 355 & \ion{He}{i/ii}, \ion{C}{iv}, \ion{Si}{iii/iv}   &   $>10^{-3}$                & ND & \multicolumn{2}{c}{} \\
        & H  & 57\,557.246 & 343 & \ion{He}{i/ii}, \ion{C}{iii/iv}, \ion{Si}{iv}   & $<10^{-10}$       & DD & $-$307 & 94 \\
152236  & H  & 56\,770.182 & 620 & \ion{He}{i}       & $>10^{-3}$      & ND & \multicolumn{2}{c}{} \\
        & H  & 56\,770.420 & 600 & \ion{He}{i}         & $<10^{-10}$       & DD & \multicolumn{2}{c}{} \\
        & H  & 57\,094.251 & 448 & \ion{He}{i}, \ion{Si}{iii/iv}       &  $2\times10^{-7}$ & DD & \multicolumn{2}{c}{} \\
152246  & H  & 56\,445.176 & 317 & \ion{He}{i/ii}, \ion{C}{iii/iv}     & $6\times10^{-7}$  & DD & \multicolumn{2}{c}{} \\
        & H  & 57\,558.144 & 494 & \ion{He}{i/ii}, \ion{C}{iv}, \ion{Si}{iii/iv}   & $<10^{-10}$   & DD & \multicolumn{2}{c}{} \\
152248  & E  & 56\,760.546       & 320 & \ion{He}{i/ii}, \ion{Si}{iv}   & $<10^{-10}$       & DD & \multicolumn{2}{c}{} \\
        & E  & 56\,761.562       & 322 & \ion{He}{i/ii}, \ion{O}{iii}    & $4\times10^{-7}$  & DD & \multicolumn{2}{c}{} \\
152408  & E  & 56\,114.347       & 384 & \ion{He}{i/ii}     & $5\times10^{-4}$  & MD & \multicolumn{2}{c}{} \\
        & E  & 56\,114.372       & 421 & \ion{He}{ii}, \ion{C}{iv}       & $<10^{-10}$       & DD & \multicolumn{2}{c}{} \\
        & E  & 56\,114.397       & 436 & \ion{He}{ii}, \ion{C}{iv}       & $<10^{-10}$       & DD & \multicolumn{2}{c}{} \\
152590  & H  & 56\,447.217       & 259 & \ion{He}{i/ii}, \ion{O}{iii}, \ion{Si}{iv}   & $1\times10^{-4}$  & MD & \multicolumn{2}{c}{} \\
        & H  & 57\,558.081       & 306 & \ion{He}{i/ii}, \ion{O}{iii}, \ion{Si}{iv}   &  $>10^{-3}$      & ND & \multicolumn{2}{c}{} \\
153919  & H  & 57\,558.198       & 505 & \ion{He}{i/ii}, \ion{C}{iv}, \ion{O}{iii}   & $>10^{-3}$ & ND & \multicolumn{2}{c}{} \\
        & E  & 55\,726.493       & 409 & \ion{He}{i/ii}, \ion{C}{iv}, \ion{O}{iii}   &   $>10^{-3}$   & ND & \multicolumn{2}{c}{} \\
155775  & H  & 56\,446.199       & 339 & \ion{He}{i}, \ion{N}{iii}, \ion{Si}{iii}     & $<10^{-10}$       & DD & $-$9 & 37 \\
164794  & H  & 57\,177.263       & 480 & \ion{He}{i/ii}, \ion{N}{iv}, \ion{O}{iii}   & $<10^{-10}$       & DD & \multicolumn{2}{c}{} \\
        & H  & 57\,557.310       & 623 & \ion{He}{ii}, \ion{N}{iv}       & $<10^{-10}$       & DD & \multicolumn{2}{c}{} \\
        & E  & 53\,540.489       & 323 & \ion{He}{i/ii}, \ion{N}{iv}, \ion{O}{iii}   & $<10^{-10}$       & DD & \multicolumn{2}{c}{} \\
166546  & H  & 56\,447.408       & 310 & \ion{He}{i}, \ion{Si}{iii}      & $<10^{-10}$       & DD & \multicolumn{2}{c}{} \\
167659  & H  & 56\,446.399       & 305 & \ion{He}{ii}        & $1\times10^{-5}$  & MD & \multicolumn{2}{c}{} \\
167971  & E  & 56\,474.363       & 356 & \ion{He}{ii}, \ion{C}{iv}, \ion{N}{iii}      & $<10^{-10}$       & DD & 1324 & 582 \\
        & E  & 57\,229.421       & 348 & \ion{He}{ii}, \ion{C}{iv}, \ion{N}{iii}      & $9\times10^{-4}$  & MD & $-$977 & 437 & blue feature \\
        & E  & 57\,229.421       & 348 & \ion{He}{ii}, \ion{C}{iv}, \ion{N}{iii} & $6\times10^{-6}$  & DD & $-$57 & 33 & central feature \\
        & E  & 57\,229.421       & 348 & \ion{He}{ii}, \ion{C}{iv}, \ion{N}{iii} &  $>10^{-3}$       & ND & \multicolumn{2}{c}{} & red feature \\
\hline
\end{tabular}
\end{table*}

\addtocounter{table}{-1}

\begin{table*}
\caption{
Continued.
}
\centering
\begin{tabular}{rllrccc r@{$\pm$}ll}
\hline
\multicolumn{1}{c}{HD} &
\multicolumn{1}{c}{Instr.} &
\multicolumn{1}{c}{MJD} &
\multicolumn{1}{c}{$S/N$} &
\multicolumn{1}{c}{Line} &
\multicolumn{1}{c}{FAP}&
\multicolumn{1}{c}{Det.} &
\multicolumn{2}{c}{$\left< B \right>_{\rm z}$} &
\multicolumn{1}{c}{Remark} \\
\multicolumn{1}{c}{number} &
\multicolumn{1}{c}{} &
\multicolumn{1}{c}{} &
\multicolumn{1}{c}{} &
\multicolumn{1}{c}{mask} &
\multicolumn{1}{c}{} &
\multicolumn{1}{c}{flag} &
\multicolumn{2}{c}{(G)} \\
\hline
190918  & E  & 54\,677.305       & 460 & \ion{He}{i/ii}     & $<10^{-10}$       & DD &  $-$355 & 44\\
        & E  & 55\,403.513       & 481 & \ion{He}{i/ii}     & $3\times10^{-4}$  & MD & 56 & 8\\
191201  & E  & 55\,724.598       & 404 & \ion{He}{i}, \ion{Si}{III/iv}     & $>10^{-3}$ & ND & \multicolumn{2}{c}{} \\
204827  & E  & 55\,726.561       & 280 & \ion{He}{i/ii}, \ion{O}{iii}    & $4\times10^{-4}$  & MD & \multicolumn{2}{c}{}\\
209481  & E  & 55\,734.550       & 446 & \ion{He}{i}        & $<10^{-10}$       & DD & \multicolumn{2}{c}{} & blue feature \\
        & E  & 55\,734.550       & 446 & \ion{He}{i}        & $3\times10^{-4}$  & MD & \multicolumn{2}{c}{} & red feature \\
        & E  & 57\,350.248       & 389 & \ion{He}{i/ii}     & $<10^{-10}$       & DD & \multicolumn{2}{c}{} \\
305524  & H  & 57\,557.966       & 198 & \ion{He}{i/ii}     & $<10^{-10}$       & DD & \multicolumn{2}{c}{} \\
\hline
\end{tabular}
\end{table*}

In the LSD technique, the line masks corresponding to the individual spectral types of the studied targets 
are generally constructed using the Vienna Atomic Line Database \citep[VALD3;][]{Kupka2011}.
As our targets are binaries or
higher order multiple systems exhibiting in their spectra very different spectral signatures depending
on the spectral classification of the individual components, special care has to be
taken to populate the line masks for each system.
Only for a few systems in our sample the Doppler separation  
between the components is large enough to allow us to search for the presence of a magnetic field in each component using
individual line masks. But before such masks for each component can be created,  the line spectra of each component in
each studied system has to be carefully surveyed to check for the presence and strength of spectral lines belonging to different elements.

The appearance of the spectra of the components usually depends on the orbital phase, the rotation rate, 
but also on the evolutionary state:
some of the stars are nitrogen enriched, exhibiting strong nitrogen lines clearly 
identified in their spectra. In some other systems the components are strongly enriched in helium 
but depleted in carbon and oxygen.
Overabundances and underabundances of different elements can also be 
caused by mass-transfer episodes expected in interacting and post-interaction systems.
Multiple systems with short orbital periods undergo tidal interaction or  mass transfer, and can suffer from X-ray 
irradiation. Obviously, combining lines of all elements together in the LSD 
line masks, without checking their presence and strength in the spectra, may lead to the dilution of the magnetic signal, or even to its 
partial cancellation. 

As mentioned above, also the shape of blended spectral lines in composite spectra can be different depending on the 
visibility of each component. Therefore for observations obtained at different epochs it is sometimes useful to carry out 
the LSD analysis using a different line mask tailored specifically for the best visible component.
In our LSD magnetic field analysis, to take into account the different spectral types of the components,
we usually constructed 
separate masks for each individual element in different ionization states, namely \ion{He}{i} and \ion{He}{ii} masks, and masks with 
lines belonging to different ionization states of CNO elements and Si. 
As different spectral lines in massive stars have different formation depths,
the use of masks constructed for individual elements 
to search for the presence of a magnetic field appears especially promising,
in particular in view of the recent work by \citet{Jarvinen2022},
who reported on different field strengths measured
using spectral lines belonging to different elements in the magnetic O-type star HD\,54879:
the measurements using helium lines that are formed higher in the stellar atmosphere compared to the formation depths of metal lines 
showed significantly lower field strengths than those obtained from the measurements using silicon and oxygen lines.
Based on these results, it is clear that an LSD analysis using masks
containing individual elements is extremely useful to select masks substantially 
contributing to the magnetic signal.

There are other factors playing an important role in the magnetic field measurements of massive O stars, such as
the much lower number of spectral lines in comparison to less massive stars, causing a relatively high noise level in the LSD profiles.
In systems with severe blending between the line profiles of the system components, it is sometimes 
possible to select unaffected or to a lesser extent blended lines
if the spectral types of the components are different. Yet, already a small degree of
contamination can have an impact on the magnetic field measurements. This is also true if some lines are contaminated by emission.  
Sometimes it is also not clear whether the observed features are of stellar origin,
or whether they are caused by a nebular contamination around the targets. 
Massive O stars can be responsible for the ionization of gas in a surrounding nebula,
but also for the creation of interstellar bubbles.

The results of our analysis of 36 binary and multiple systems using 61 HARPS\-pol and ESPaDOnS observations 
are presented in Table~\ref{tab:obsall}, where we indicate
the spectropolarimeter used for the observations, the MJD values at the middle of the exposures, $S/N$ values measured in the Stokes~$I$ 
spectra in the spectral region around 5000\,\AA{}, the applied line mask, and the ${\rm FAP}$ values with the 
detection flags DD, MD, or ND.
For the few systems with components appearing clearly separated in the recorded spectra,
we also present in the last column of this table the results of our measurements of the
mean longitudinal magnetic field. Notably, in contrast to chemical abundance studies, spectrum disentangling is
usually not needed to detect the presence of a magnetic field as Zeeman features should clearly stand out
in high $S/N$ LSD Stokes~$V$ spectra,
if one of the binary or multiple system components possesses a magnetic field.
However, if several spectropolarimetric observations phased over the orbital and rotation period for each target
are available, an iterative disentangling process (e.g.\ \citealt{Gonzalez2017})
should be applied to properly estimate the field strength and 
deduce the most likely field configuration.

In the following, we give a brief description of the targets
and discuss the results of the magnetic field measurements for each target individually.
The studied systems are divided into five different groups:
the first group is composed of PACWBs (Sect.~\ref{subsect:pacwb}),
the second of X-ray binaries with colliding winds (Sect.~\ref{subsect:xrb}),
the third group includes all other systems with definite detections of magnetic fields (Sect.~\ref{subsect:det}),
the fourth group contains all other systems with marginal field detections (Sect.~\ref{subsect:marg}). 
and the fifth group contains the remainder,
all other systems with non-detections (Sect.~\ref{subsect:nodet}). 
Mosaics with the plots presenting the LSD Stokes~$I$, $V$, and diagnostic null $N$ spectra calculated for
the available HARPS\-pol and ESPaDOnS observations for the first and second group are presented in 
Figs.~\ref{fig:collid} and \ref{fig:xraycoll}, respectively.
The LSD spectra for the third group are presented in Fig.~\ref{fig:hddef}
and in Figs.~\ref{fig:marg} and \ref{fig:nodet} for the fourth and fifth groups.


\section{Results of our LSD analysis for individual targets}
\label{sect:field}

\subsection{Particle-accelerating colliding-wind binaries}
\label{subsect:pacwb}

\begin{figure*}
\centering 
\includegraphics[width=0.220\textwidth]{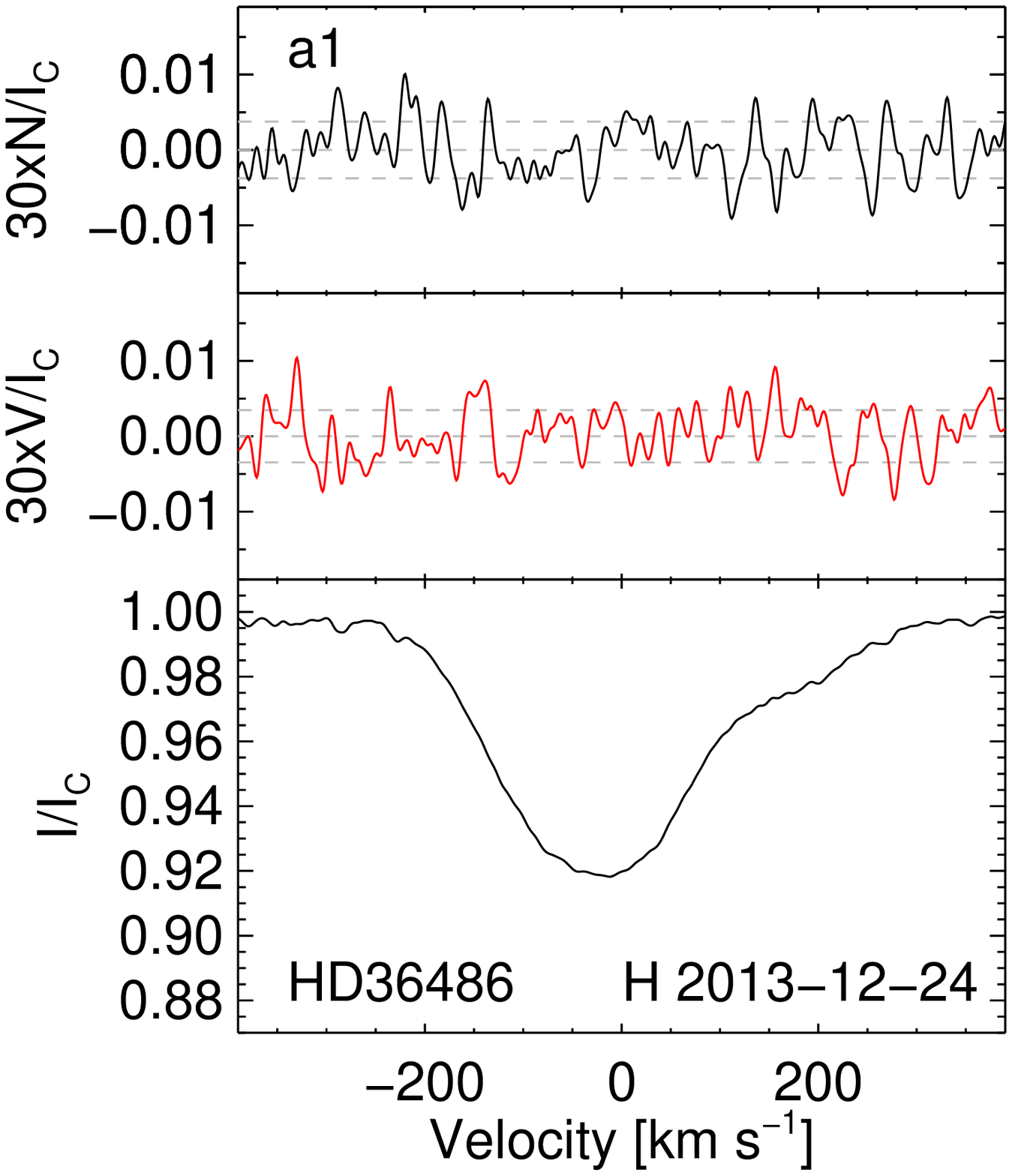}
\includegraphics[width=0.220\textwidth]{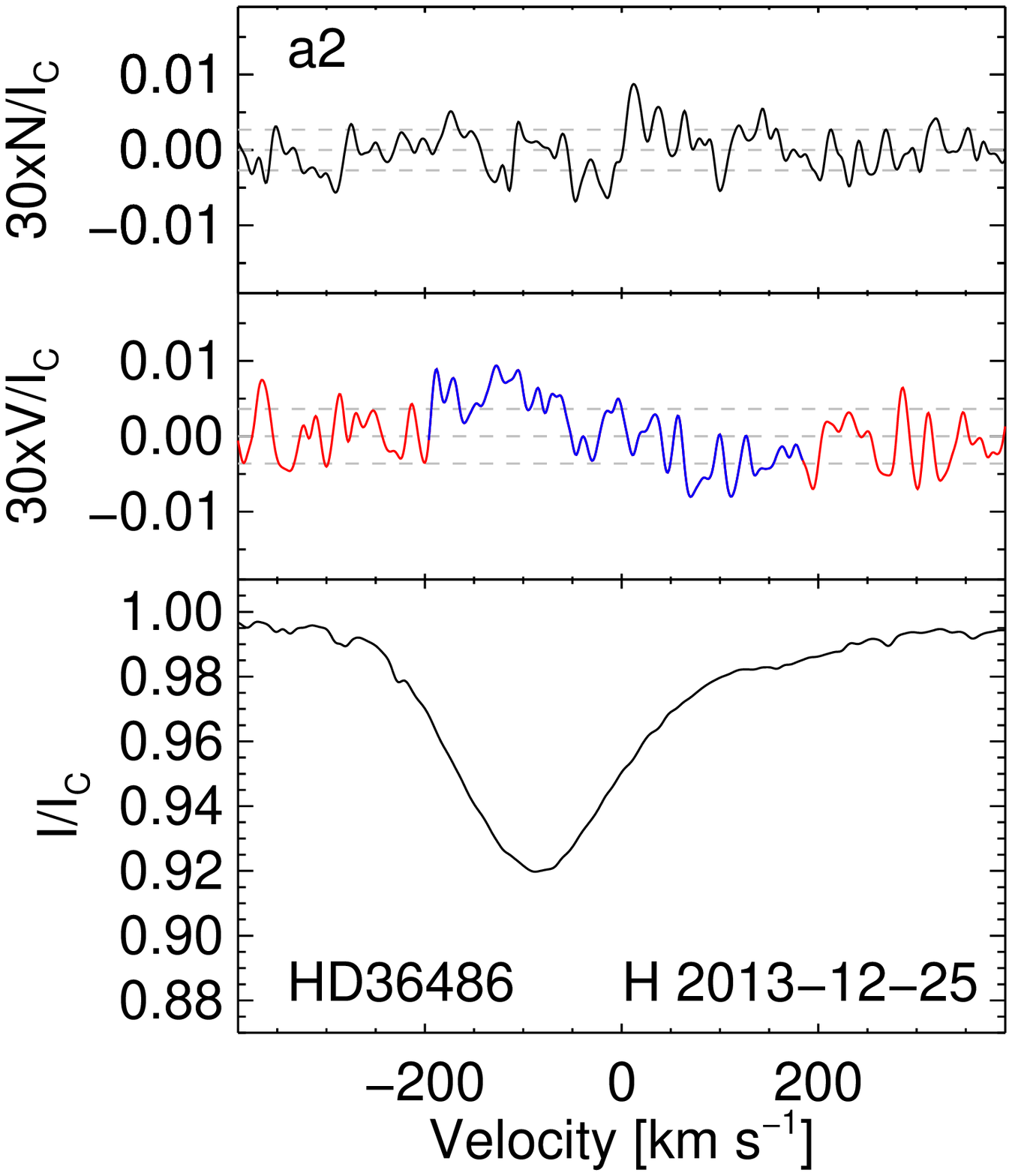}
\includegraphics[width=0.220\textwidth]{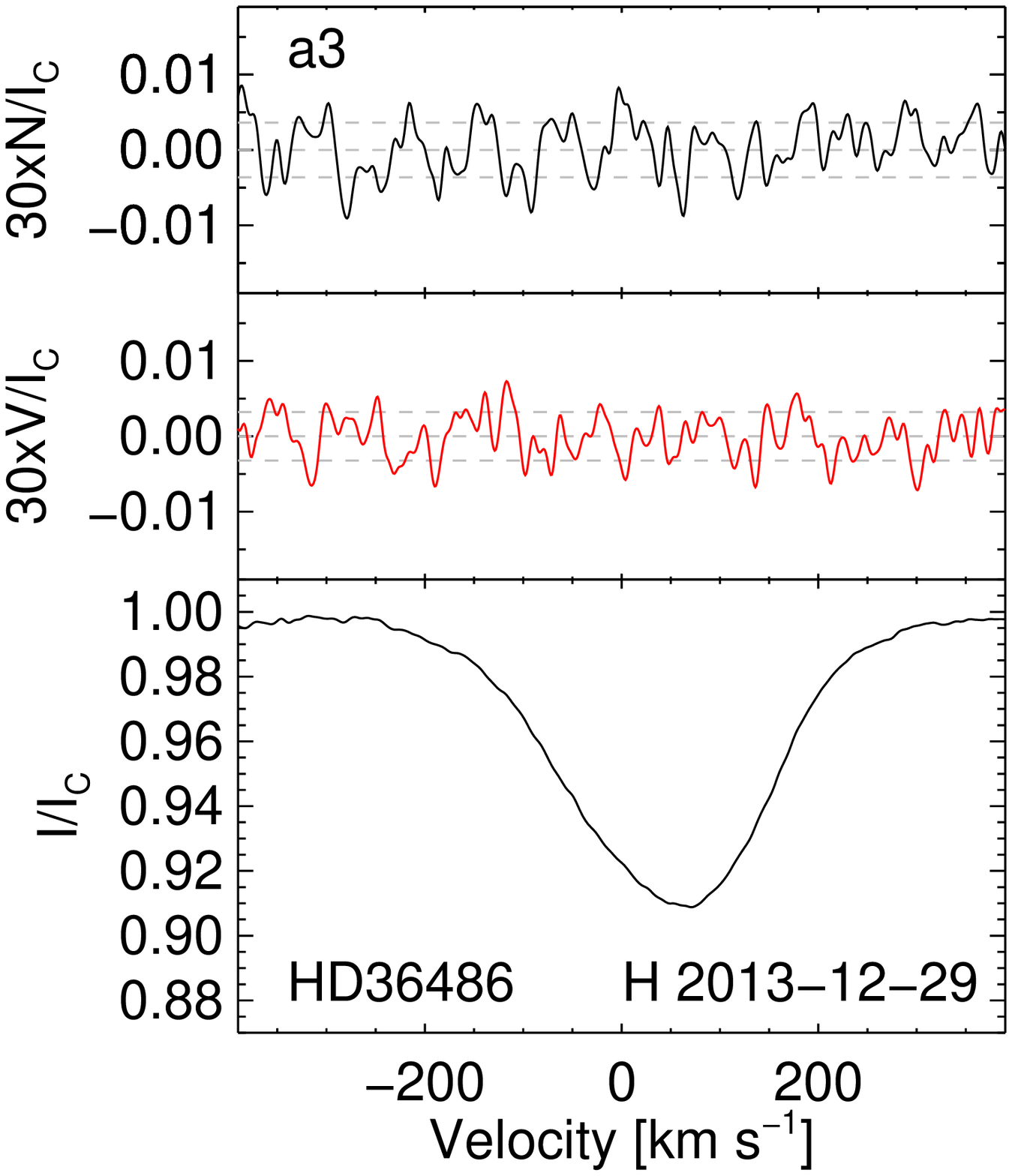}
\includegraphics[width=0.220\textwidth]{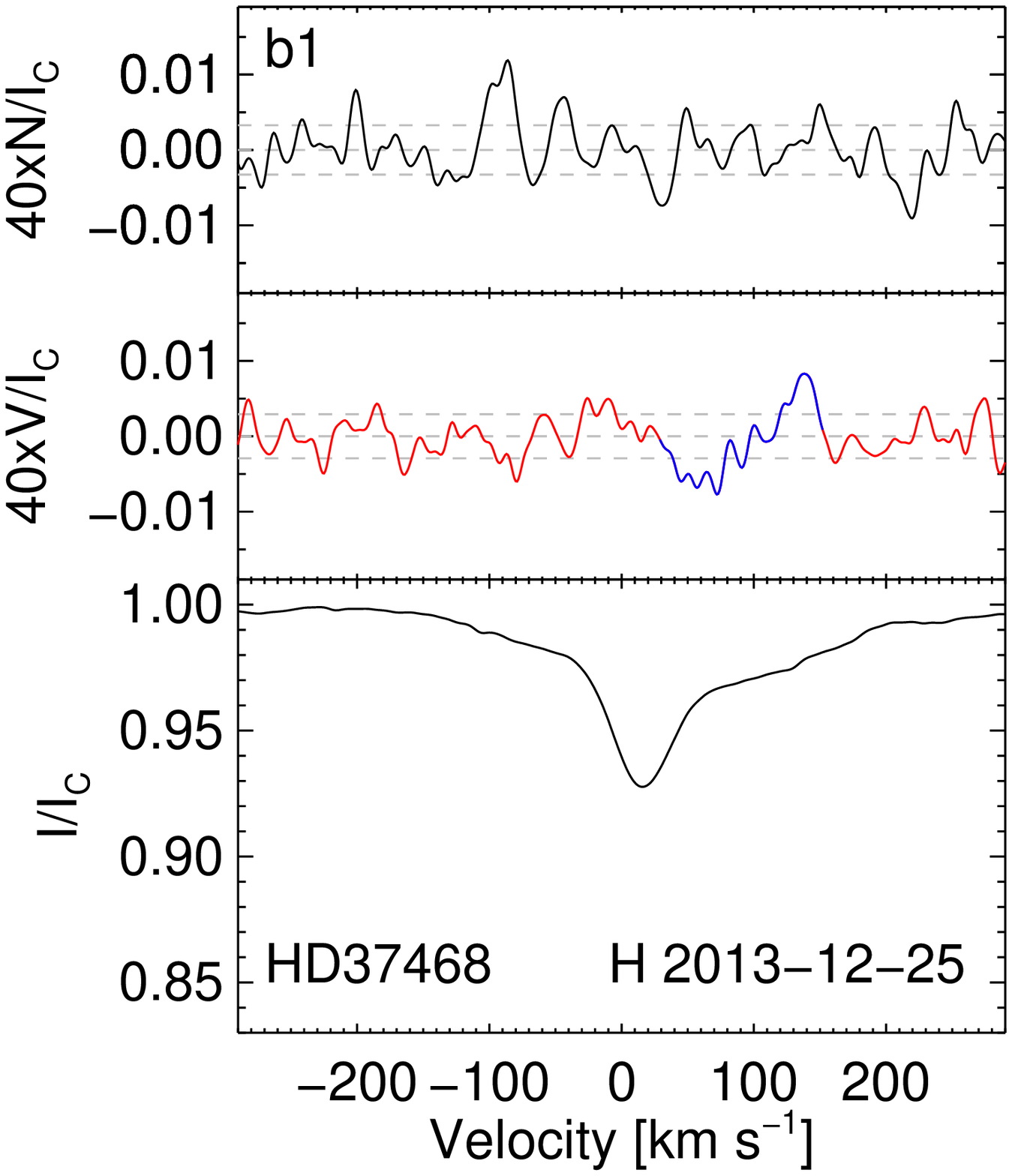}
\includegraphics[width=0.220\textwidth]{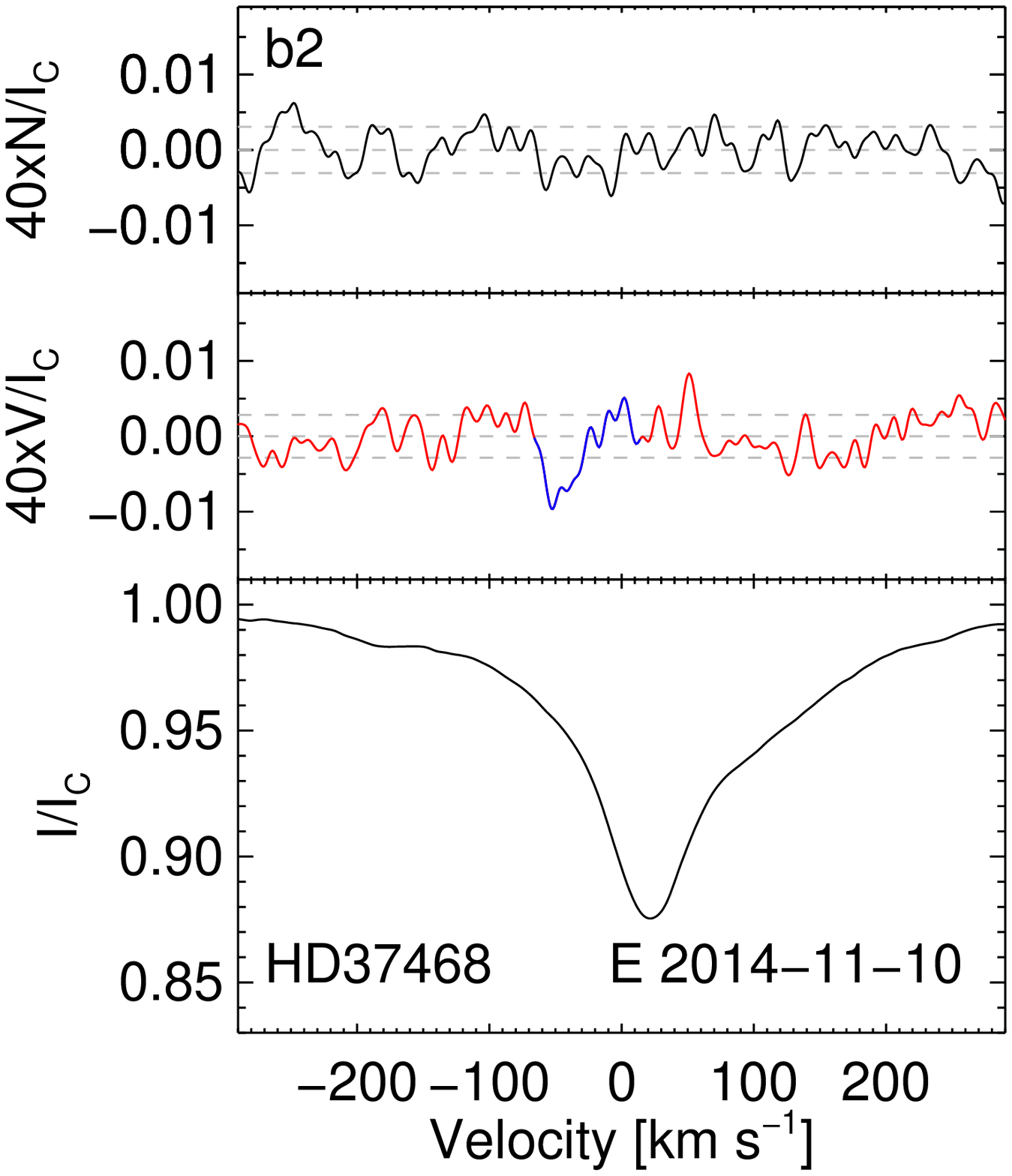}
\includegraphics[width=0.220\textwidth]{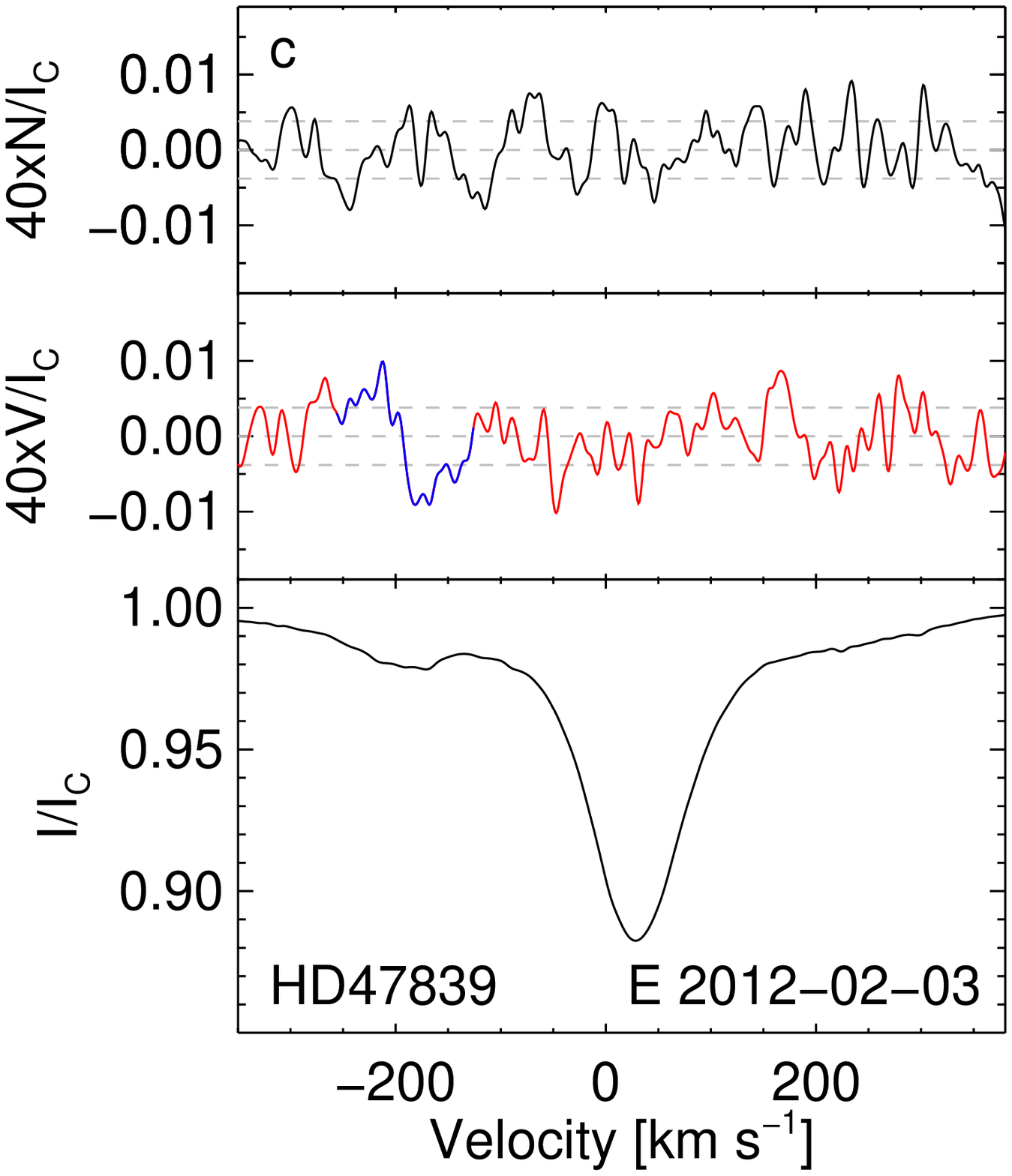}
\includegraphics[width=0.220\textwidth]{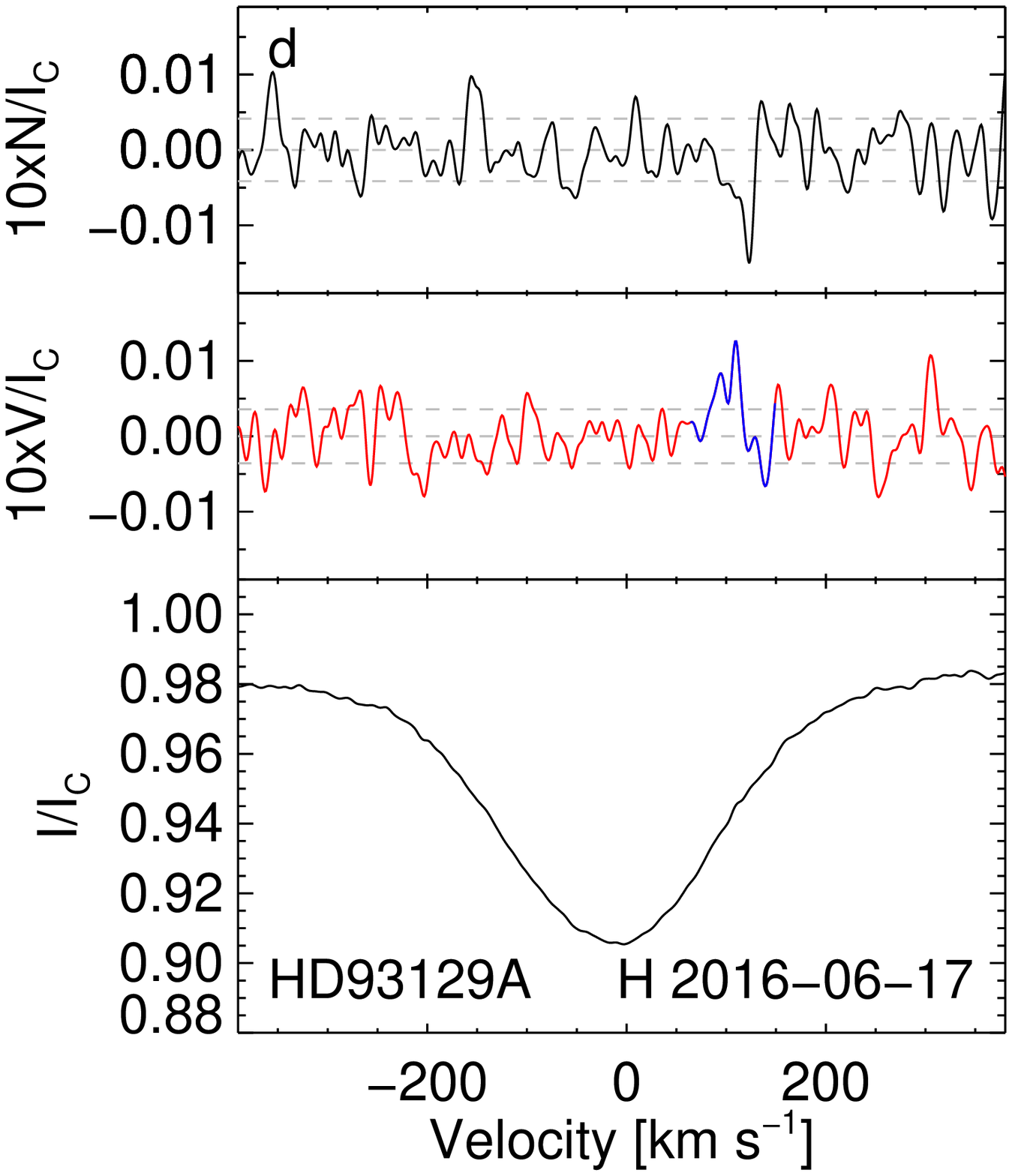}
\includegraphics[width=0.220\textwidth]{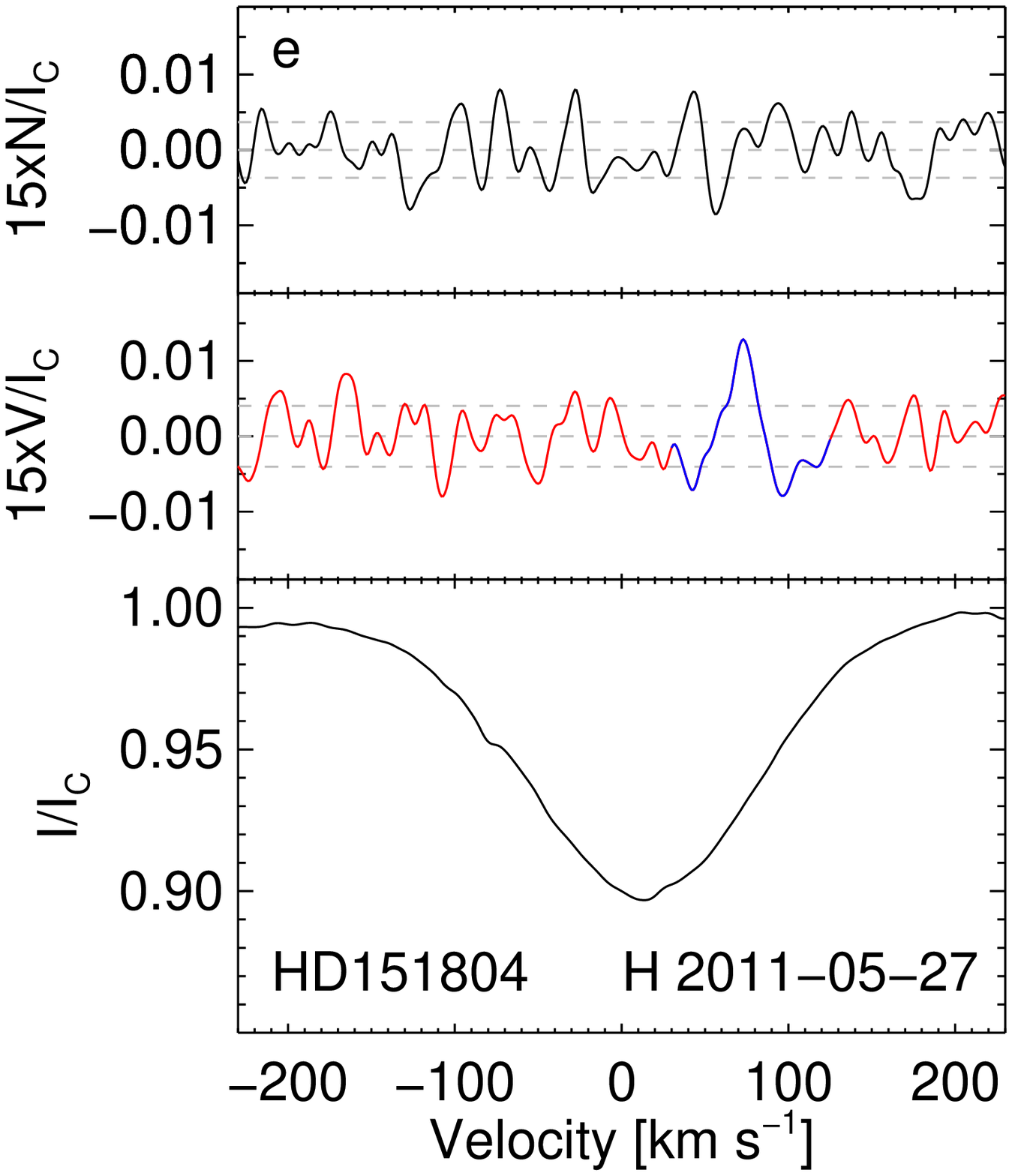}
\includegraphics[width=0.220\textwidth]{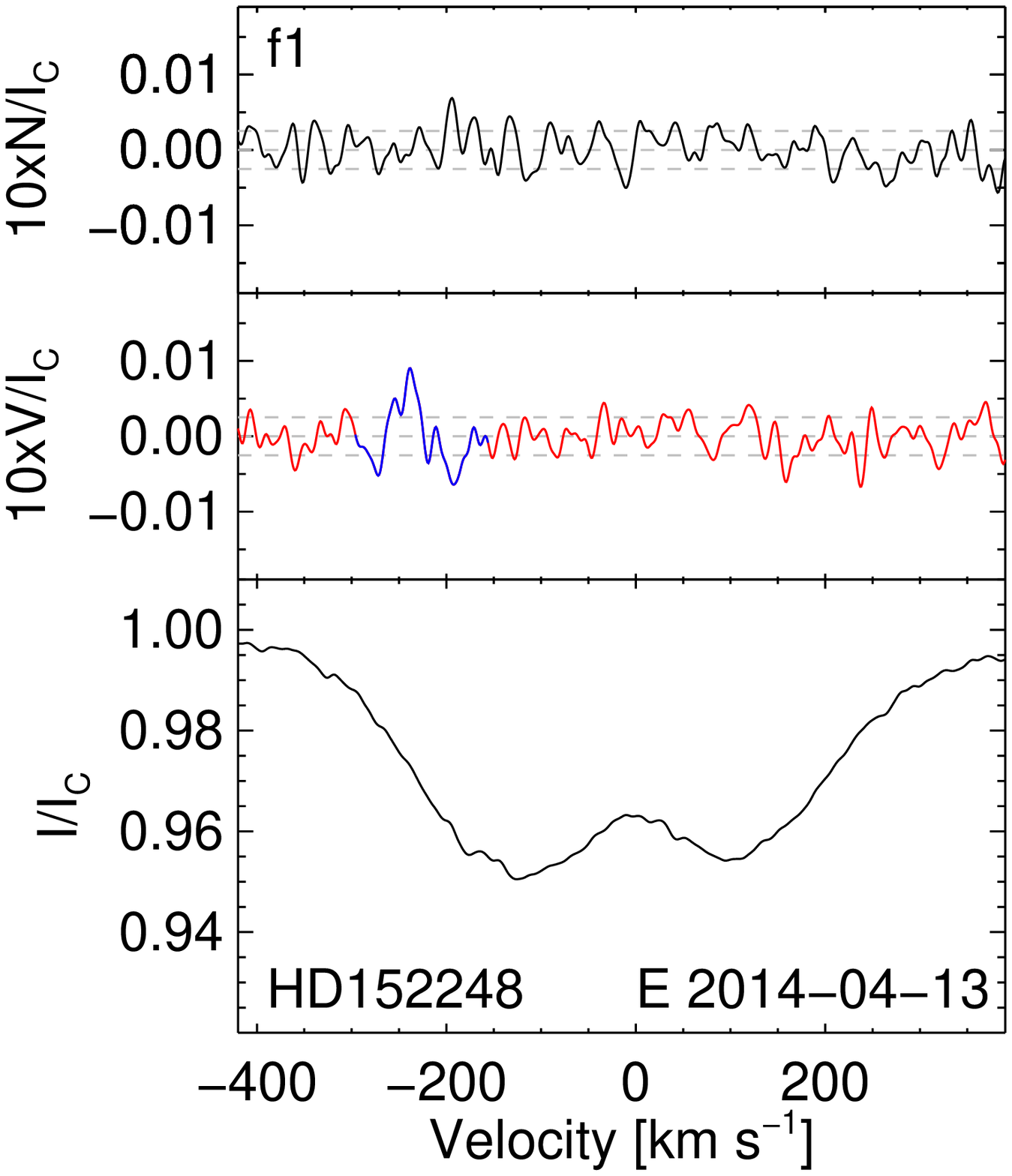}
\includegraphics[width=0.220\textwidth]{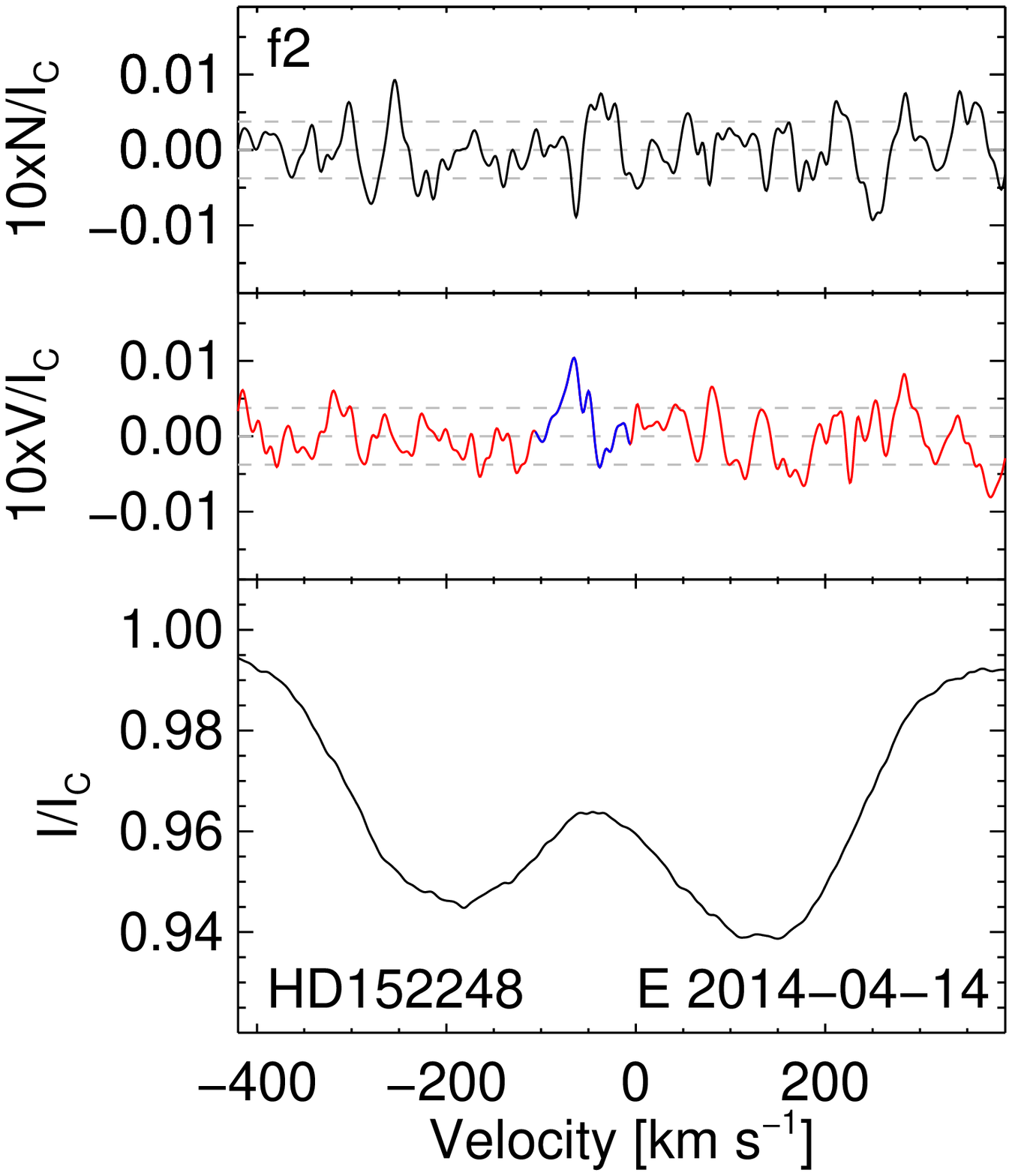}
\includegraphics[width=0.220\textwidth]{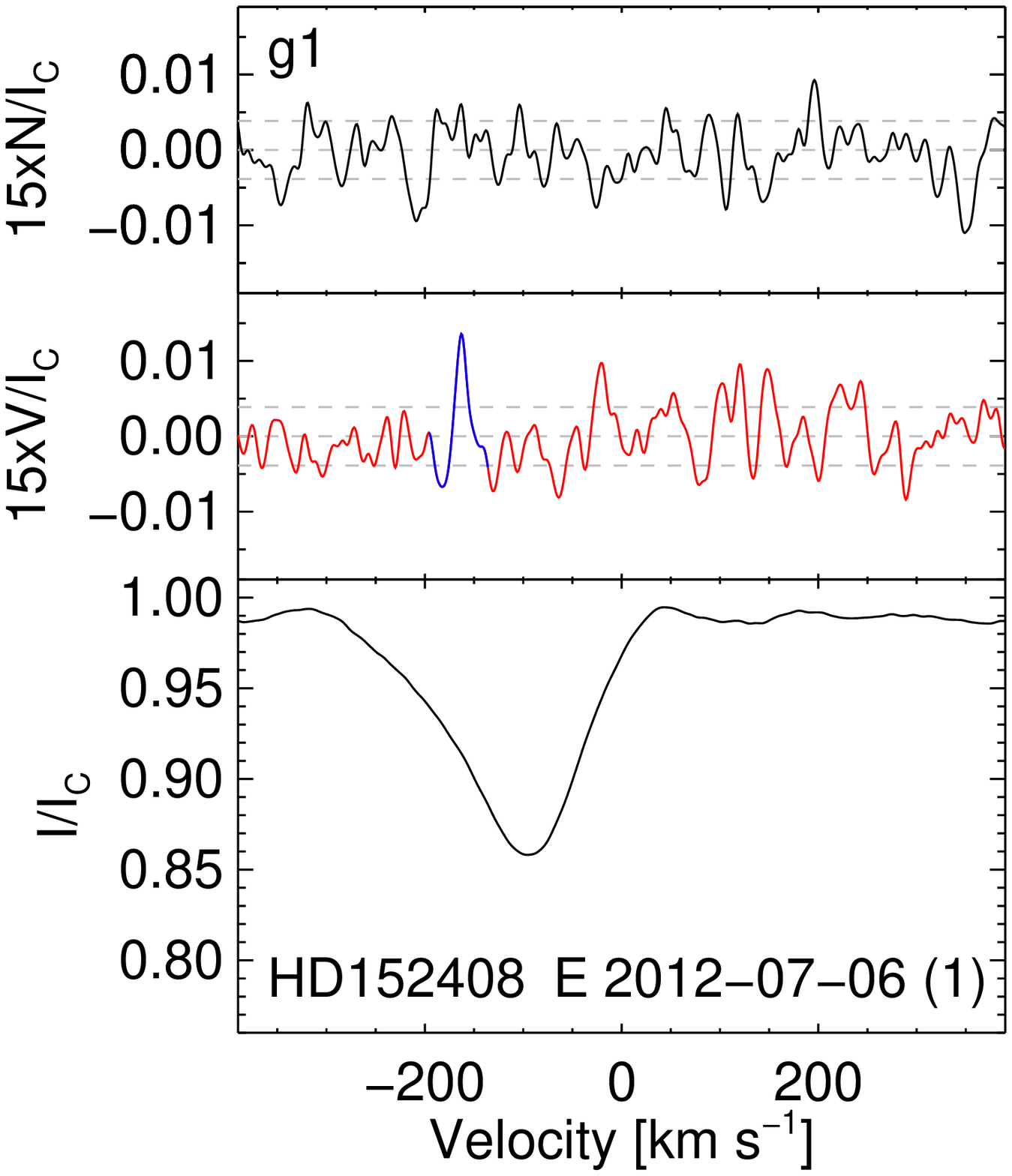}
\includegraphics[width=0.220\textwidth]{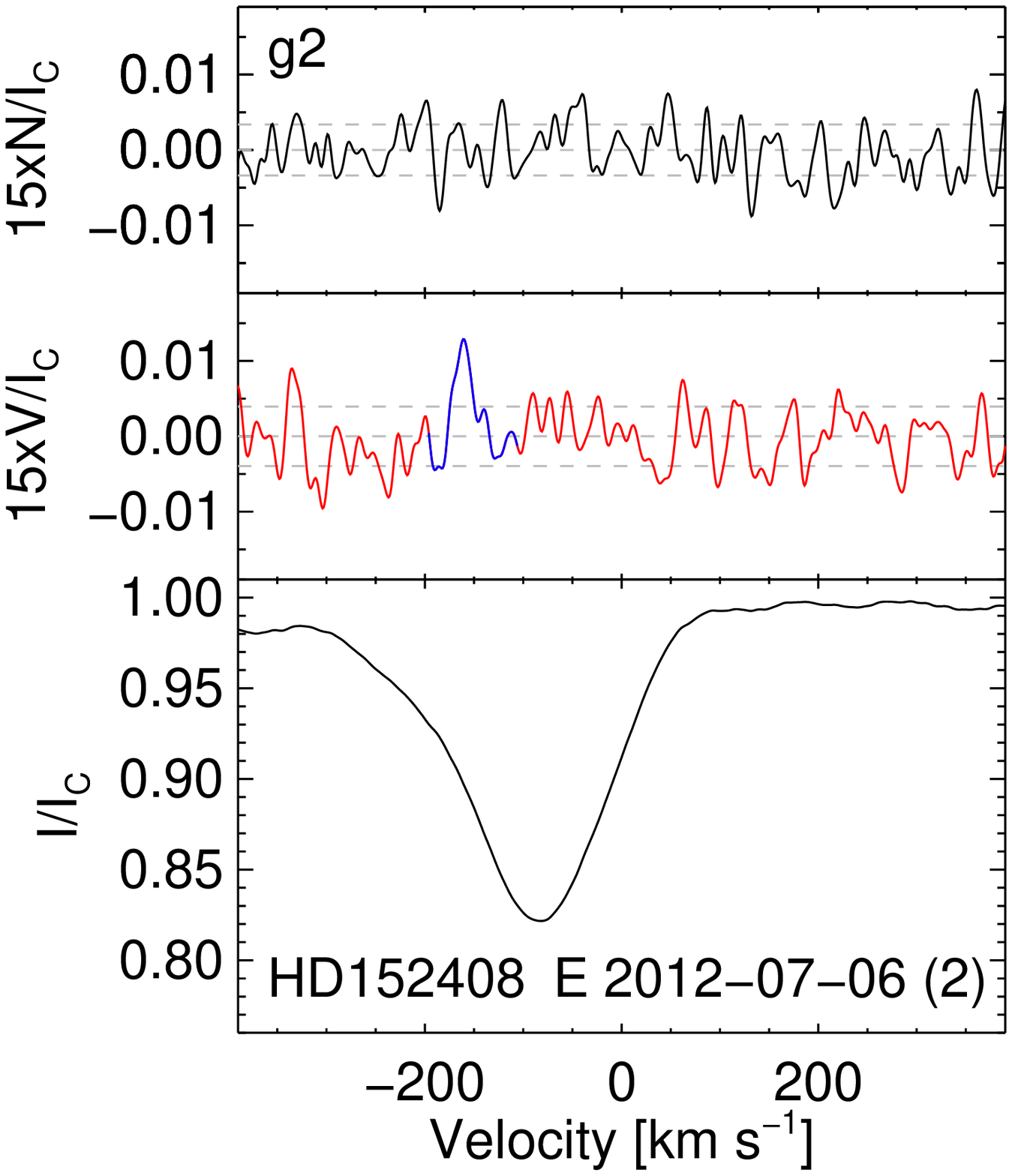}
\includegraphics[width=0.220\textwidth]{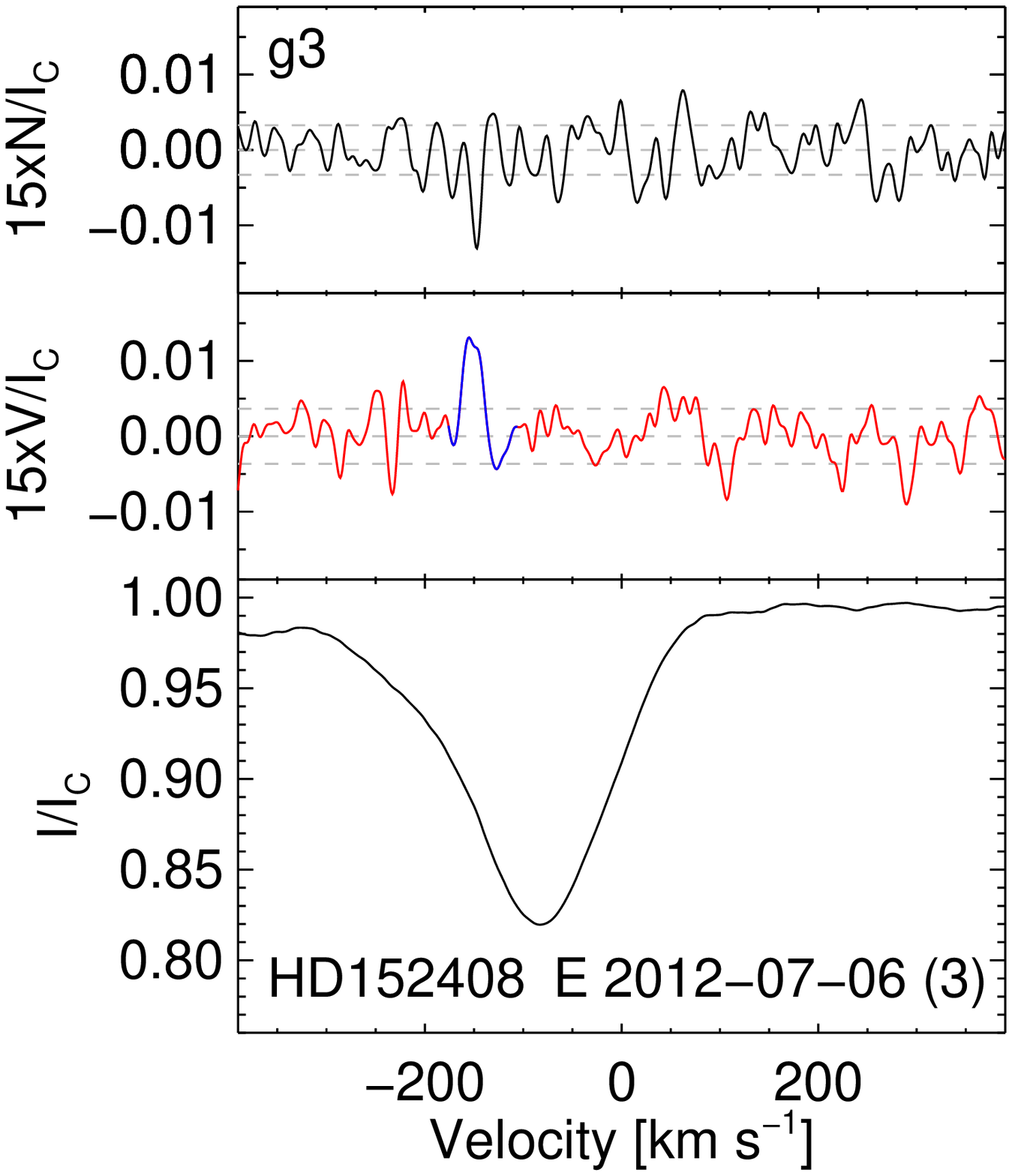}
\includegraphics[width=0.220\textwidth]{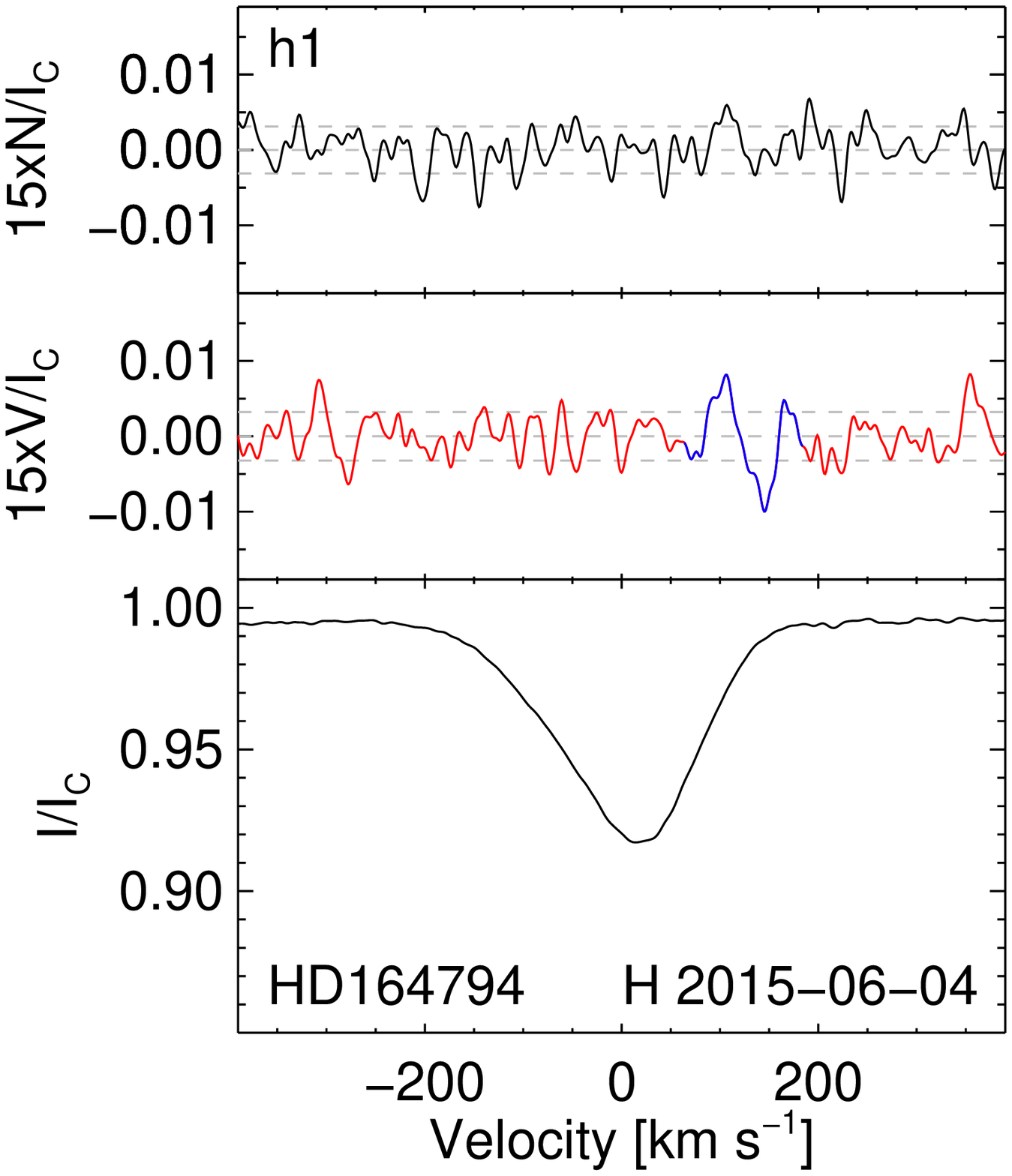}
\includegraphics[width=0.220\textwidth]{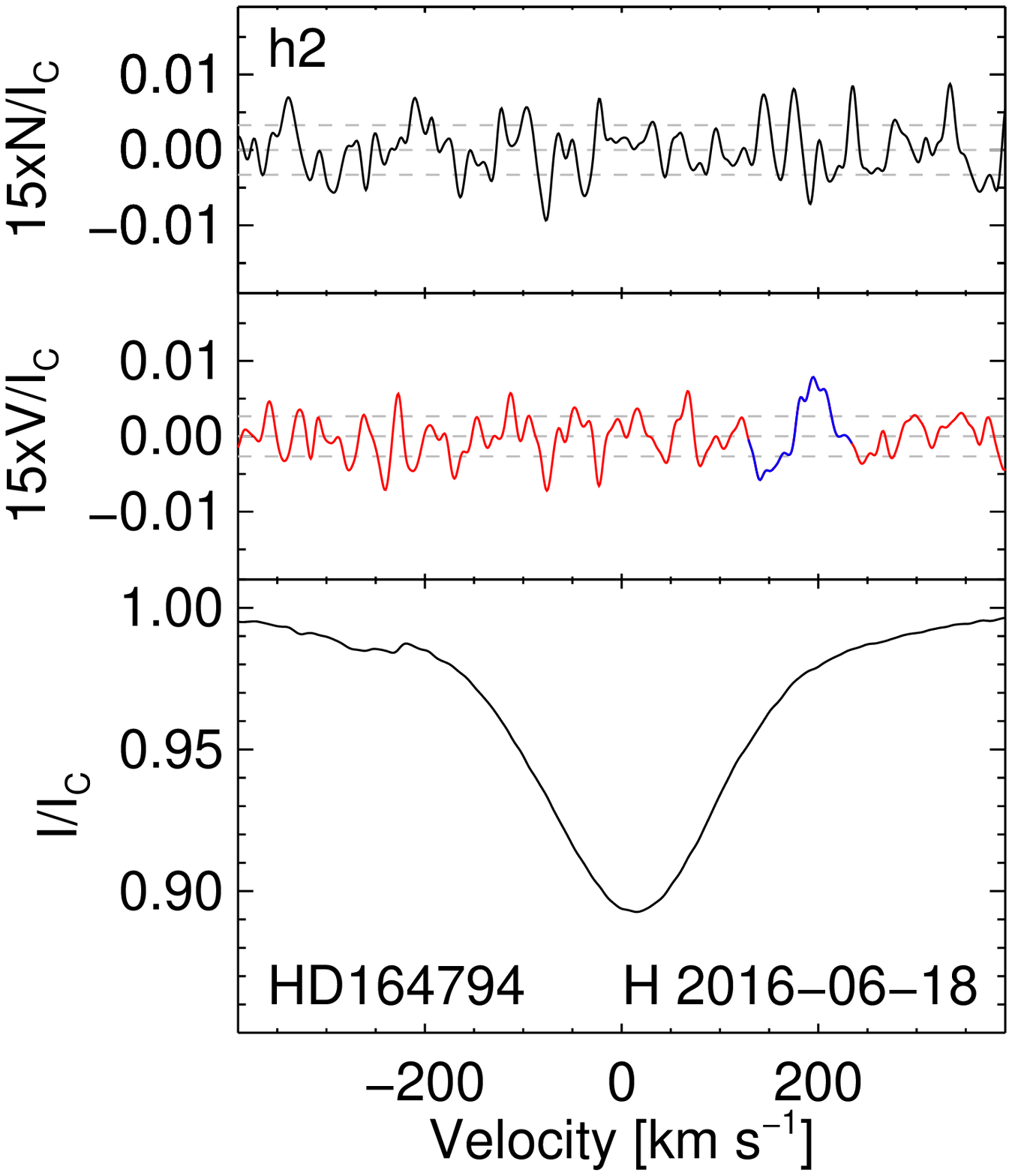}
\includegraphics[width=0.220\textwidth]{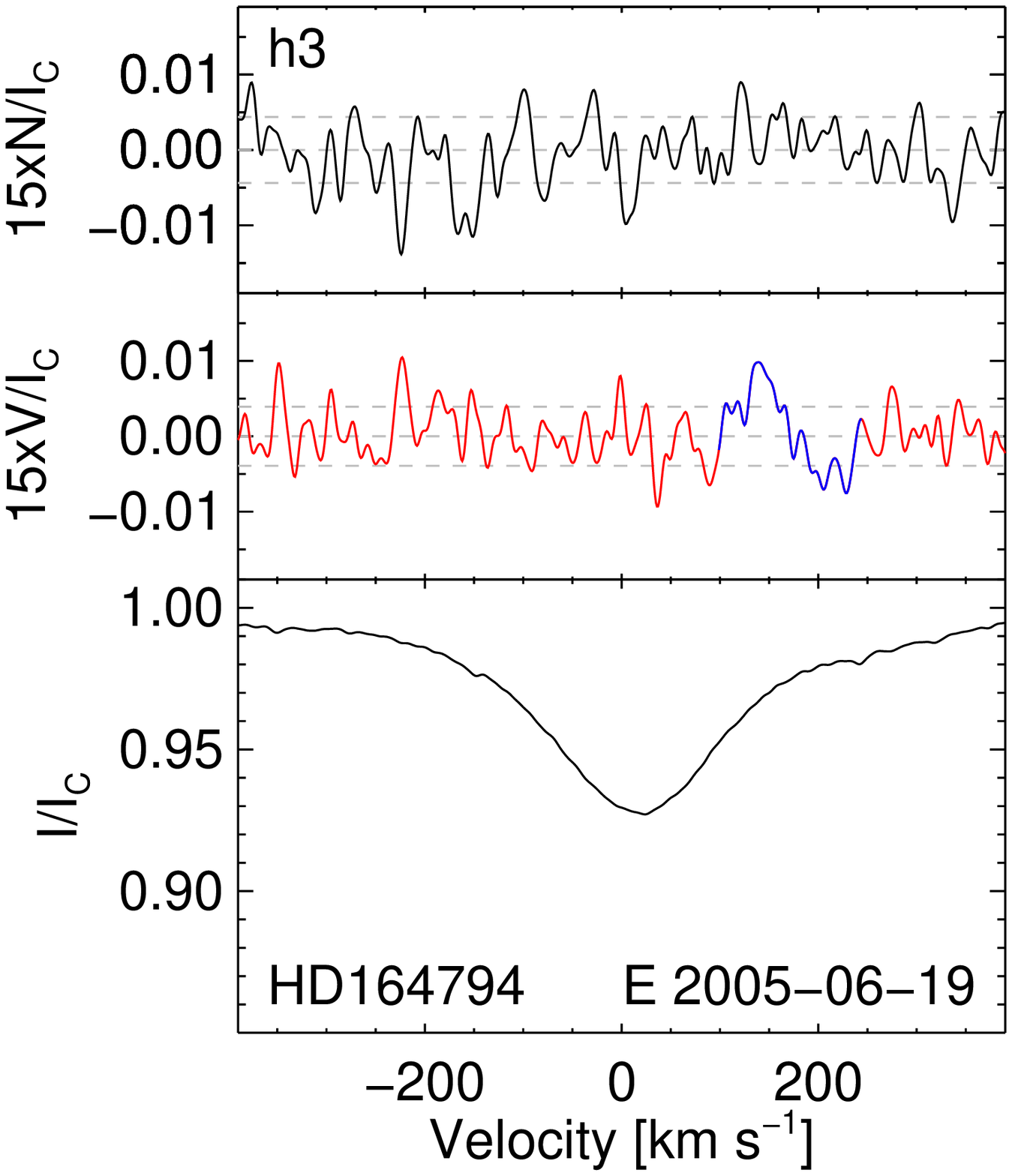}
\caption{
LSD Stokes~$I$, $V$, and diagnostic null $N$ spectra (from bottom to top)
calculated for particle-accelerating colliding-wind binaries (PACWBs).
The  Stokes~$V$ and $N$ spectra are magnified individually for better visibility.
Recognizable features identified in the Stokes~$V$ spectra are highlighted in blue.
a1-a3) LSD analysis results for HD\,36486 using
three HARPS\-pol observations obtained in 2013;
b1/b2) LSD analysis results for HD\,37468 using one HARPS\-pol observation from 2013
and one ESPaDOnS observation from 2014;
c) LSD analysis results for HD\,47839 using one ESPaDOnS observation from 2012;
d) LSD analysis results for HD\,93129A using one HARPS\-pol observation from 2016;
e) LSD analysis results for HD\,151804 using one HARPS\-pol observation of 2011;
f1/f2) LSD analysis results for HD\,152248 using two ESPaDOnS observations acquired in 2014;
g1-g3) LSD analysis results for HD\,152408 using three ESPaDOnS observations acquired in 2012 during the same night 
with a time lapse of 0.6\,h;
h1-h3) LSD analysis results for HD\,164794 using two HARPS\,pol observations acquired in 2015 and 2016, and one ESPaDOnS observations
from 2005.
}
\label{fig:collid}
\end{figure*}

\addtocounter{figure}{-1}

\begin{figure*}
\centering 
\includegraphics[width=0.220\textwidth]{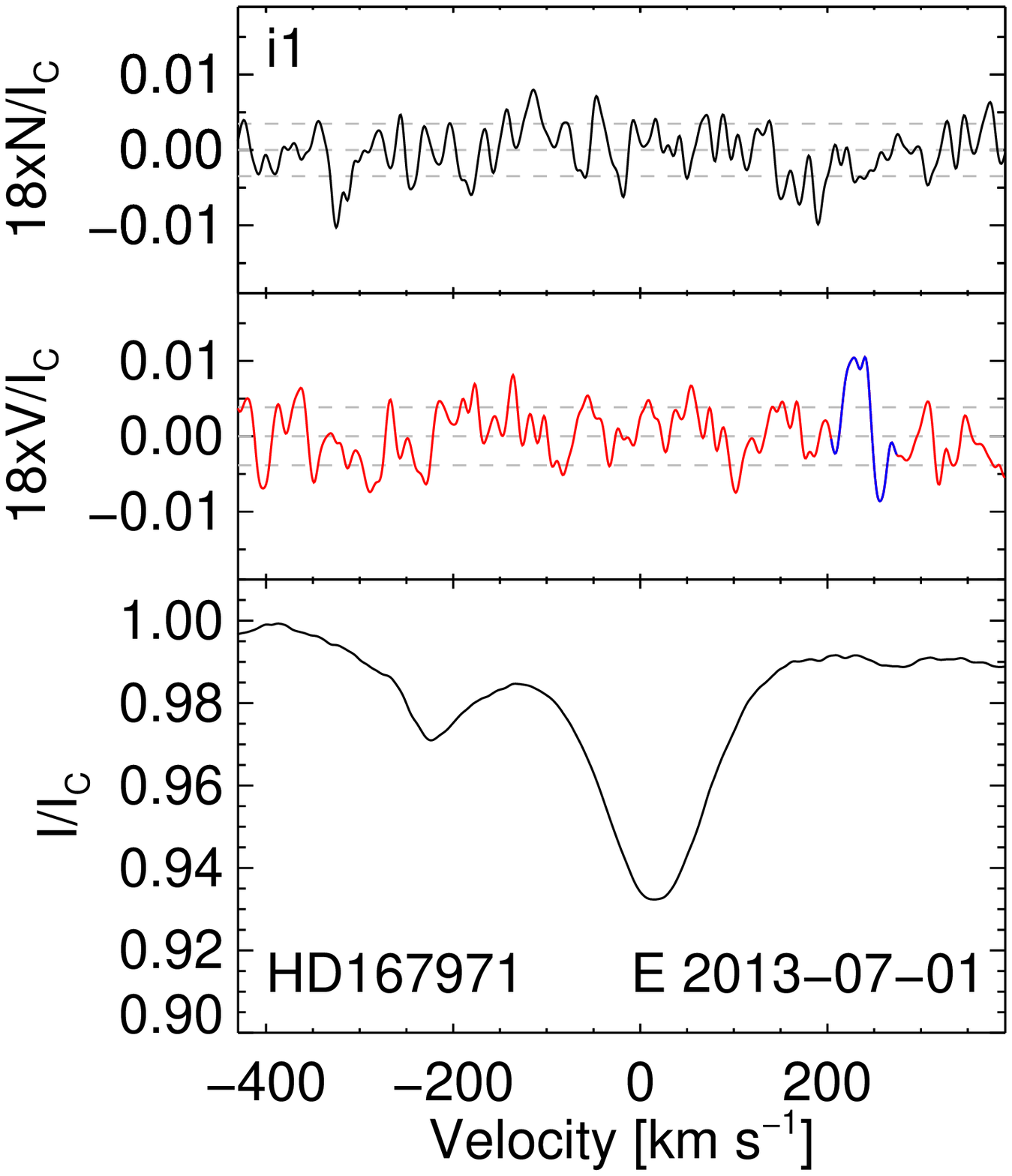}
\includegraphics[width=0.220\textwidth]{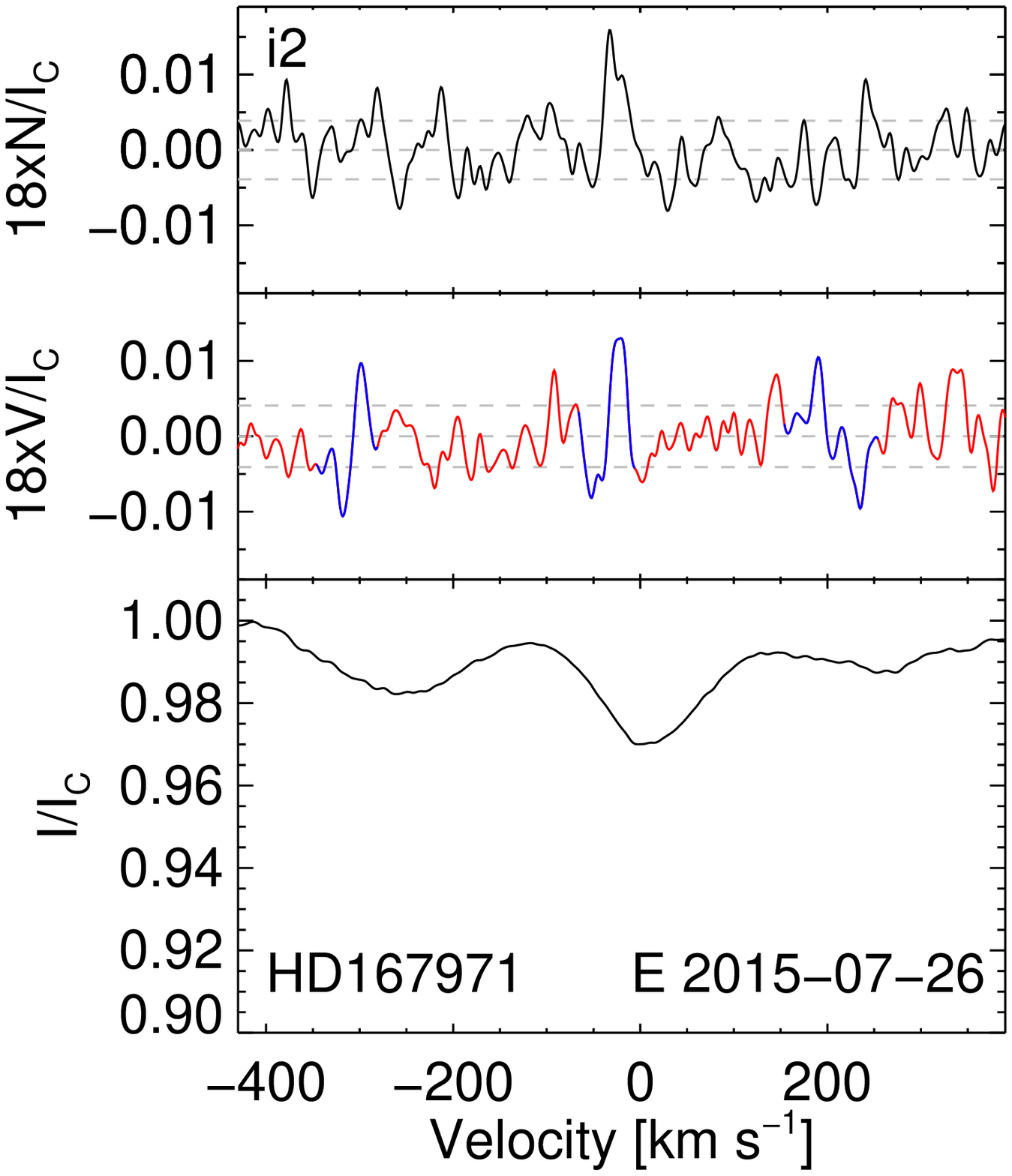}
\includegraphics[width=0.220\textwidth]{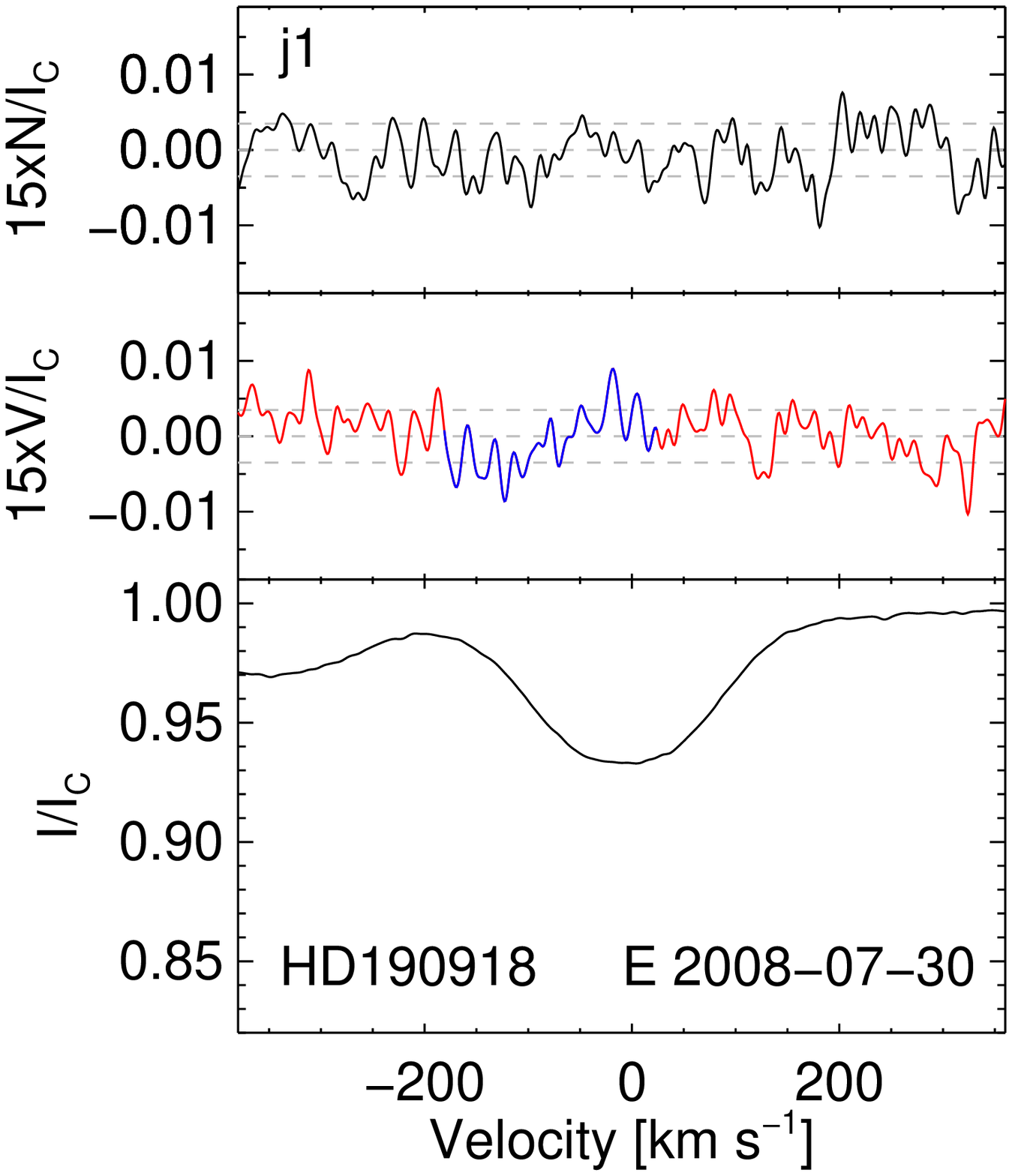}
\includegraphics[width=0.220\textwidth]{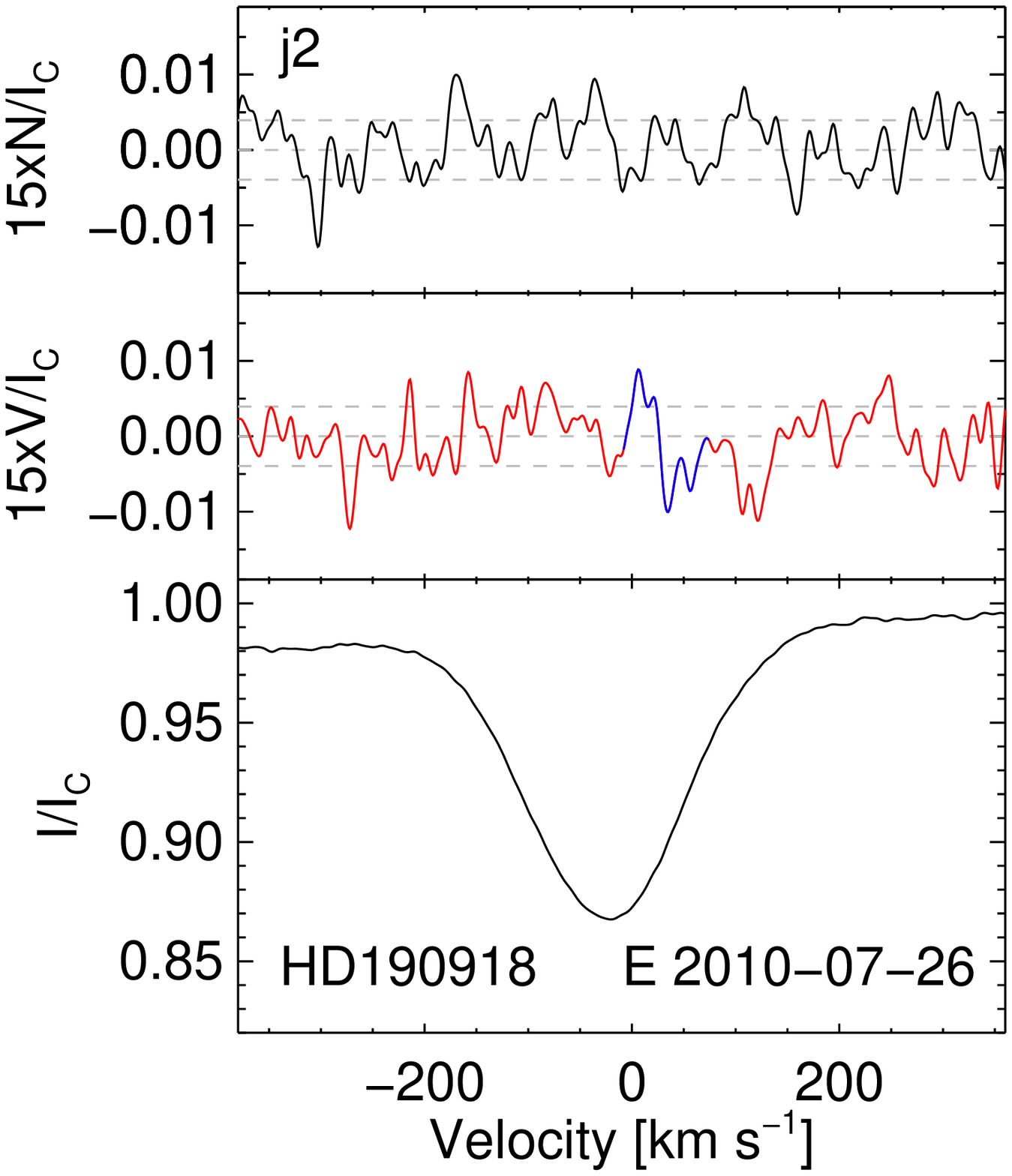}
\caption{
Continued:
i1/i2) LSD analysis results for HD\,167971 using two ESPaDOnS observations acquired in 2013 and 2015;
j1/j2) LSD analysis results for HD\,190918 using two ESPaDOnS observations of HD\,190918 from the years 2008 and 2010.
}
\label{fig:collid2}
\end{figure*}

\paragraph*{HD\,36486 ($\delta$\,Ori):}

According to \citet{Harvin2002}, this system consists of an eclipsing
binary with an orbital period of $P_{\rm orb}=5.732$\,d and a distant early B-type tertiary
at an angular separation of about 0.3\,arcsec relative to the binary with an orbital period of the order of several thousand days.
The authors suggested that the close binary may have suffered extensive mass loss through binary interaction.
The system was classified as a PACWB exhibiting synchrotron 
radio emission \citep{DeBeckerRaucq2013}. 
\citet{Neiner2015} analysed two Narval spectropolarimetric observations of HD\,36486, but reported no magnetic field detection.

This system was observed three time with HARPS\-pol in 2013 December, where the first two observations  were acquired on two
consecutive nights.
The LSD analysis of these spectropolarimetric observations
was carried out using a mask containing only \ion{He}{i} lines. 
As presented in Fig.~\ref{fig:collid} and Table~\ref{tab:obsall}, we detect a broad Zeeman feature in the second observation,
achieving a marginal detection
$\left< B_{\rm z} \right>=358\pm17$\,G with ${\rm FAP}=4\times10^{-4}$.
Taking into account figure~2 in \citet{Harvin2002}, this feature most probably 
corresponds to the O9.5\,II primary component.

\paragraph*{HD\,37468 ($\sigma$\,Ori):}

According to \citeauthor{MaizApellaniz2019} (\citeyear{MaizApellaniz2019}, and references therein),
this system is a SB2 system with a visual B0.2 V(n)
companion. The distant component is probably also an SB2 system \citep{MaizApellaniz2021}
in a nearly circular orbit around the close AaAb system with a period of 159.90\,yr and a current separation of 0.26\,arcsec.
The AaAb has an orbital period of 143.198\,d (\citealt{MaizApellaniz2018}, and references therein).
The system was mentioned as a PACWB exhibiting synchrotron 
radio emission  by \citet{DeBeckerRaucq2013}. \citet{Neiner2015} reported the absence
of a magnetic field in this system using one ESPaDOnS observation acquired in 2008. 

We confirm the field non-detection
from the same archival observations, but achieve marginal magnetic field detections
in more recent HARPS\-pol and ESPaDOnS observations.
As presented in Fig.~\ref{fig:collid} and Table~\ref{tab:obsall}, the LSD analysis of the HARPS\-pol observation acquired in 2013 
shows the presence of a Zeeman feature with ${\rm FAP}=8\times10^{-5}$
in the Stokes~$V$ profile calculated with a mask containing \ion{He}{i}, \ion{He}{ii}, \ion{Si}{iii}, and \ion{Si}{iv} lines.
For the ESPaDOnS observation acquired in 2014, we detect a feature with ${\rm FAP}=9\times10^{-4}$
in the Stokes~$V$ spectrum using a mask with \ion{He}{i} and \ion{He}{ii} lines.
Since the observed composite spectra show insufficient velocity separation of the system components, 
it is not clear to which component the detected Stokes~$V$ features belong, although their shifted positions relative to the
center of the corresponding Stokes~$I$ profiles seem to indicate the magnetic nature of the secondary B-type component. 

\paragraph*{HD\,47839 (15\,Mon):}

According to \citeauthor{MaizApellaniz2019} (\citeyear{MaizApellaniz2019}, and references therein), this system is a 
long-period SB1 binary in the cluster NGC\,2264. From the analysis of TESS data, \citet{Burssens2020}
found in HD\,47839 the presence of stochastic low-frequency (SLF) photometric and $\beta$\,Cep variability,
where the $\beta$\,Cep pulsations most probably belong to the
fast-rotating early-type B component. \citet{Kaper1996}
and \citet{Walborn2006} reported that this system exhibits distinct peculiarities in its spectrum,
which could be typical for stars possessing magnetic fields.
HD\,47839 was classified as a PACWB exhibiting synchrotron 
radio emission \citep{DeBeckerRaucq2013}. \citet{Neiner2015} used several Narval and one ESPaDOnS observations
to search for the presence of 
a magnetic field and reported that no field was detected. 

We downloaded from the CFHT archive the same ESPaDOnS
observation from 2012 and detect, using a mask containing \ion{He}{i}, \ion{He}{ii}, and \ion{C}{iv} lines, a clear
Zeeman feature with ${\rm FAP}<10^{-10}$ corresponding to the B1.5/2V component.
As presented in Table~\ref{tab:obsall} and in Fig.~\ref{fig:collid}, we achieve a definite detection
of the mean longitudinal magnetic field $\left< B_{\rm z} \right>=282\pm58$\,G.
The previous  measurements of the longitudinal magnetic field in HD\,47839 by \citet{Hubrig2013b} using the full
FORS\,1 spectrum coverage  showed a magnetic field of positive polarity
at a 2.6$\sigma$ significance level, $\left< B_{\rm z} \right>=134\pm52$\,G,
whereas a magnetic field of the order
of $-$160\,G  at a significance level of 4.4$\sigma$ was reported using observations with
the SOFIN spectropolarimeter attached to the 2.6\,m Nordic Optical Telescope.

\paragraph*{HD\,93129A:}

According to \citet{BenagliaKoribalski2004}, this visual multiple system stands out as
the earliest, most luminous, and most massive star known in our Galaxy.
Only 2.6\,arcsec from HD\,93129A, \citet{Walborn1973} discovered a second star, HD\,93129B, of 
spectral type O3.5\,V((f)).
The most recent study by \citet{Sana2014} showed that the separation between the Aa and Ab components is about 30\,mas.
This system dominates the core of Trumpler 14 and, in total, six  companions within  5\,arcsec from the central  star
were reported in the literature (e.g.\ \citealt{Sana2014}).
HD\,93129 was identified as a PACWB by \citet{DeBeckerRaucq2013}.

The LSD analysis of the single HARPS\-pol observation of this system 
obtained in 2016 was carried out using a mask containing \ion{He}{ii} lines.
As presented in Table~\ref{tab:obsall} and in Fig.~\ref{fig:collid}, we achieve a definite detection
with ${\rm FAP}<10^{-10}$. Given the location of the detected Zeeman feature in the Stokes~$V$ spectrum relative to the
broad Stokes~$I$ line profile presenting a blend composed of the Aa and Ab components, it is possible that
the secondary component with spectral type O3.5V((f)) possesses a magnetic field.

\paragraph*{HD\,151804 (HR\,6245):} 

This system was classified as a PACWB exhibiting synchrotron 
radio emission by \citet{DeBeckerRaucq2013}.
\citet{Kervella2022} show a significant detection of a companion using Gaia EDR3 data ($S/N = 5.2$)
but not using Gaia DR2 data 
($S/N = 0.82$). This star falls into the 9 per cent of their sample that has an improved $S/N$ after the release of EDR3.
According to \citet{Burssens2020}, this system shows SLF photometrical  variability.

HD\,151804 was previously observed with FORS\,1 in spectropolarimetric mode by \citet{Hubrig2008}.
However, no field detection at a significance level of 3$\sigma$ was achieved in this study. 
Also, no magnetic field was detected in the study of \citet{Neiner2015}, who analysed one HARPS\-pol observation acquired in 2011.
As presented in Table~\ref{tab:obsall} and in Fig.~\ref{fig:collid}, analysing the same observation, we achieved a marginal detection with
${\rm FAP}=6\times10^{-4}$ using a 
mask containing \ion{He}{i} and \ion{He}{ii} lines.
Given the location of the detected feature in the Stokes~$V$ spectrum relative to the
broad Stokes~$I$ line profile, it is not clear
whether this feature is due to the presence of a background or foreground object or an unresolved second body in the system. 

\paragraph*{HD\,152248 (V1007\,Sco):}

According to \citet{Sana2001}, this PACWB eccentric eclipsing SB2 binary
has an orbital period close to six days \citep{Hill1974}.
It belongs to the young rich open cluster NGC\,6231.
Both components are close to filling their Roche lobe at periastron passage \citep{Sana2001}.
However, the authors do not find  any  evidence  that  mass transfer is currently  taking  place in this system. On the other hand,
it cannot be excluded that such a process has already been taking place in the past.

As presented in Table~\ref{tab:obsall} and in Fig.~\ref{fig:collid}, using ESPaDOnS observations of HD\,152248 obtained in
2013 and 2014, we detect distinct features in the LSD Stokes~$V$ spectra with ${\rm FAP}<10^{-10}$ using a mask
with \ion{He}{i}, \ion{He}{ii}, and \ion{Si}{iv} lines for the first observation
and ${\rm FAP}=4\times10^{-7}$  using a mask containing \ion{He}{i}, \ion{He}{ii},
and \ion{O}{iii} lines for the second observation. Given the change of the position of the detected features in 
velocity space from about $-$230 to $-$50\,km\,s$^{-1}$ just over one day, and the fact that these features do not correspond to
the position of the LSD Stokes~$I$ profiles of any of the O-type components, it is possible that a yet undetected third component
is present in this system.

\paragraph*{HD\,152408 (HR\,6272, WR79a):} 

This target is a member of NGC\,6231 and was classified as O8Iape by \cite{Sota2014}.
It is listed in the Seventh Catalogue of Galactic Wolf-Rayet stars presented by \citet{vanderHucht2001}.
\citeauthor{Skinner2010} (\citeyear{Skinner2010}, and references therein) reported on the X-ray detection in HD\,152408 with XMM-Newton,
emphasising that this detection is of special importance as it is achieved in the object belonging to the latest WN subtype
in the WN\,7--9 range. The authors also reported on the existence of a fainter near-IR source (2MASSJ165458.09-410903.6) at an
offset of 4.8\,arcsec from HD\,152408 (see also \citealt{Mason1998}). 
According to \citet{DeBeckerRaucq2013}, this target is a PACWB.
No magnetic field was detected  by \citet{Neiner2015} using three consecutive ESPaDOnS observations
obtained during the same night on 2012 July~6.

\citet{Hubrig2008}
reported on the detection of a mean longitudinal magnetic field of the order of 90\,G at a significance level of 3$\sigma$ using 
low-resolution FORS\,1 spectropolarimetric observations.
For our LSD analysis we used the same three consecutive ESPaDOnS observations with a time lapse of 0.56\,h over 1.7\,h
during the same night as \citet{Neiner2015} and a mask
containing \ion{He}{i} and \ion{He}{ii} lines for the first observation and a mask containing \ion{He}{ii} and
\ion{C}{iv} lines for the second and the third observations. As presented in Table~\ref{tab:obsall} and
in Fig.~\ref{fig:collid}, we detect in the LSD Stokes~$V$ spectra small Zeeman features  with ${\rm FAP}=5\times10^{-4}$
for the first observations and ${\rm FAP}<10^{-10}$ for both the second and third observations, indicating the magnetic nature of
this target.
Given the slightly decentered position of the detected features relative to the observed LSD Stokes~$I$ profile,
it is quite possible that the detected Zeeman features correspond to an unresolved second body.

\paragraph*{HD\,164794 (9 Sgr):}

According to \citet{Fabry2021}, this target is an astrometric SB2 system in the Lagoon Nebula with a 53\,\msun{}
primary and a 39\,\msun{} secondary with an orbital
period of 3261\,d. The system was classified as a PACWB exhibiting synchrotron 
radio emission by \citet{DeBeckerRaucq2013}. \citet{Rauw2016} reported that a maximum in the X-ray
emission during the periastron passage is coming from shocked gas in the interaction zone of the stellar winds.
\citet{Neiner2015} have not detected the presence of a magnetic field in this system using several
ESPaDOnS observations and one HARPS\-pol observation.

The presence of a weak magnetic field of about 200\,G was reported by \citet{Hubrig2008}
using low-resolution spectropolarimetric observations with FORS\,1 at the VLT. Also subsequent high-resolution SOFIN
and HARPS\-pol observations analysed using the moment technique indicated the presence of a magnetic field of a similar 
order \citep{Hubrig2013b}.
As presented in Table~\ref{tab:obsall} and
in Fig.~\ref{fig:collid}, our LSD analysis of the two high-resolution HARPS\-pol observations acquired in 2015 and 2016
and one ESPaDOnS observation from 2005 shows for all these observations definite detections  of clear Zeeman features of 
positive and negative field polarities  with ${\rm FAP}<10^{-10}$.
For the first HARPS\-pol observation and the ESPaDOnS observation we used a mask containing 
\ion{He}{i}, \ion{He}{ii}, \ion{N}{iv}, and \ion{O}{iii} lines whereas for the second HARPS\-pol observation we used a mask
containing \ion{He}{ii} and \ion{N}{iv} lines.
As the components in HD\,164794 appear overlapped in all LSD Stokes~$I$ spectra and their $K{\rm 1}$ and $K{\rm 2}$ 
amplitudes are rather low, 25\,km\,s$^{-1}$ for the primary and 39\,km\,s$^{-1}$ for the secondary \citep{Rauw2016},  
it is difficult to decide about the origin
of these features shifted  to the red relative to the centers of the Stokes~$I$ profiles. We believe that the origin of these
features is not related to a magnetic background or foreground object:
as is shown in  Fig.~\ref{fig:collid}, their position in velocity space is slightly changing over the years
in the range $150-180$\,km\,s$^{-1}$ and thus we cannot exclude the presence of an additional component in this system.

\paragraph*{HD\,167971 (MY\,Ser):} 

According to \citet{Ibanoglu2013}
this SB3 system is one of the rare massive O-type triple systems where the secondary and the tertiary components compose an eclipsing 
binary.
It is the brightest member in the NGC\,6604 cluster embedded in the Ser\,OB2
association. The brighter component of
this triple system is an overcontact eclipsing binary with an orbital period of 3.32\,d, just filling the entire outer contact surface
with a fill-out factor of 0.99. The outer pair has a period of about 22\,yr.
HD\,167971  was classified as a PACWB exhibiting synchrotron 
radio emission by \citet{DeBeckerRaucq2013}. \citet{Neiner2015} reported a magnetic field non-detection 
using one ESPaDOnS spectrum acquired in 2013.

For our LSD analysis we used two archival ESPaDOnS spectra, one from 2013 and another one from 2015.
As presented in Table~\ref{tab:obsall} and
in Fig.~\ref{fig:collid2}, all three components are well visible in both ESPaDOnS observations.
For the first observation we detect the presence of a Zeeman signature with ${\rm FAP}<10^{-10}$ only in the weakest
red component and measure a very strong magnetic field $\left< B_{\rm z} \right>= 1324\pm582$\,G.
Given the 
rather large uncertainty affecting this measurement, it is necessary to confirm the presence of such a strong magnetic field
with future observations obtained at higher $S/N$.
Zeeman features for all three components are detected in the second ESPaDOnS observation where a marginal
detection with ${\rm FAP}=9\times10^{-4}$ corresponding to $\left< B_{\rm z} \right>=-977\pm437$\,G is achieved for the blue component,
and a definite detection with ${\rm FAP}=6\times10^{-6}$ corresponding to $\left< B_{\rm z} \right>=-57\pm33$\,G
is achieved for the central component showing the deepest intensity profile in the Stokes~$I$ spectrum. No detection is achieved 
for the third component. 
We also detect a structure  
in the diagnostic null $N$ spectrum calculated for the second observation. The origin of this structure corresponding to the central
component is probably related to the presence of pulsations.
For both observations we used a line mask containing \ion{He}{ii}, \ion{C}{iv}, and \ion{N}{iii} lines.

\paragraph*{HD\,190918 (WR\,133):} 

This SB2 system, also known as WR\,133, was classified as a PACWB exhibiting synchrotron 
radio emission by \citet{DeBeckerRaucq2013}. 
According to \citet{Richardson2021},
the orbital period is 112.8\,d with a moderate eccentricity of 0.36,
and a separation of 0.79\,mas on the sky. 
Since the derived masses are low compared to the spectral types of the components, the authors 
suggest that WR\,133 should have been formed through binary interactions.

This system was mentioned as a non-magnetic PACWB by \citet{Neiner2015}, who used one ESPaDOnS observation acquired in 2010.
We used for our LSD analysis the same ESPaDOnS observation from 2010 and another one from 2008 using a mask with 
\ion{He}{i} and \ion{He}{ii} lines. As presented in Table~\ref{tab:obsall} and
in Fig.~\ref{fig:collid2}, there is evidence for the presence of a definite Zeeman feature in the observation acquired in 2008
with ${\rm FAP}<10^{-10}$ corresponding to $\left< B_{\rm z} \right>=-355\pm44$\,G. 
A marginal detection with ${\rm FAP}=3\times10^{-4}$ corresponding to
$\left< B_{\rm z} \right>=56\pm8$\,G is achieved for the
observation acquired in 2010. 
Since the components of this system are overlapped in the
LSD Stokes~$I$ spectra, it is not clear to which component the detected features belong.
Notably, studies of systems with WR components are of special interest: according to \citet{Dsilva2022}, the multiplicity properties of WR stars, 
which are the immediate progenitors of black holes, directly affect the properties of
black-hole binaries  (e.g.\ through  black-hole  kicks in a core-collapse  scenario), and
hence the predictions for gravitational-wave progenitors.

\subsection{X-ray binaries with colliding winds}
\label{subsect:xrb}

\begin{figure*}
\centering 
\includegraphics[width=0.220\textwidth]{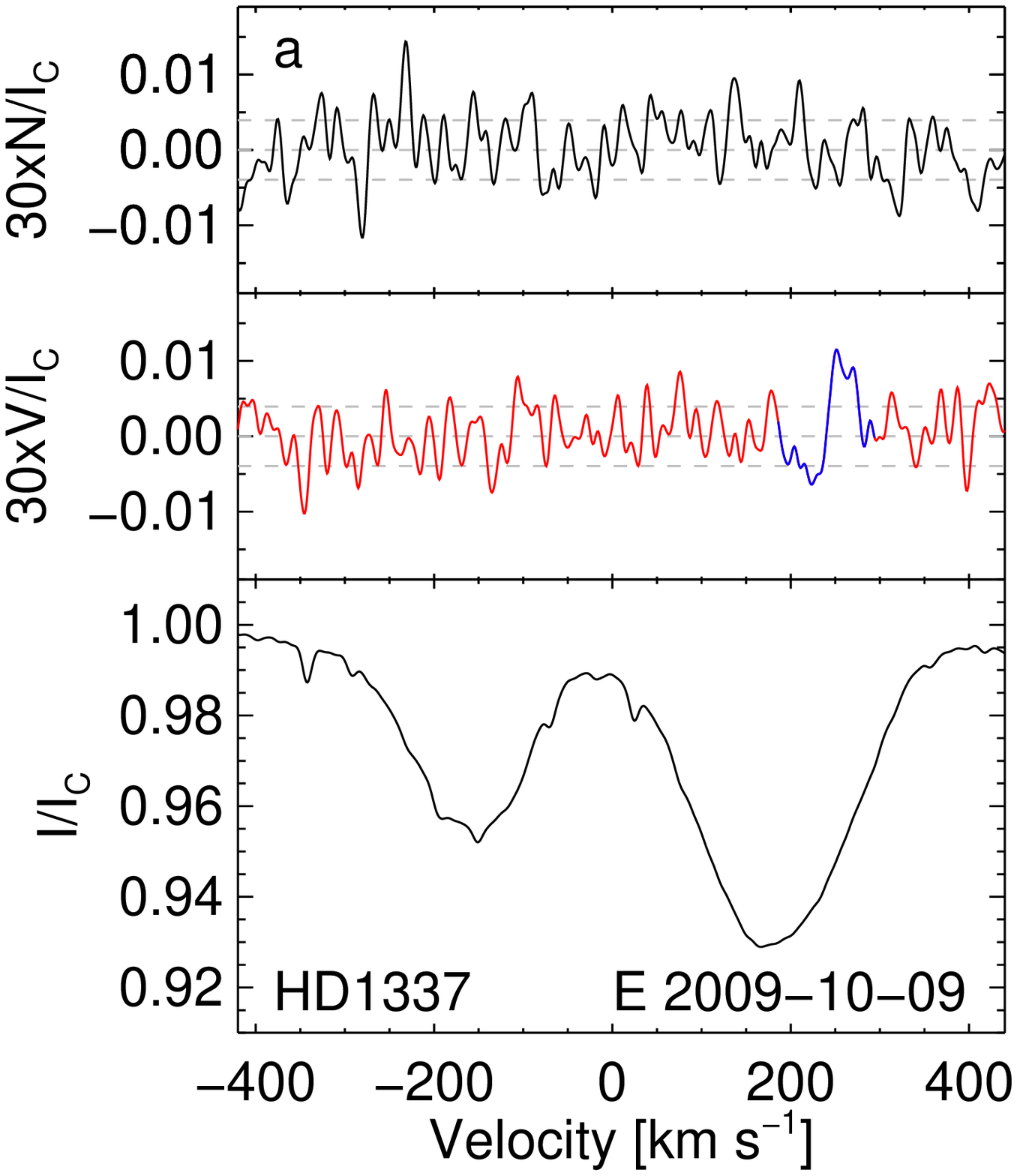}
\includegraphics[width=0.220\textwidth]{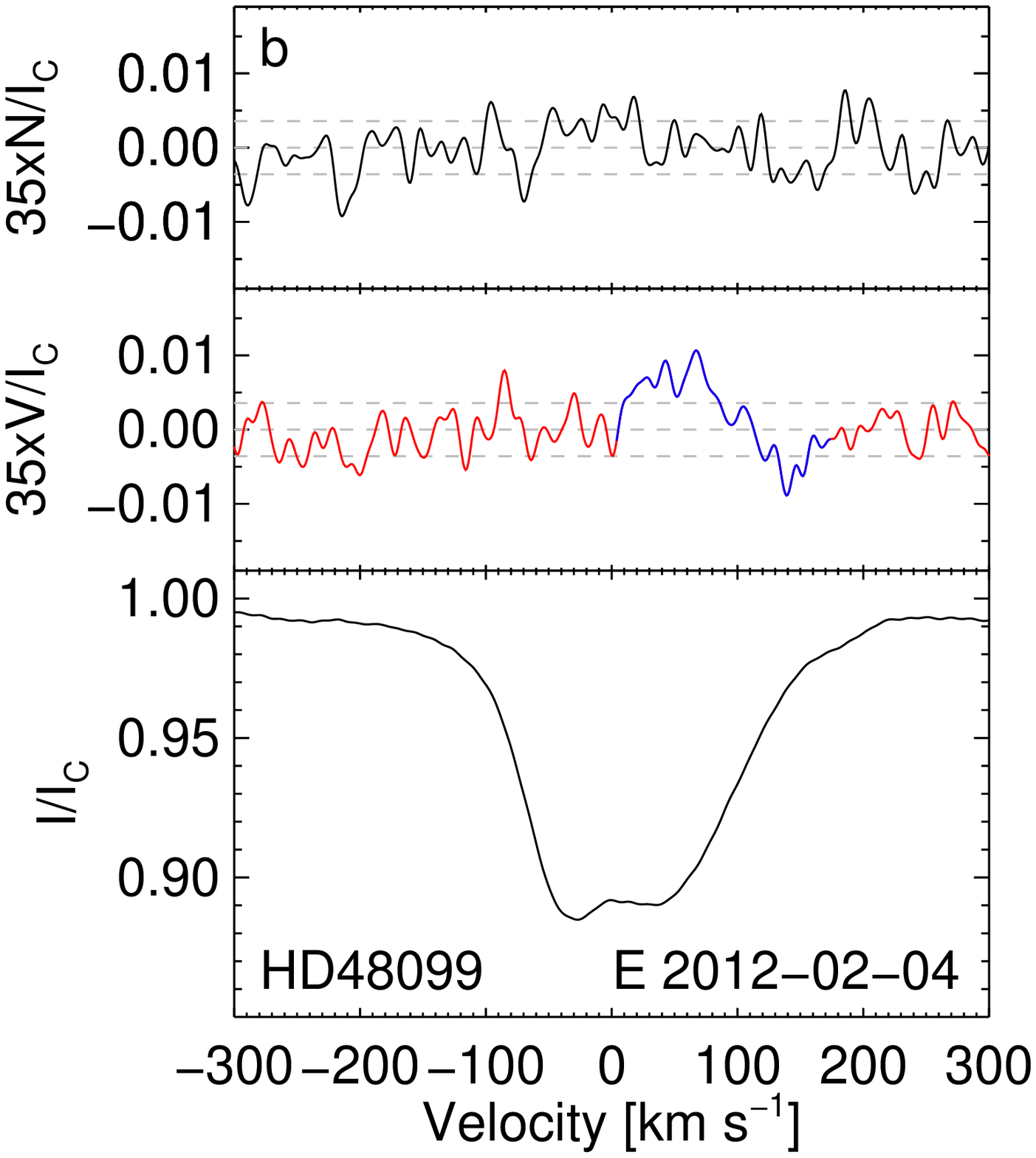}
\includegraphics[width=0.220\textwidth]{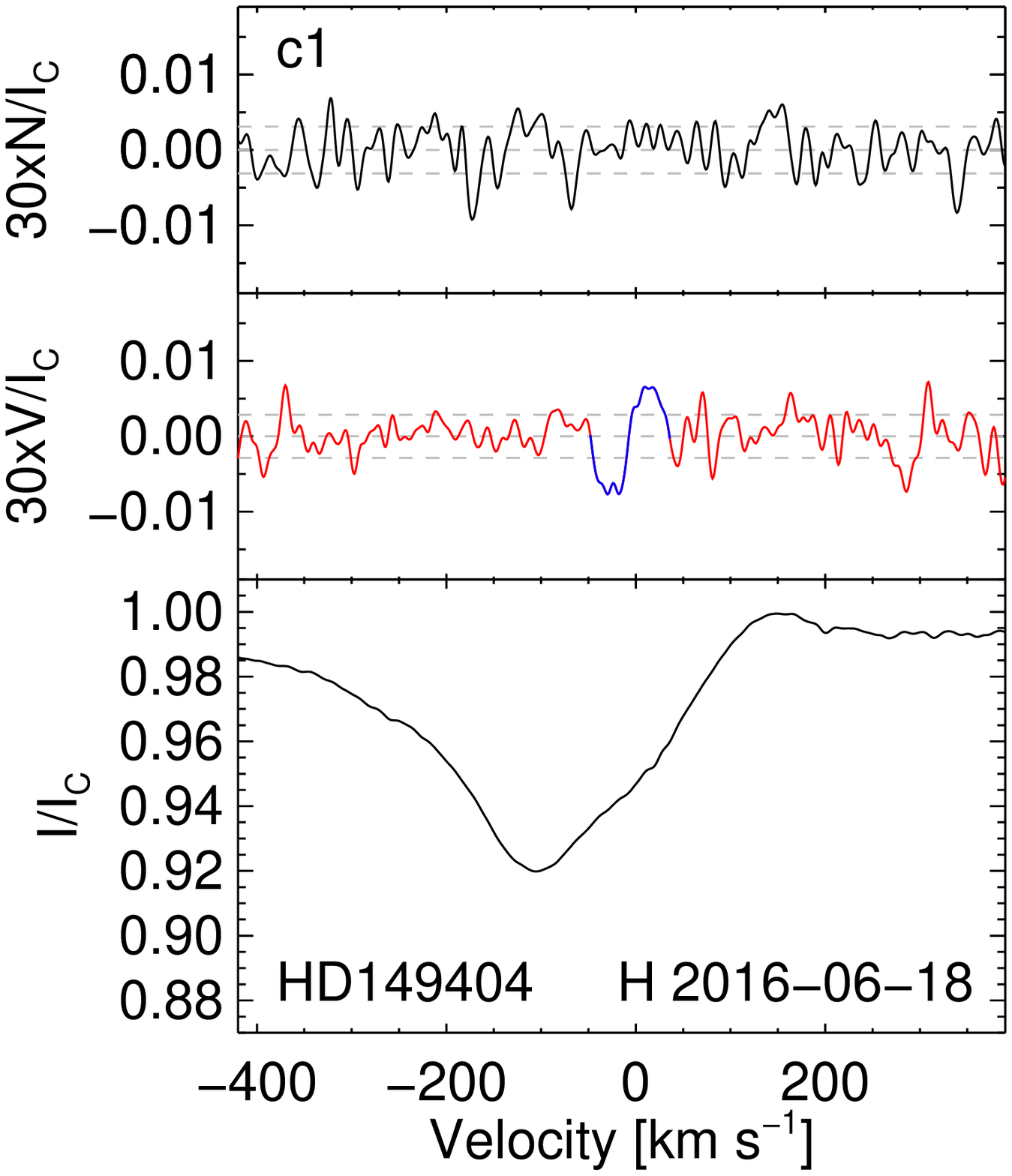}
\includegraphics[width=0.220\textwidth]{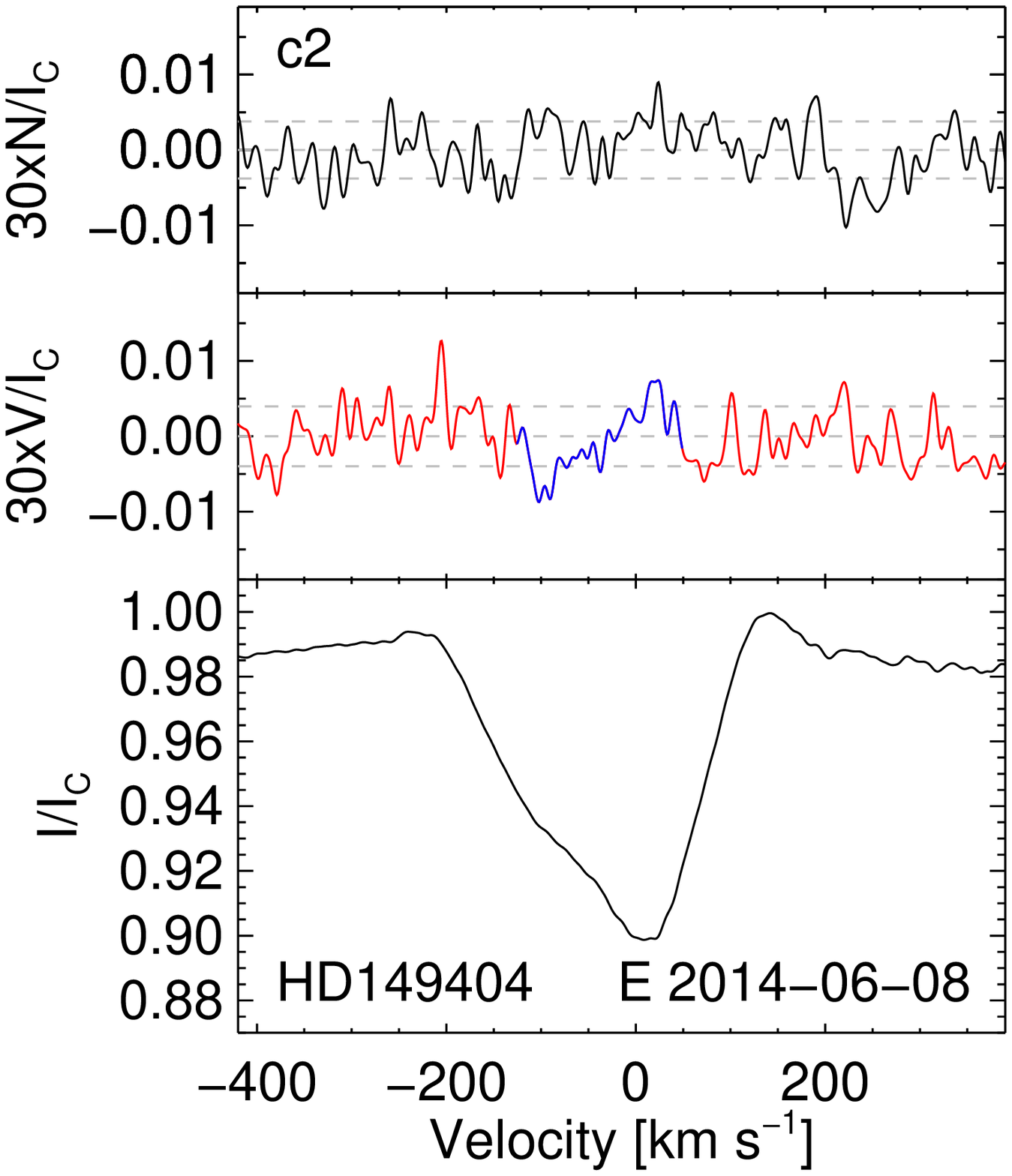}
\includegraphics[width=0.220\textwidth]{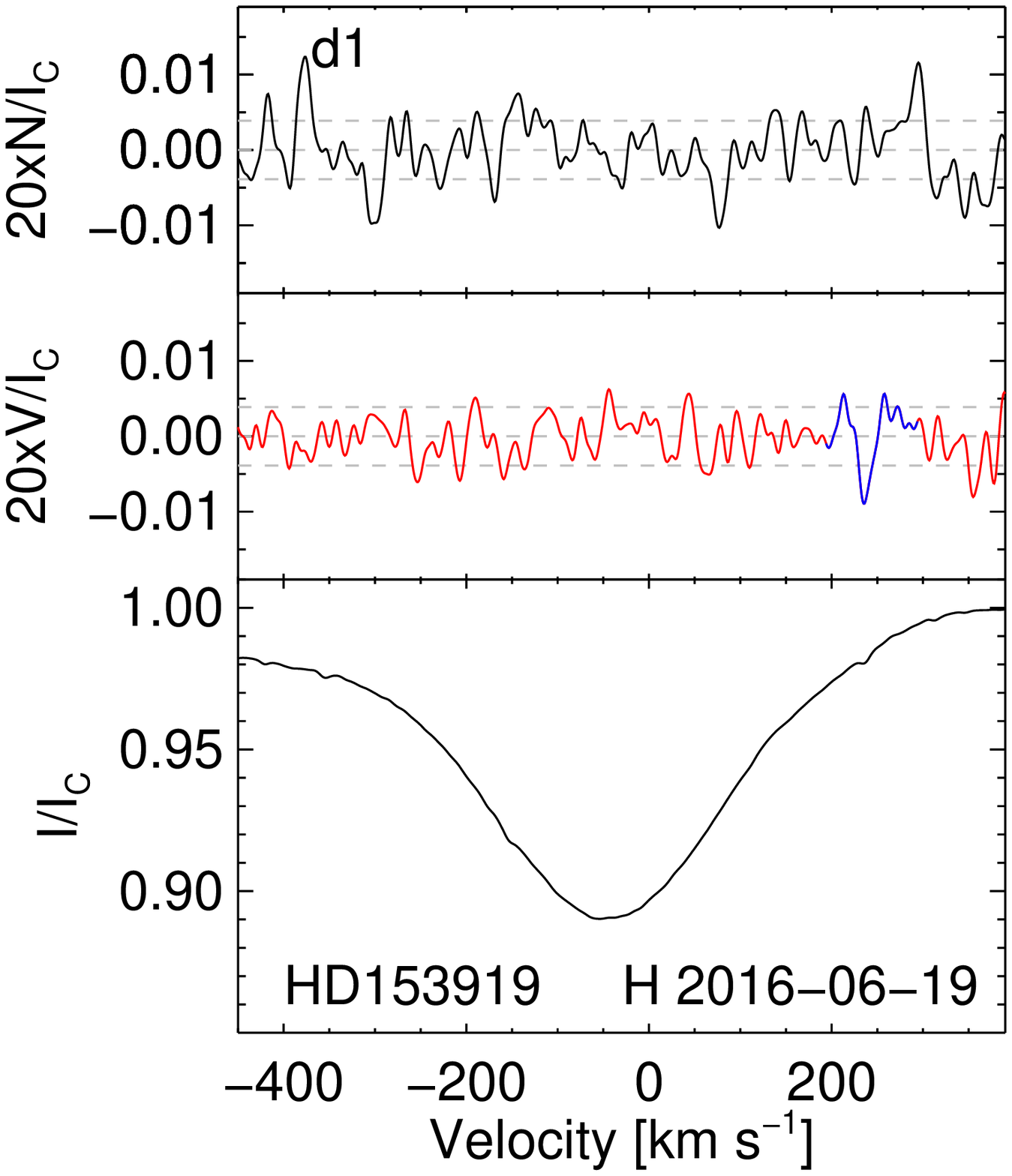}
\includegraphics[width=0.220\textwidth]{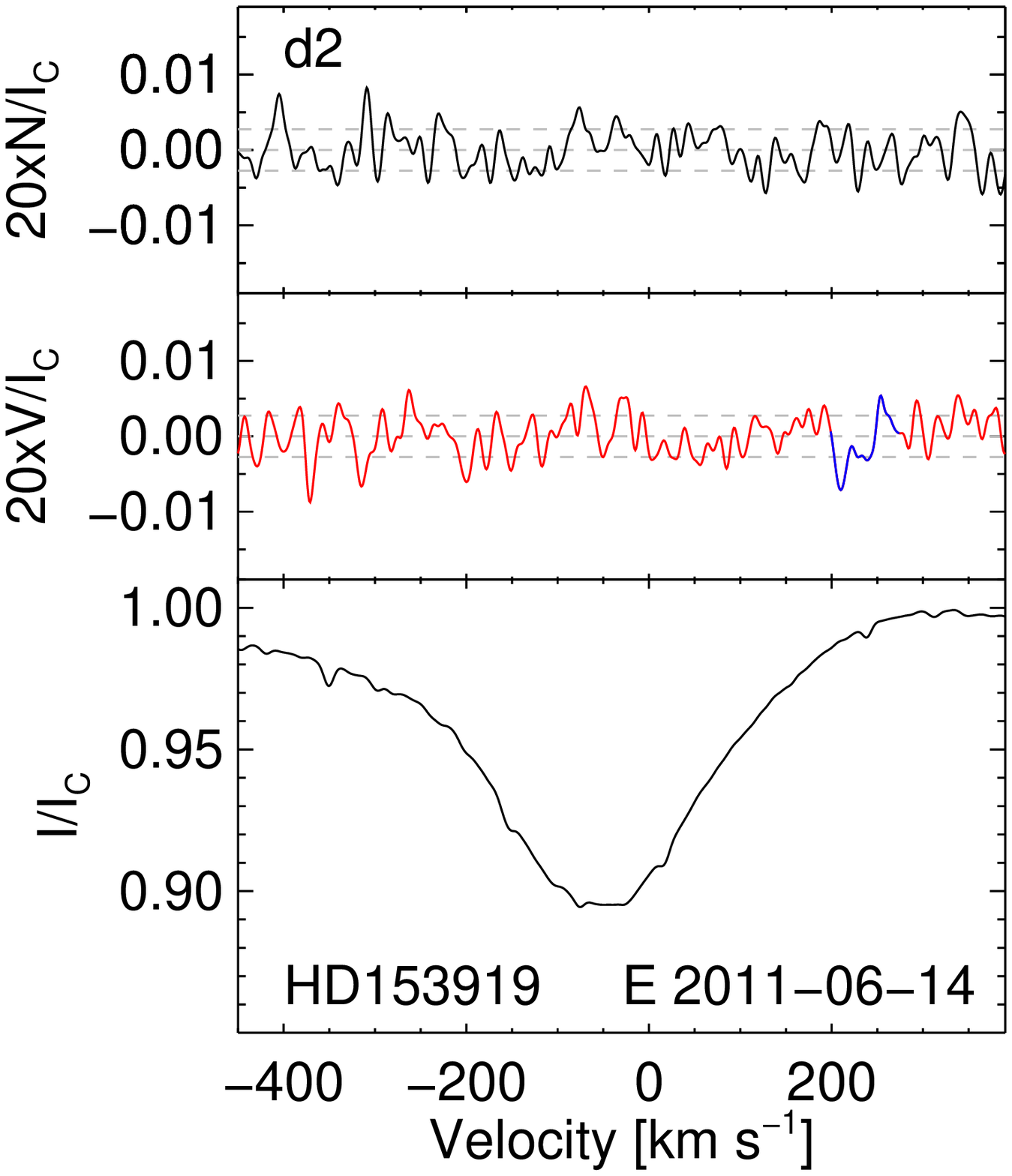}
\caption{
As Fig.~\ref{fig:collid} for systems with X-rays and colliding winds:
a) LSD analysis results for HD\,1337 using one ESPaDOnS observation from 2009;
b) LSD analysis results for HD\,48099 using one ESPaDOnS observation from 2012;
c1/c2) LSD analysis results for HD\,149404 using one HARPS\-pol observation from 2016  and one ESPaDOnS observation from 2014;
d1/d2) LSD analysis results for HD\,153919 using one HARPS\-pol observation from 2016 and one ESPaDOnS observation from 2011.
}
   \label{fig:xraycoll}
\end{figure*}

\paragraph*{HD\,1337 (AO\,Cas):} 

According to \citet{Linder2008}, this O9.2II+O8V((f))
system is probably an interacting eclipsing binary in a semidetached configuration with an orbital period of
3.5\,d, where the secondary is probably impacted by an accretion stream.
\citet{PortegiesZwart2002} listed this target among the systems with X-ray colliding-winds, where X-rays are produced
in the collisions between the winds.
Using Hipparcos photometry,  \citet{Lefevre2009}
reported that HD\,1337 is variable with a mean amplitude of 0.198\,mag.

The LSD analysis of the single ESPaDOnS observation of HD\,1337 
obtained in 2009 was carried out using a mask containing \ion{He}{i}, \ion{He}{ii}, and 
\ion{O}{iii} lines. As presented in Table~\ref{tab:obsall} and in Fig.~\ref{fig:xraycoll}, we
detect a clear Zeeman feature with ${\rm FAP}<10^{-10}$ in the LSD Stokes~$V$ spectrum achieving a definite detection
of the mean longitudinal magnetic field with a strength $\left< B_{\rm z} \right>=-274\pm52$\,G in the more luminous O9.2\,II component.

\paragraph*{HD\,48099:}

This SB2 system with a 3.1\,d orbital period belongs to the Mon\,OB2 association \citep{Garmany1980}.
The detection of a strongly clumped wind and a nitrogen abundance of about 8 times the solar value in the primary
probably indicate that the system went through an evolutionary phase of
mass exchange or mass loss through Roche lobe overflow \citep{Mahy2010}.
\citet{Berdyugin2016} investigated the structure of the  binary system by measuring linear polarization 
that arises due to a light scattering process. Their model fit suggested that light is scattered on a 
cloud produced by the colliding stellar winds.  The available X-ray data provided additional evidence for the 
existence of colliding stellar winds in this system.
\citet{Burssens2020} reported that one-sector TESS observations of HD\,48099 show SLF variability.

Two components are clearly visible in
the LSD Stokes~$I$ profile presented in Fig.~\ref{fig:xraycoll}. The secondary component shows a narrow-lined profile,
while the primary is rotating fast with a $v\,\sin\,i$ of about 91\,km\,s$^{-1}$ \citep{Mahy2010}.
As reported in Table~\ref{tab:obsall}, we detect a Zeeman feature with ${\rm FAP}=7\times10^{-6}$ in the LSD Stokes~$V$ profile 
calculated for the single available ESPaDOnS observation obtained in 2012, using a line mask 
containing  \ion{He}{i}, \ion{He}{ii}, \ion{C}{iv}, and \ion{Si}{iv} lines.
The position
of the Zeeman feature relative to the composite Stokes~$I$ spectrum indicates that the magnetic field
is detected in the broad-lined O5V primary component. 

\paragraph*{HD\,149404 (V918\,Sco):} 

According to \citet{Rauw2001}, the secondary in this system with an orbital period of 9.81\,d seems to be the
most evolved component and its current evolutionary status could best be 
explained if the system has undergone a Roche lobe overflow episode during the past. 
In \citeyear{Thaller1998}, \citeauthor{Thaller1998} suggested that the observed double-peaked structure of
the H$\alpha$ emission  is
due to a colliding wind interaction. Also \citet{Rauw2001} favour a model where the
H$\alpha$ emissions arise in the arms of a colliding wind shock region.

We used for the LSD analysis of one HARPS\-pol observation obtained in 2016 a mask
containing \ion{He}{i}, \ion{He}{ii}, \ion{C}{iv}, \ion{Si}{iii},
and \ion{Si}{iv} lines, 
whereas for the ESPaDOnS observation acquired in 2014 the mask was populated using  \ion{He}{i}, \ion{He}{ii}, \ion{Si}{iii},
and \ion{Si}{iv} lines.
As presented in Table~\ref{tab:obsall} and in Fig.~\ref{fig:xraycoll}, we achieve for both observations
marginal detections with ${\rm FAP}=5\times10^{-4}$ and ${\rm FAP}=7\times10^{-4}$, respectively.

\paragraph*{HD\,153919 (4U 1700-37):} 

According to \citet{Falanga2015}, this runaway high-mass X-ray binary (HMXB) is the
potentially most massive primary
in the known sample of HMXBs in the Galaxy. Previous studies suggested that the companion in
the HD\,153919 system with $P_{\rm orb}=3.41$\,d is either a massive neutron star or a low-mass black hole
(\citealt{Clark2002}; \citealt{Abbott2020}).
\citet{Ankay2001} proposed that HD\,153919
originates from the OB association Sco\,OB1 of
which NGC\,6231 is the suggested core. The authors also reported that the progenitor of this system went
into supernova within 6\,Myr.
According to \citet{vanderMeij2021}, HD\,153919
might be a prototype in the Milky Way for the progenitor of gravitational wave events such as GW190412.
Also a presence of X-ray flares was reported in different studies (e.g.\ \citealt{Boroson2003}). 
Based on a model of an entirely wind-accreting system, \citet{Brinkmann1981} suggested that the flares
were associated with accretion from the magnetotail of a neutron star. 

Using one low-resolution FORS\,2 spectropolarimetric observation of HD\,153919 obtained in 2010,
\citet{Hubrig2011} reported
the presence of a mean longitudinal magnetic field $\left< B_{\rm z} \right>=213\pm68$\,G. In contrast, no field
was detected in the single FORS\,2 observation acquired in 2013 \citep{Hubrig2013b}.
As presented in Table~\ref{tab:obsall} and in Fig.~\ref{fig:xraycoll}, no obvious Zeeman features 
were detected in the HARPS\-pol observation obtained in 2016 and in
the ESPaDOnS observation from 2011, using a line mask containing 
\ion{He}{i}, \ion{He}{ii}, \ion{C}{iv}, and \ion{O}{iii} lines. It is possible that the tiny features visible in both
observations close to 250\,km\,s$^{-1}$ in velocity space are due to a background or
a foreground object.


\subsection{Other targets with definitely detected magnetic fields}
\label{subsect:det}

\begin{figure*}
\centering 
\includegraphics[width=0.220\textwidth]{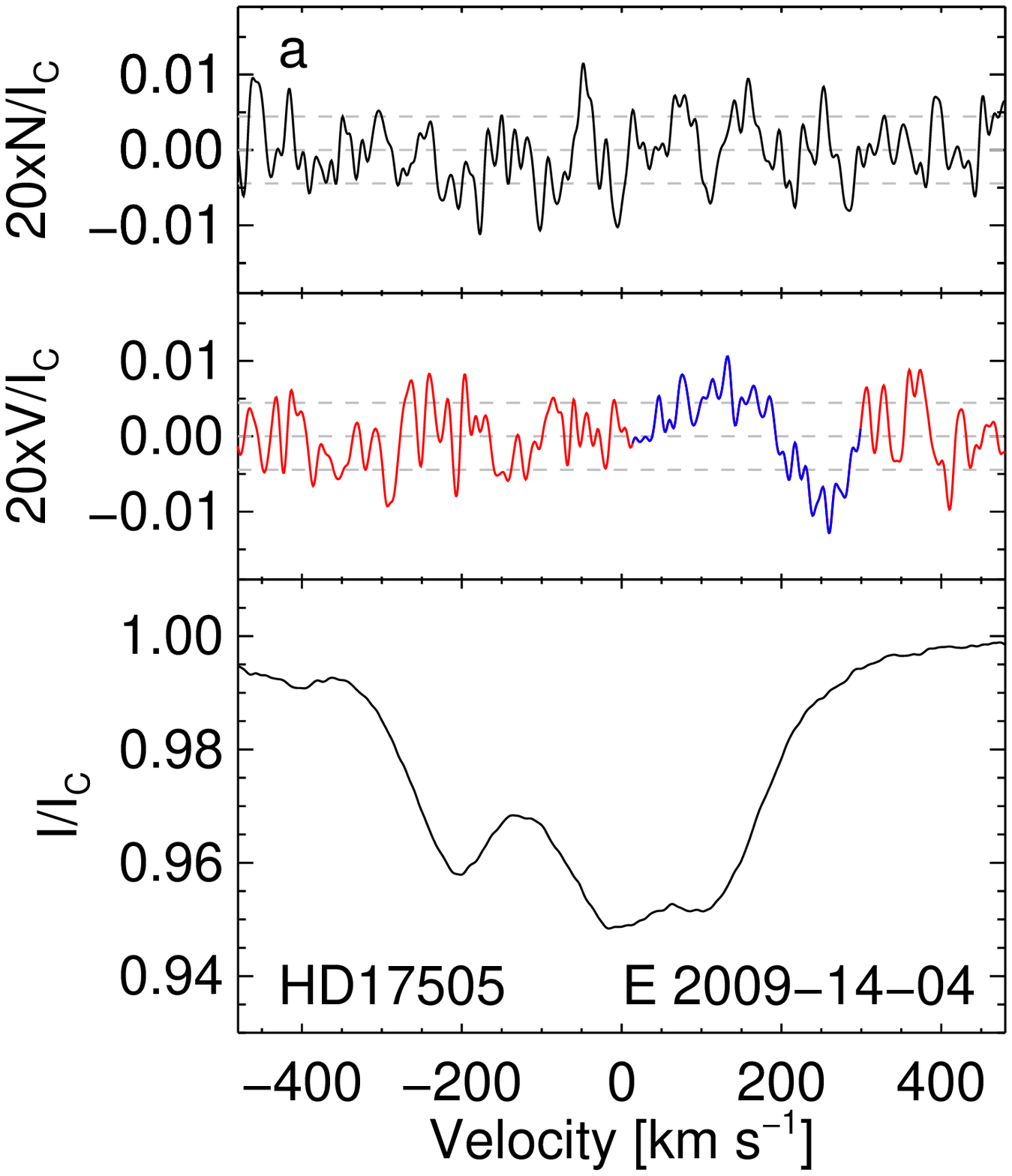}
\includegraphics[width=0.220\textwidth]{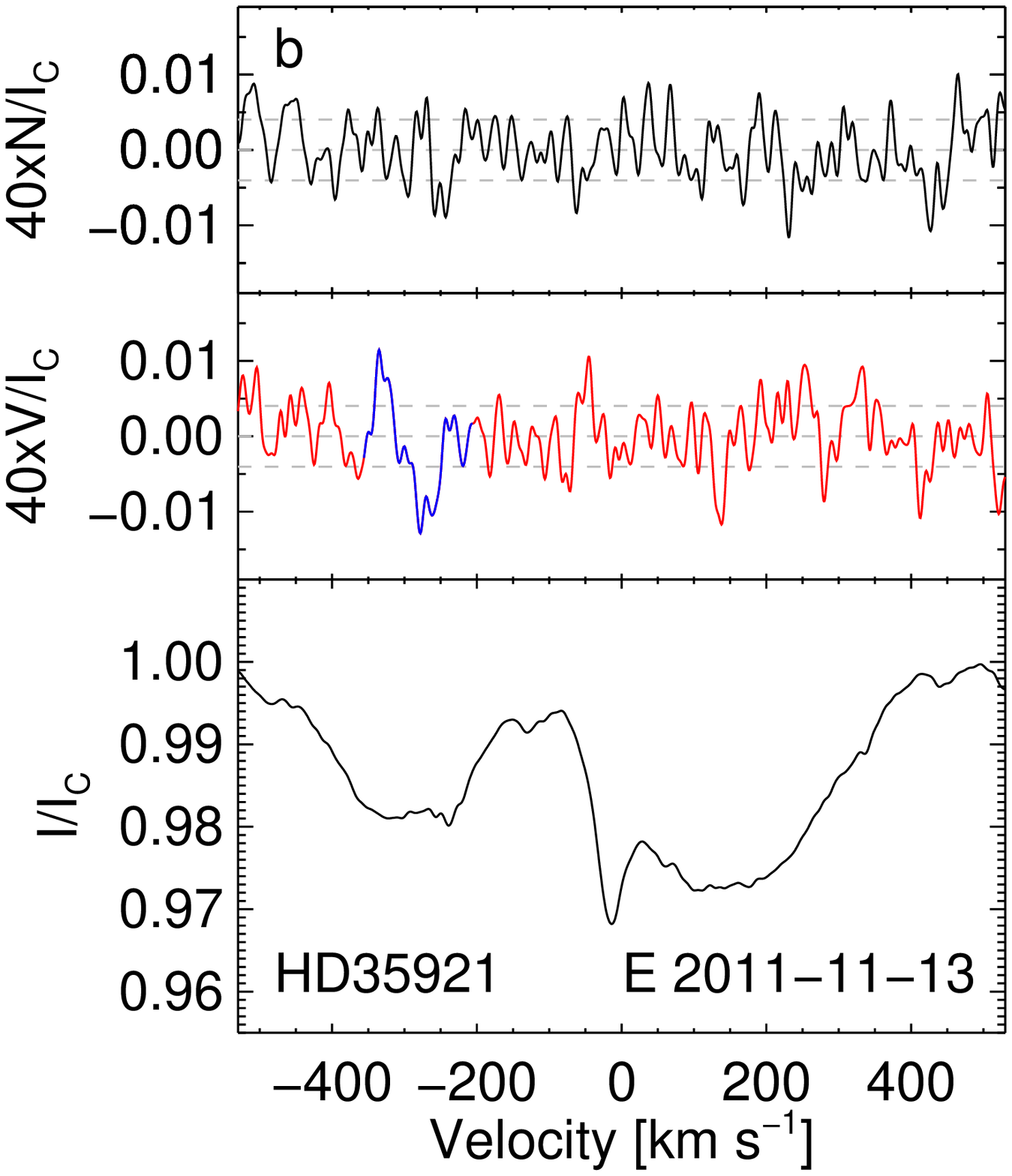}
\includegraphics[width=0.220\textwidth]{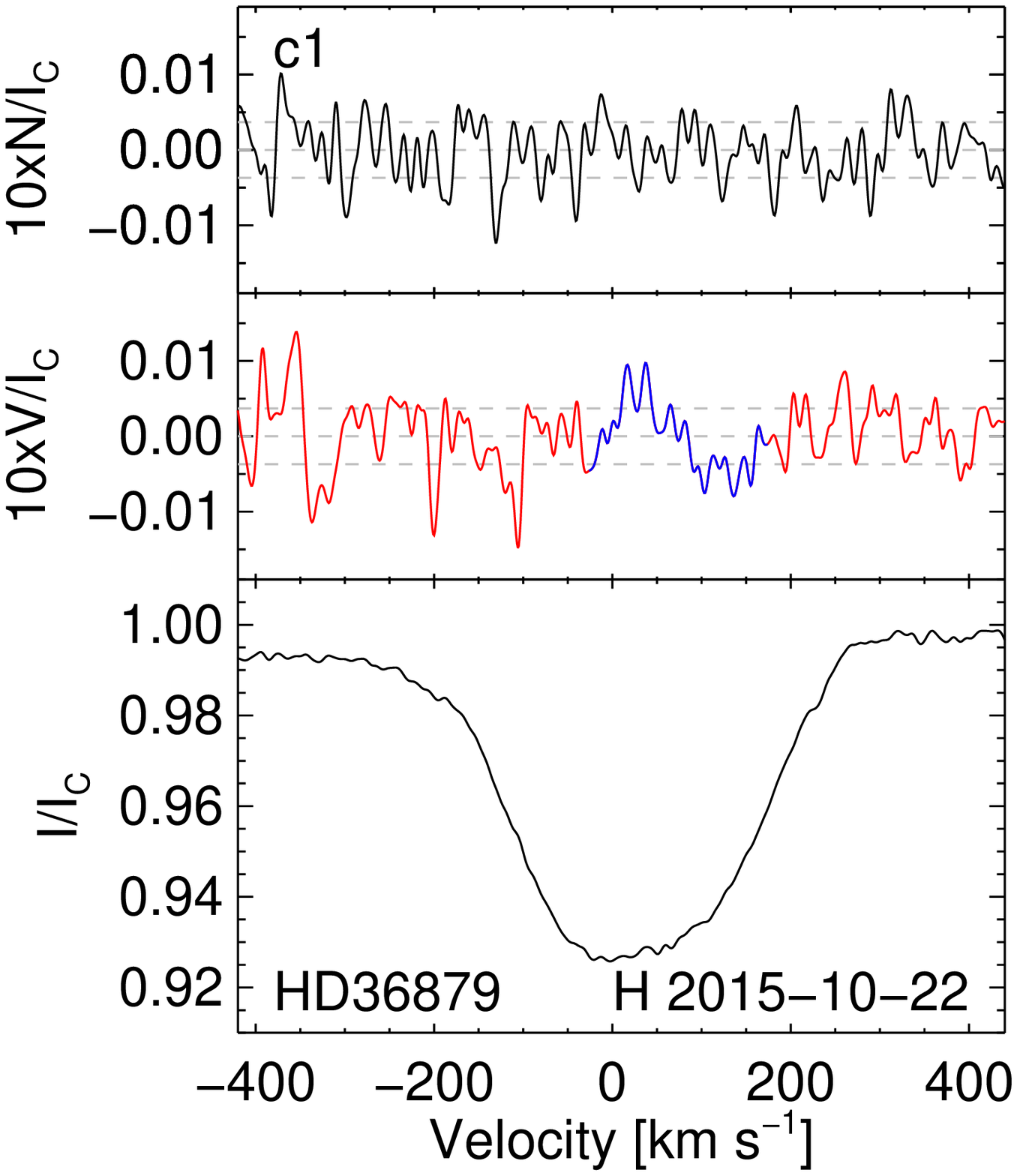}
\includegraphics[width=0.220\textwidth]{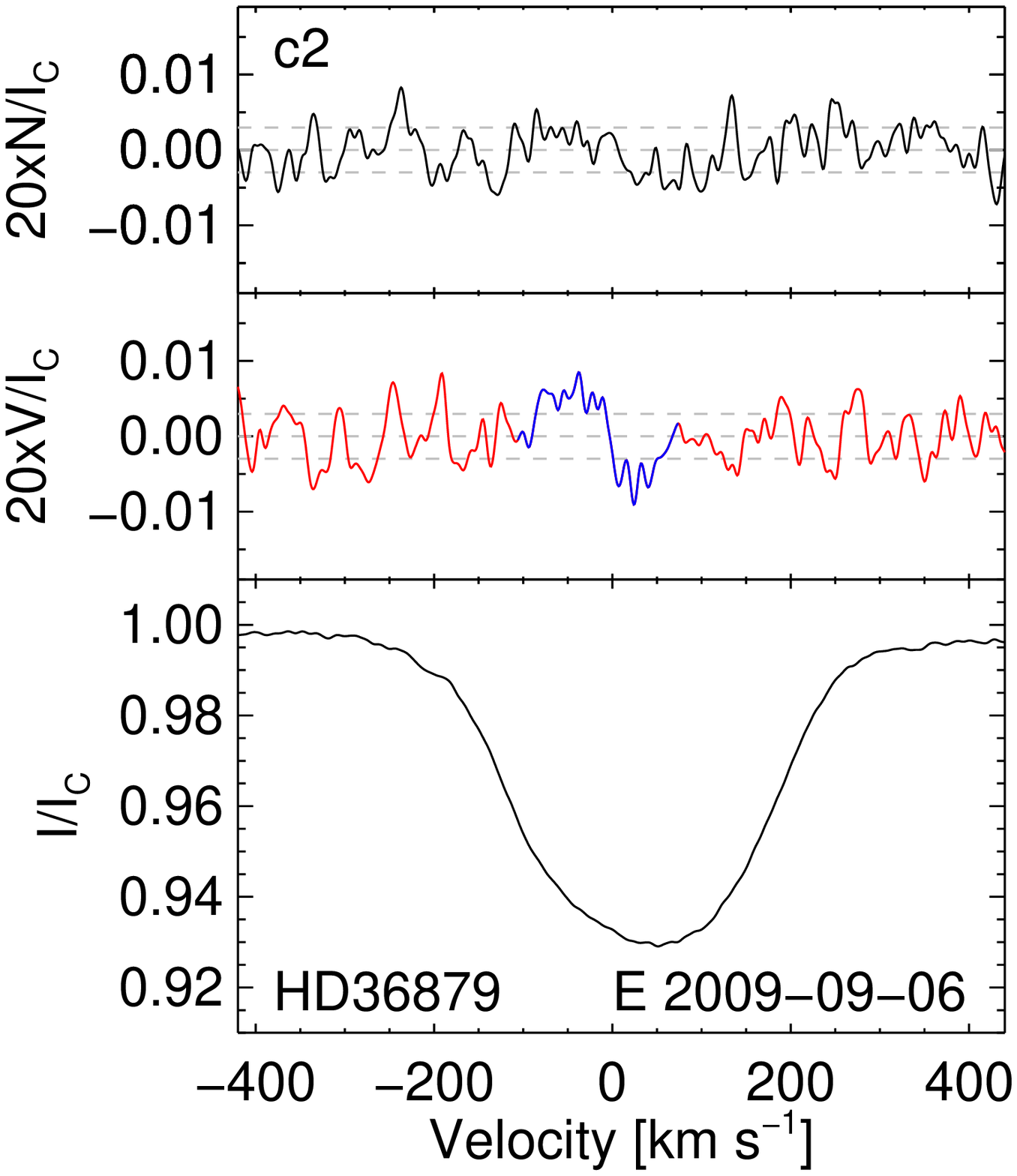}
\includegraphics[width=0.220\textwidth]{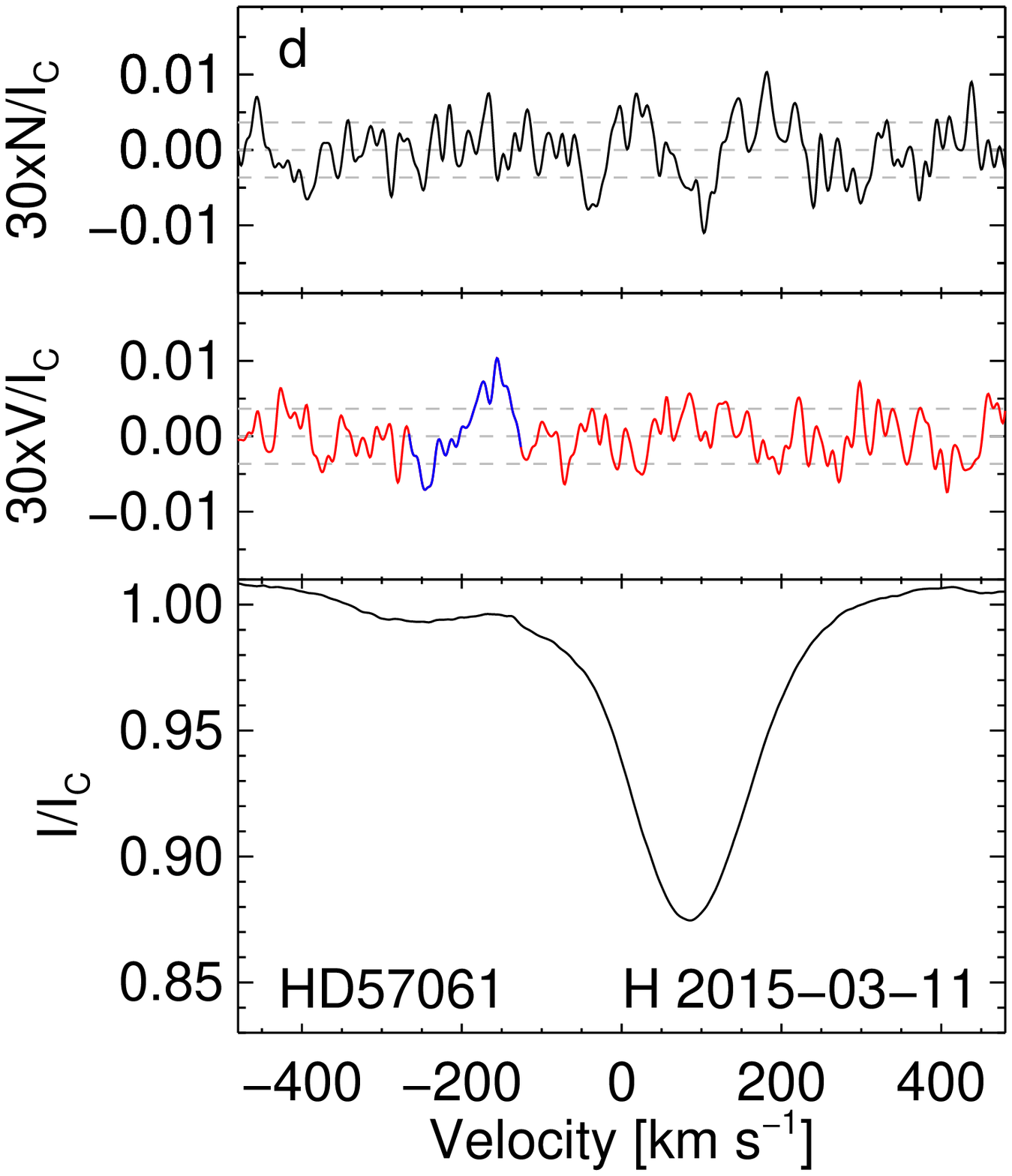}
\includegraphics[width=0.220\textwidth]{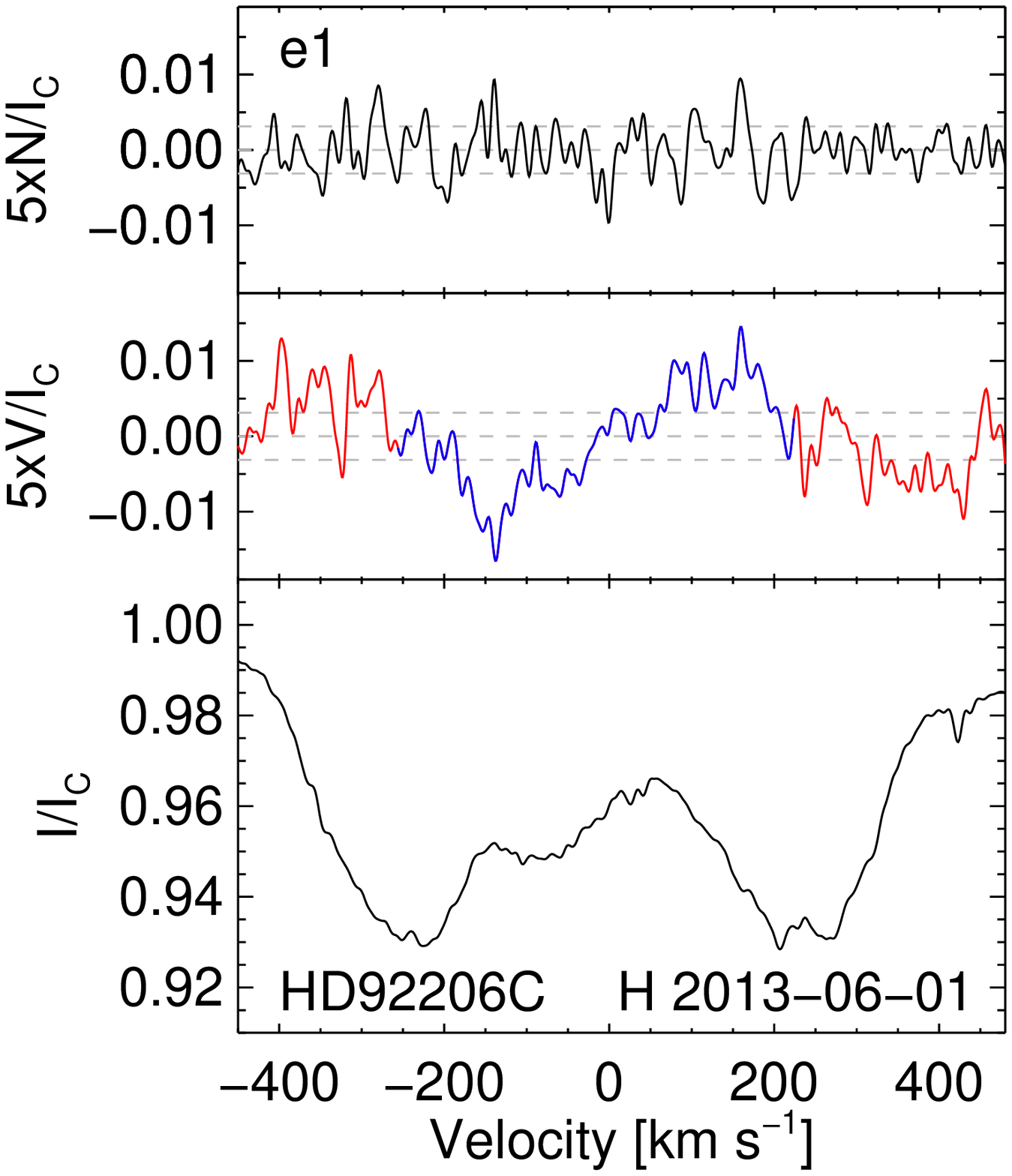}
\includegraphics[width=0.220\textwidth]{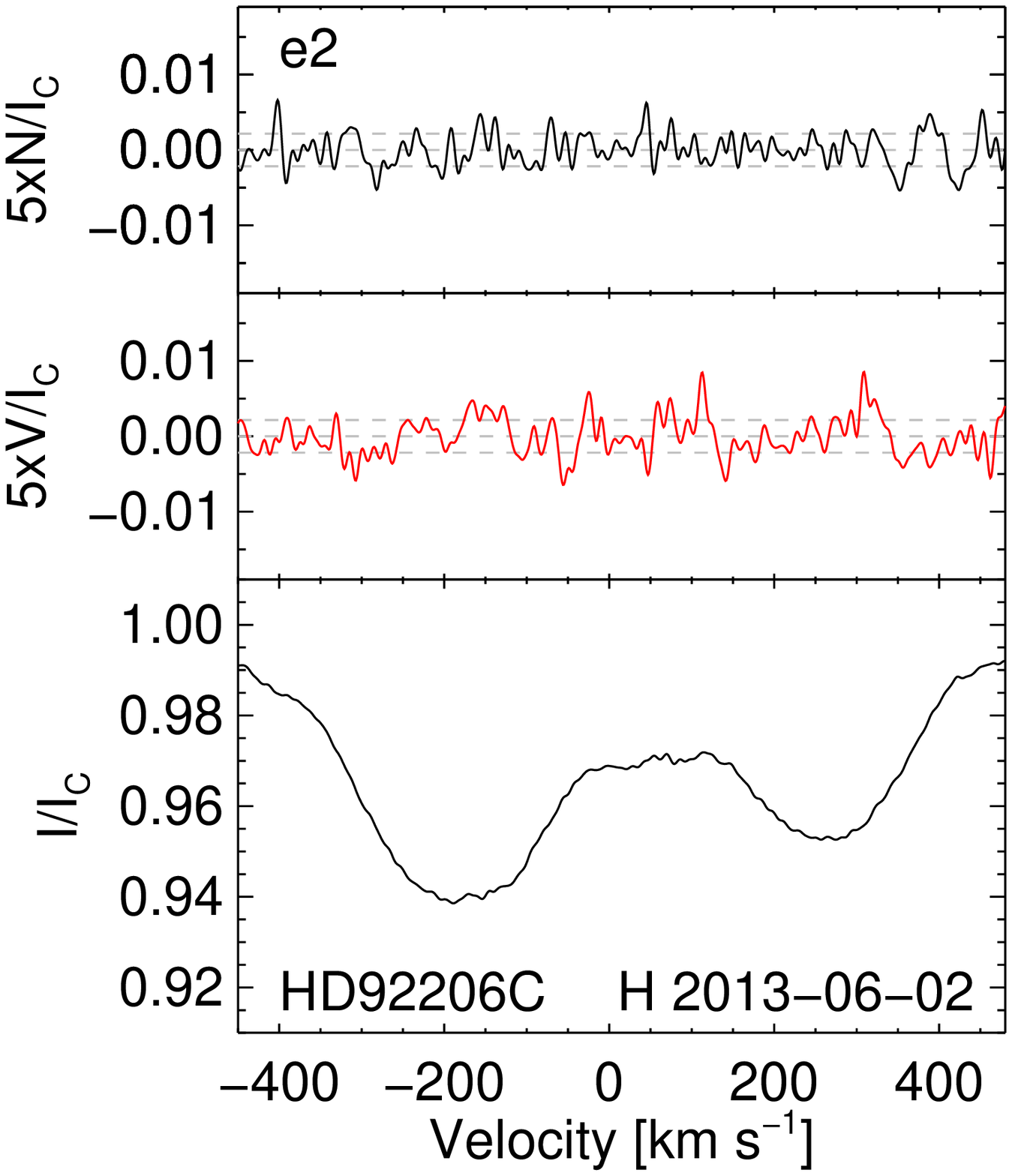}
\includegraphics[width=0.220\textwidth]{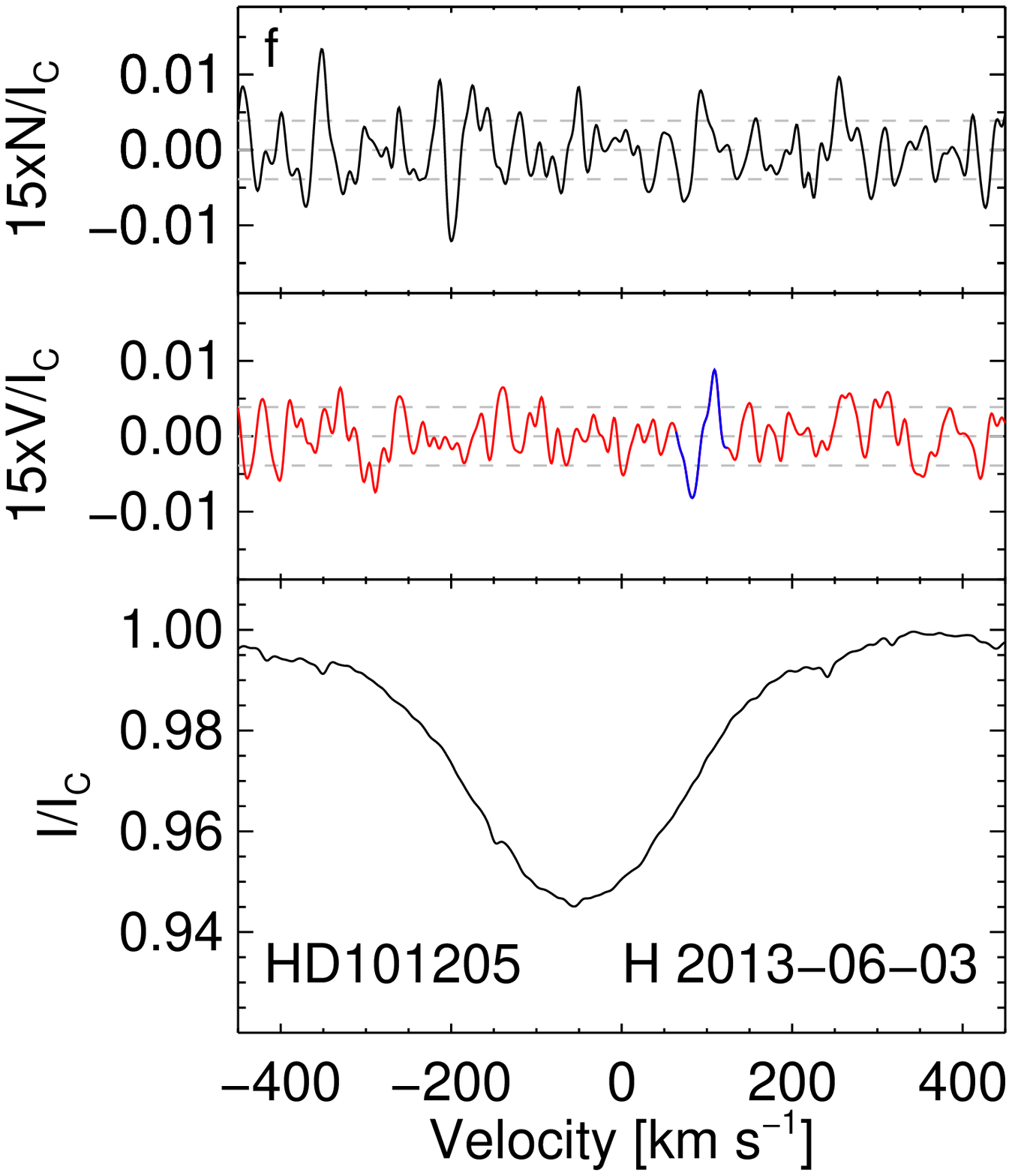}
\includegraphics[width=0.220\textwidth]{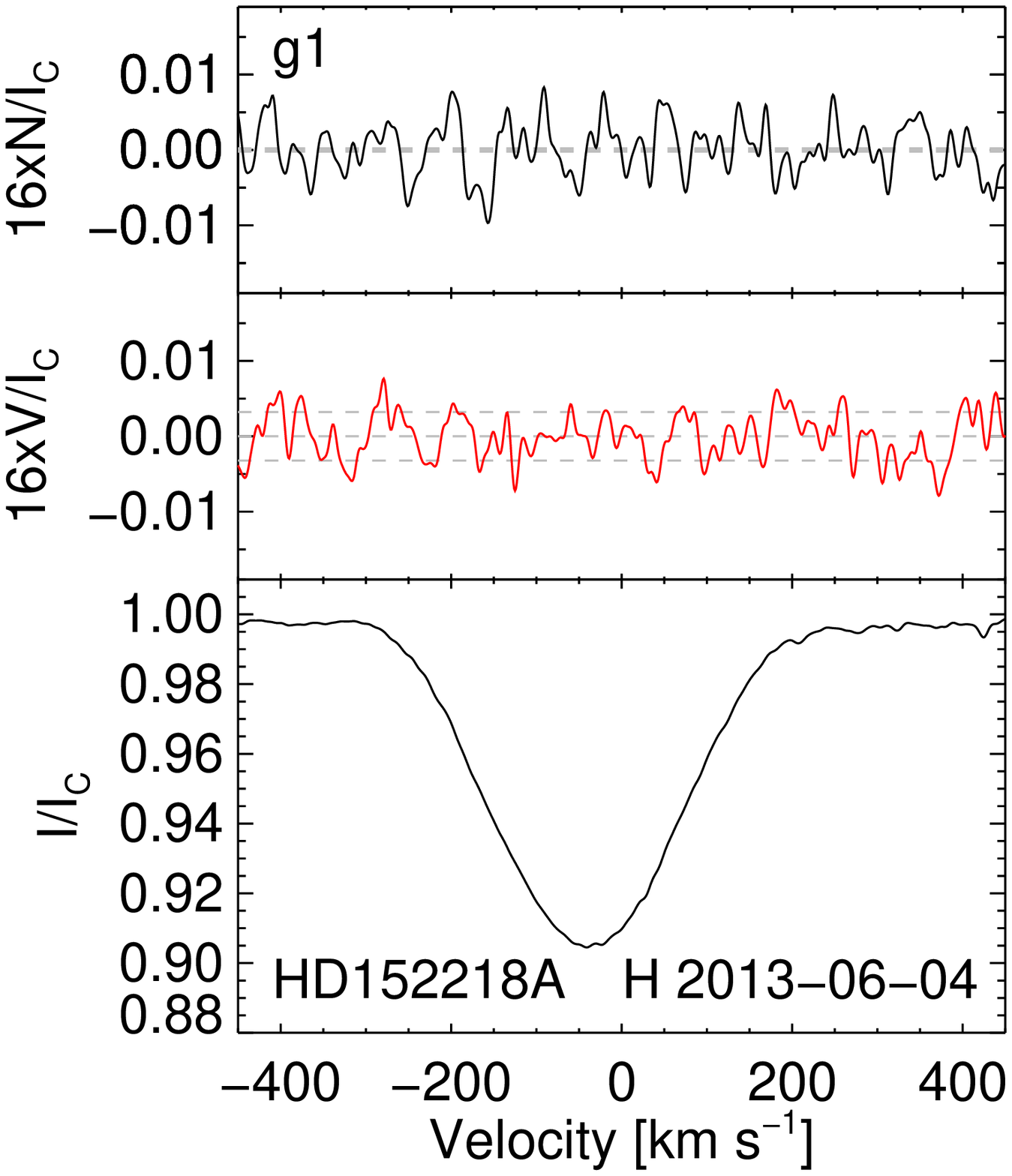}
\includegraphics[width=0.220\textwidth]{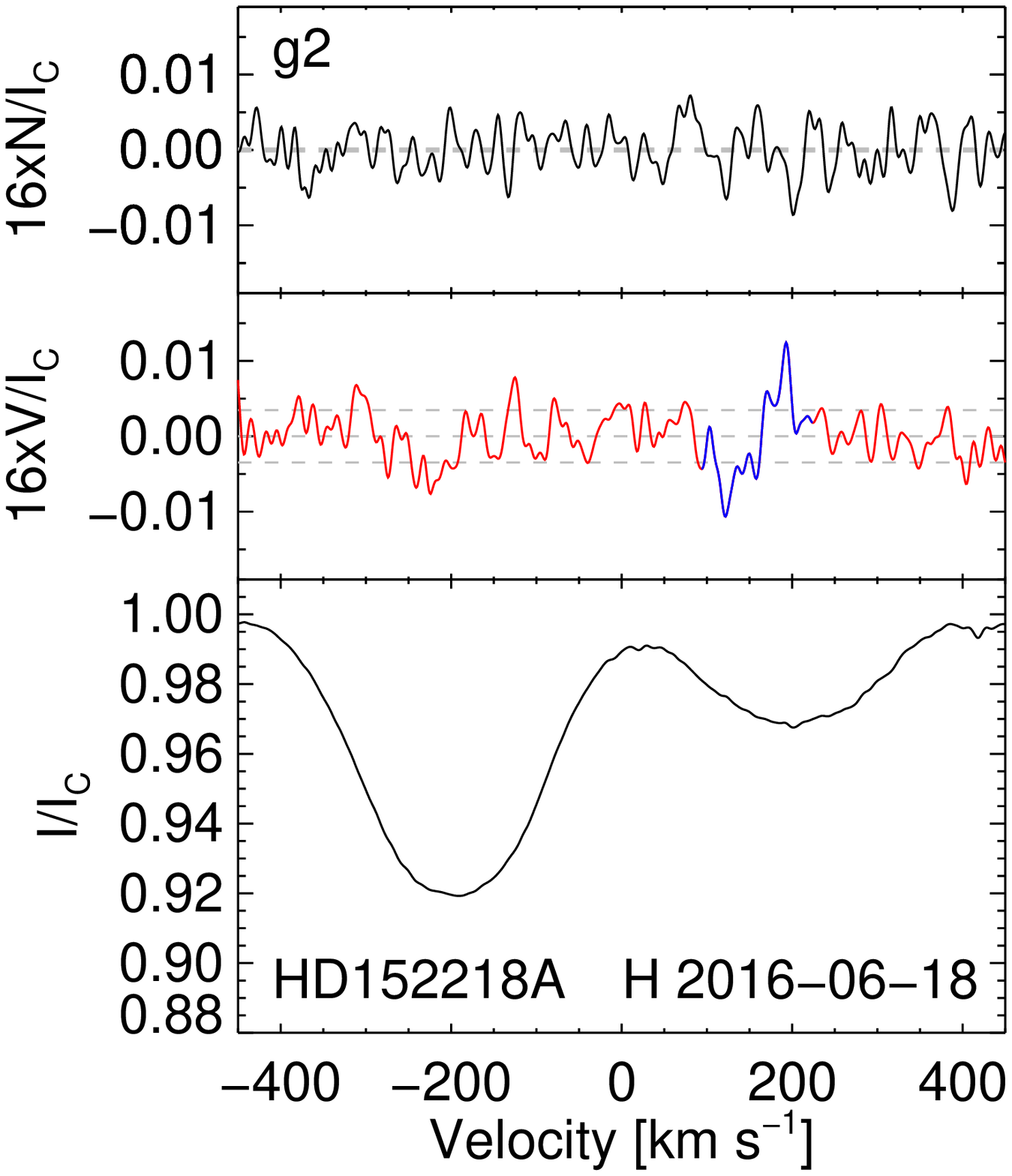}
\includegraphics[width=0.220\textwidth]{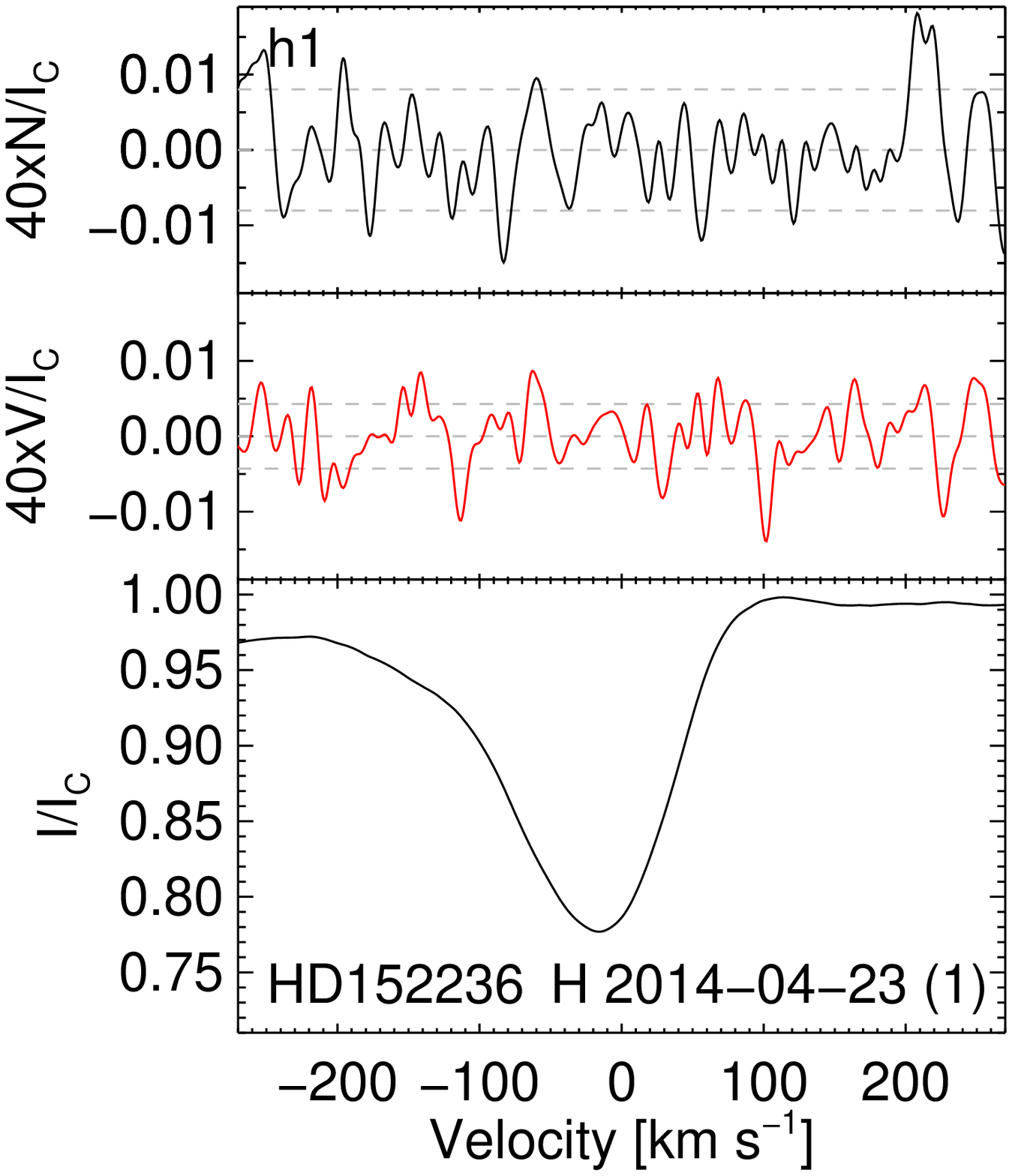}
\includegraphics[width=0.220\textwidth]{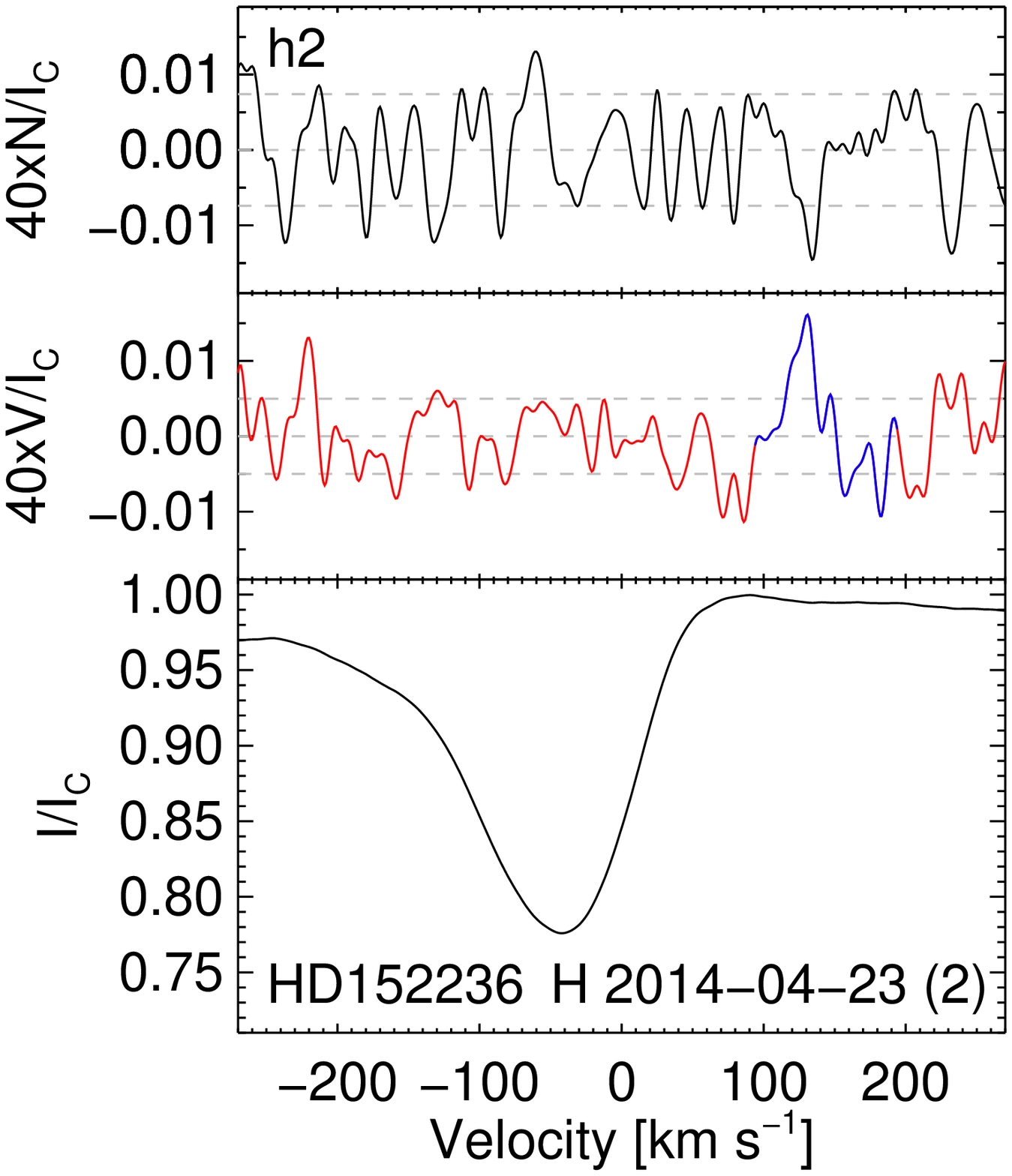}
\includegraphics[width=0.220\textwidth]{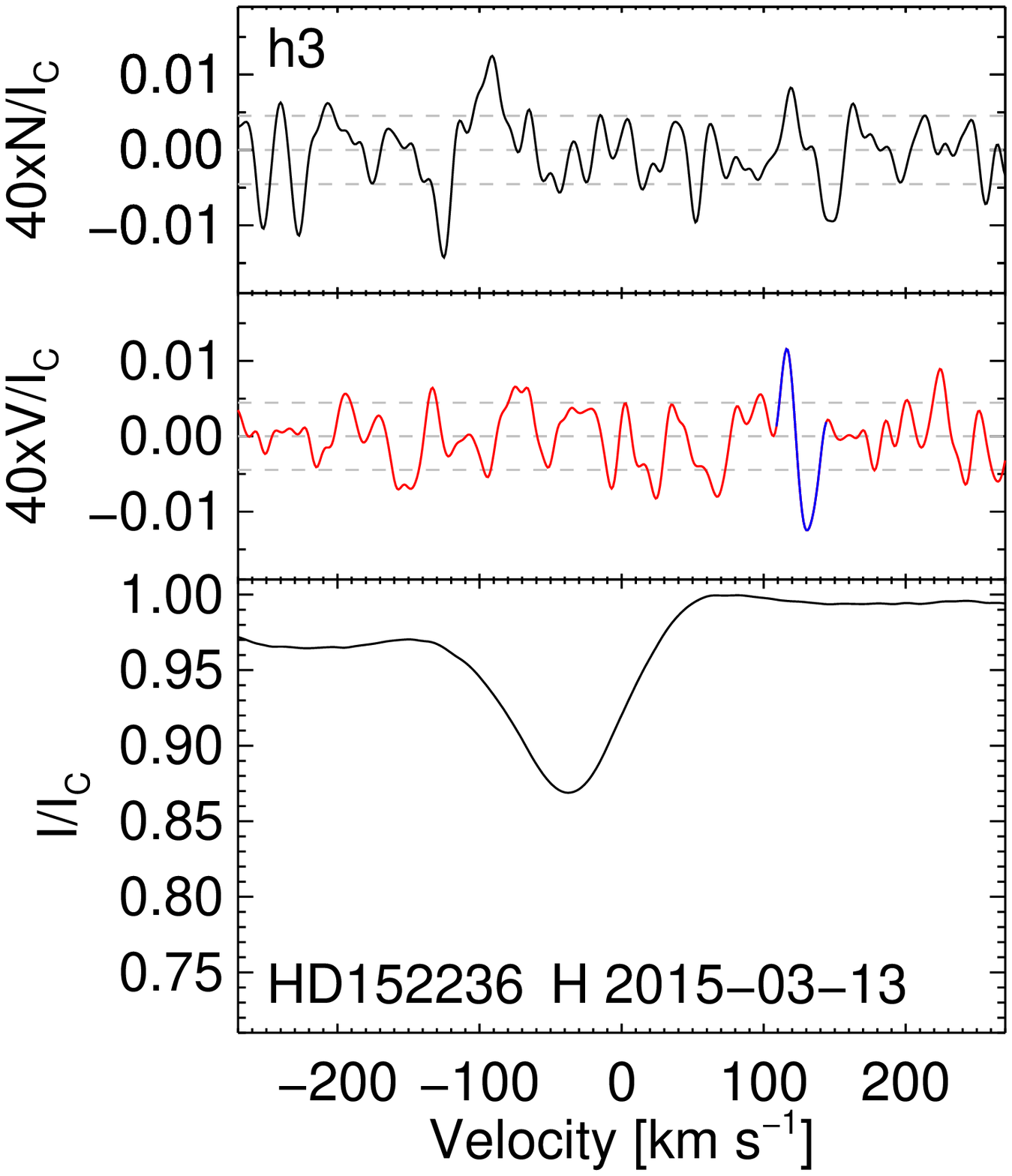}
\includegraphics[width=0.2200\textwidth]{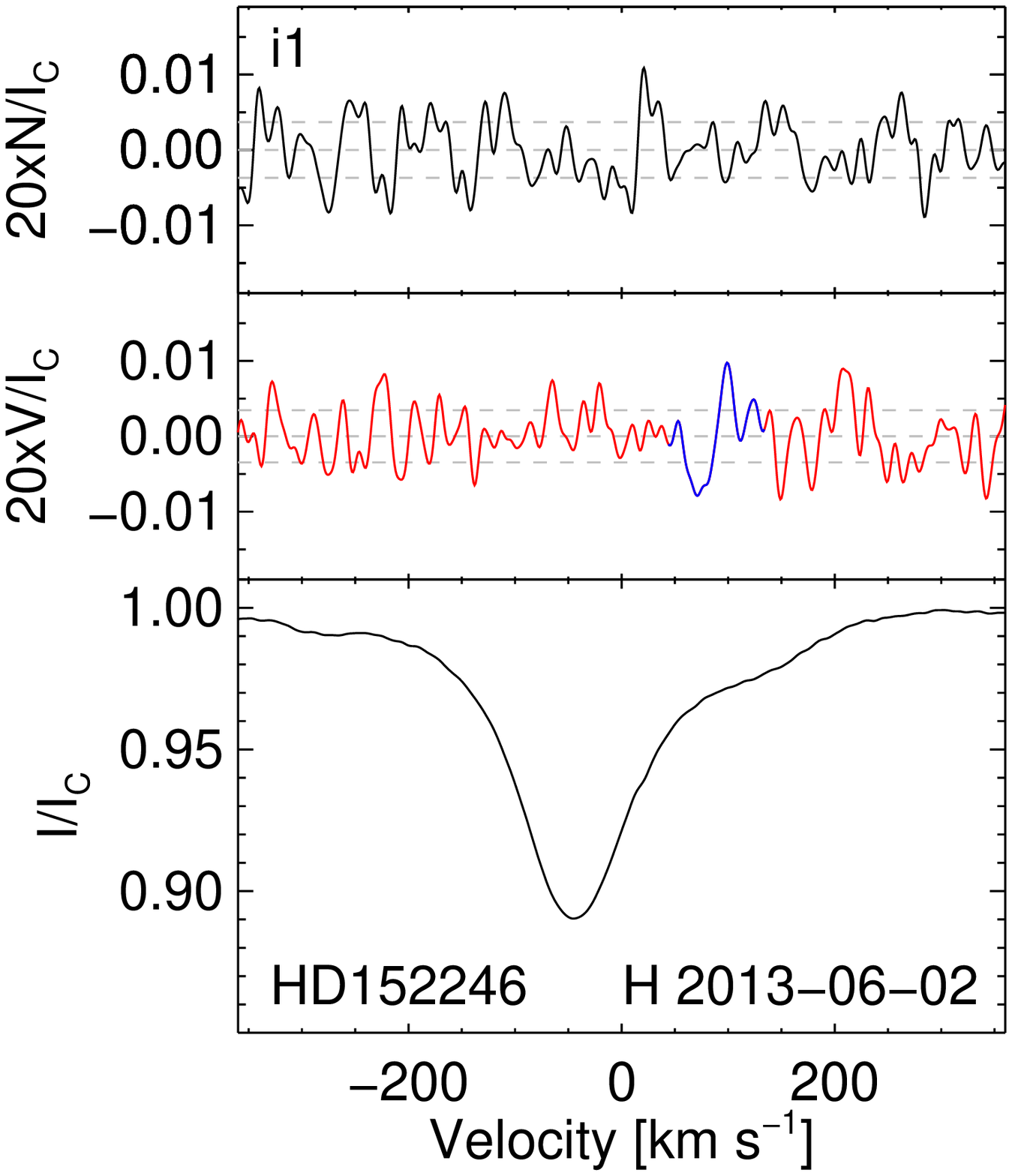}
\includegraphics[width=0.220\textwidth]{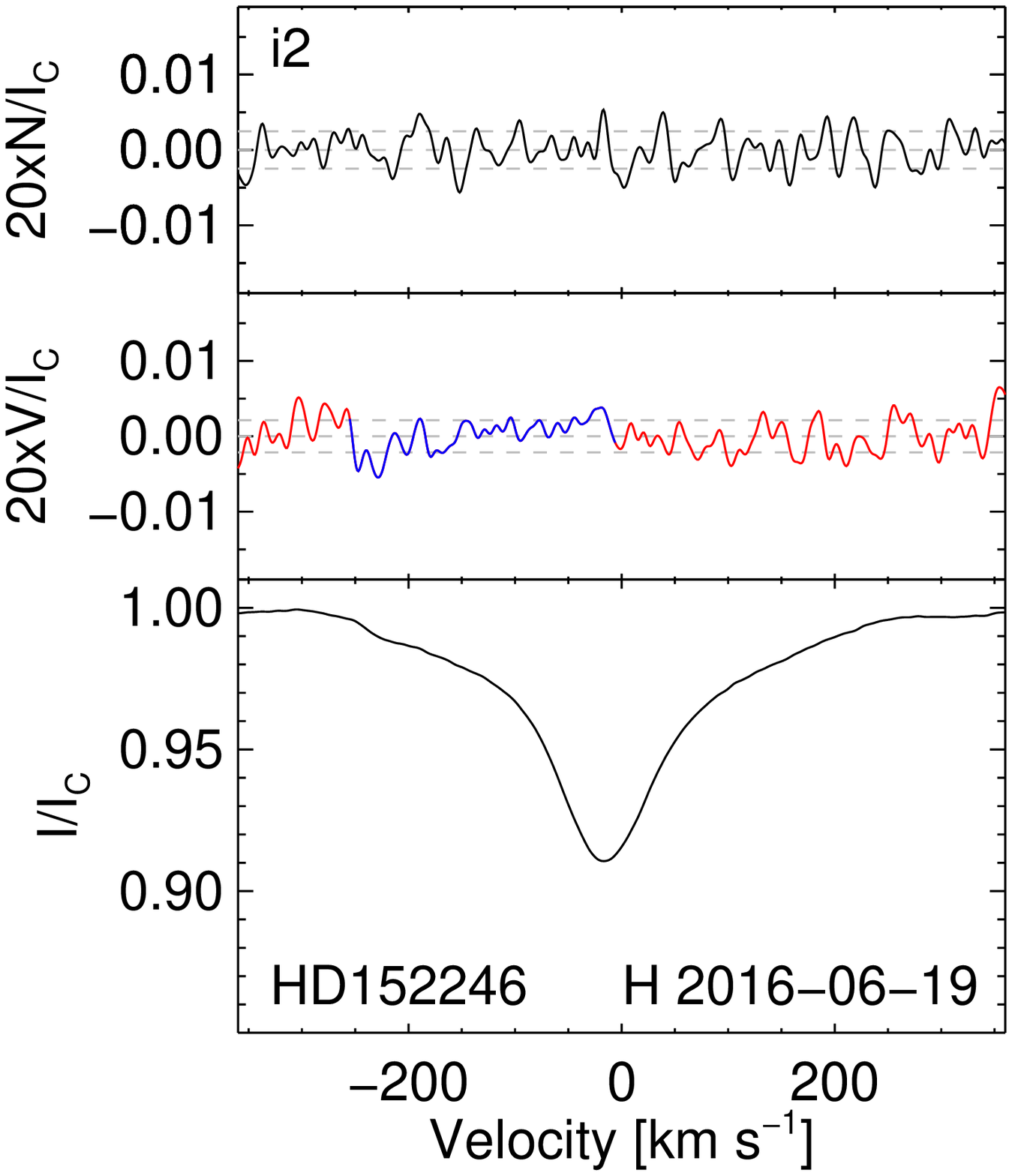}
\includegraphics[width=0.220\textwidth]{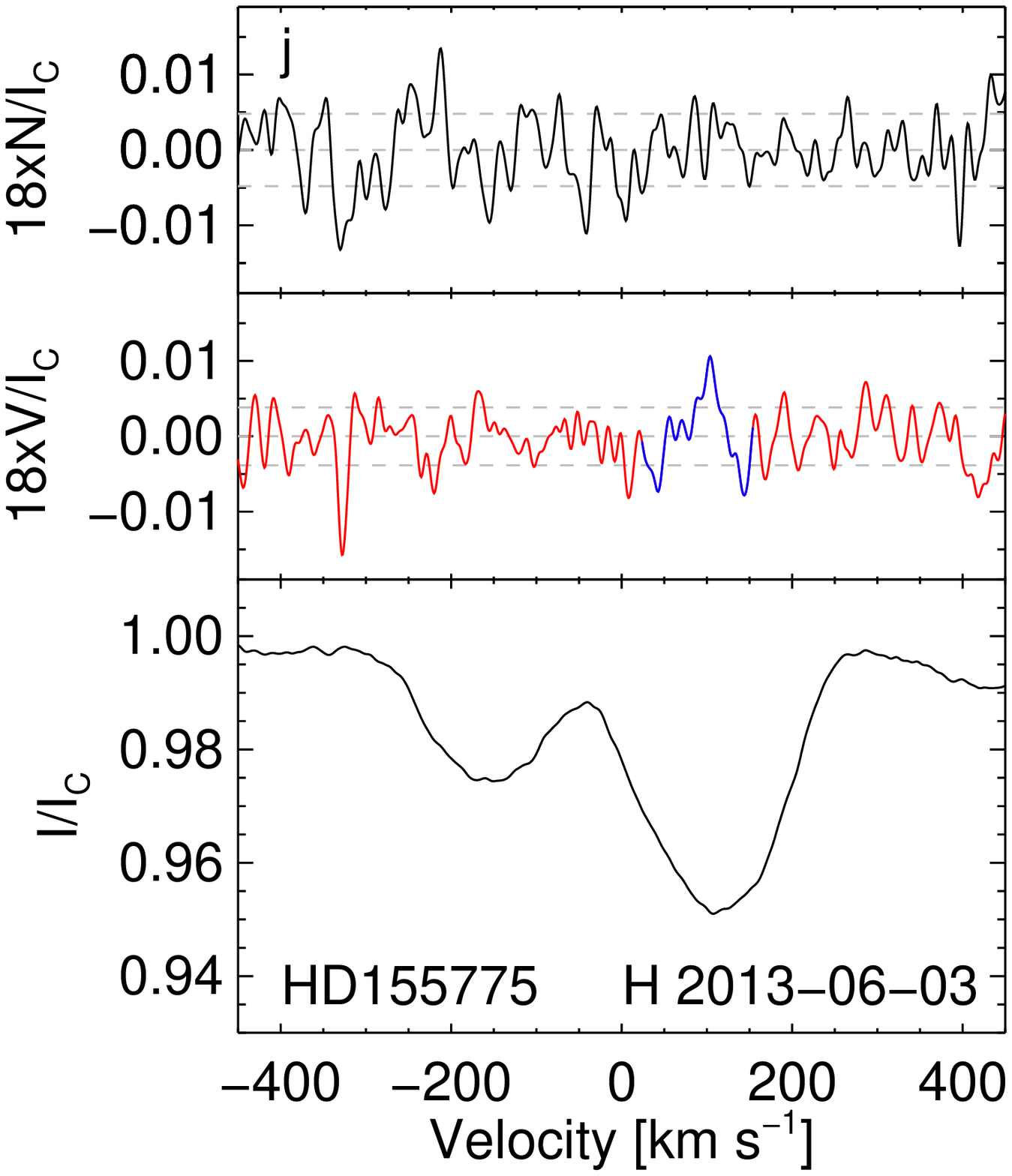}
\caption{
As Fig.~\ref{fig:collid} for the other systems with definitely detected magnetic fields:
a) LSD analysis results for HD\,17505 using one ESPaDOnS observation from 2009;
b) LSD analysis results for HD\,35921 using one ESPaDOnS observation from 2011;
c1/c2) LSD analysis results for HD\,36879 using one HARPS\-pol observation from 2015 and one ESPaDOnS observation from 2009;
d) LSD analysis results for HD\,57061 using one HARPS\-pol observation from 2015;
e1/e2) LSD analysis results for HD\,92206c using two HARPS\-pol observations obtained in 2013 on two consecutive nights;
f) LSD analysis results for HD\,101205 using one HARPS\-pol observation from 2013;
g1/g2) LSD analysis results for HD\,152218A using two HARPS\-pol observations acquired in 2013 and 2016;
h1-h3) LSD analysis results for HD\,152236 using three HARPS\-pol observations acquired in 2014 and 2015; 
i1/i2) LSD analysis results for HD\,152246 using two HARPS\-pol observations acquired in 2013 and 2016;
j) LSD analysis results for HD\,155775 using one HARPS\-pol observation from 2013.
}
   \label{fig:hddef}
\end{figure*}

\addtocounter{figure}{-1}

\begin{figure*}
\centering 
\includegraphics[width=0.220\textwidth]{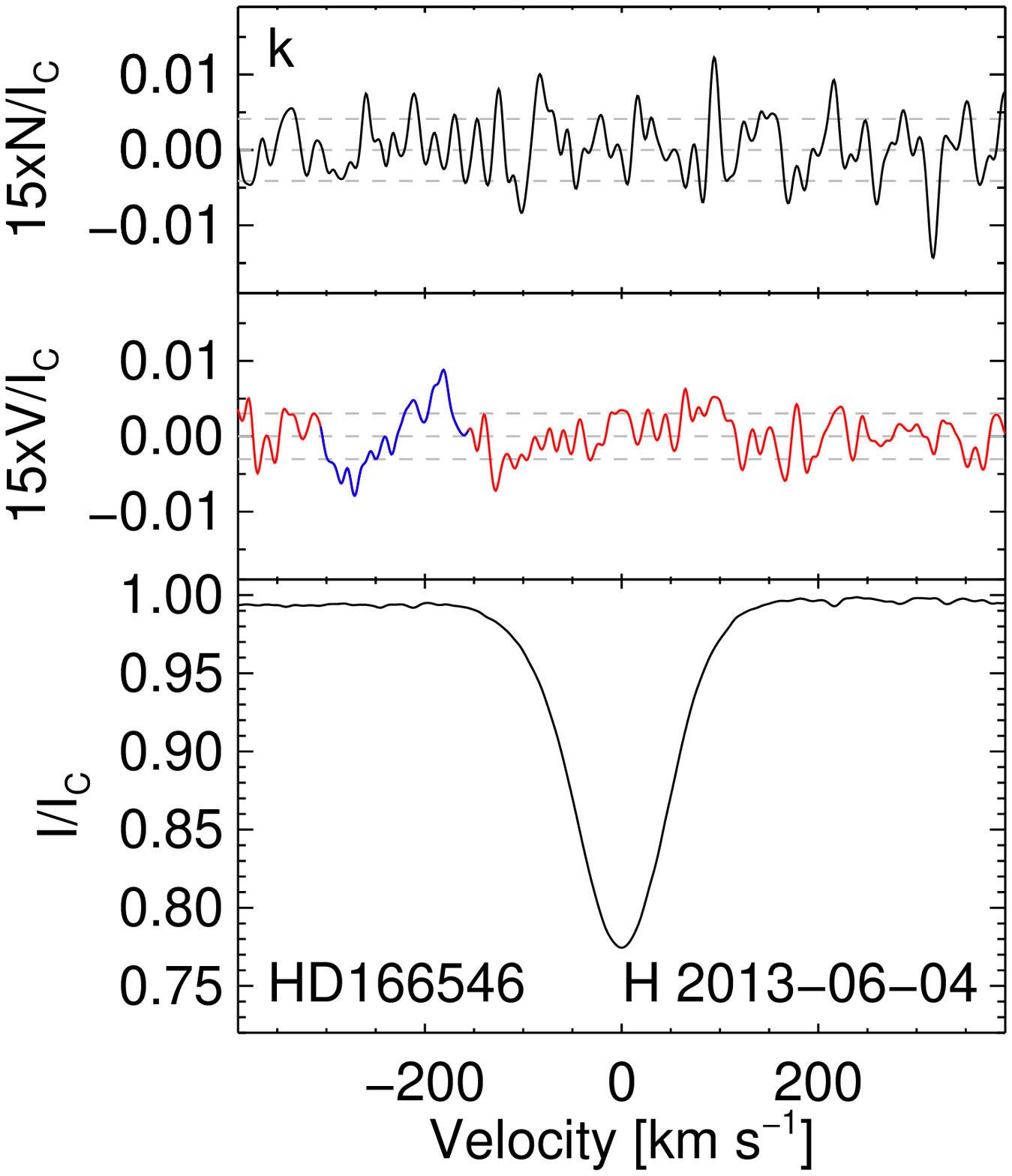}
\includegraphics[width=0.220\textwidth]{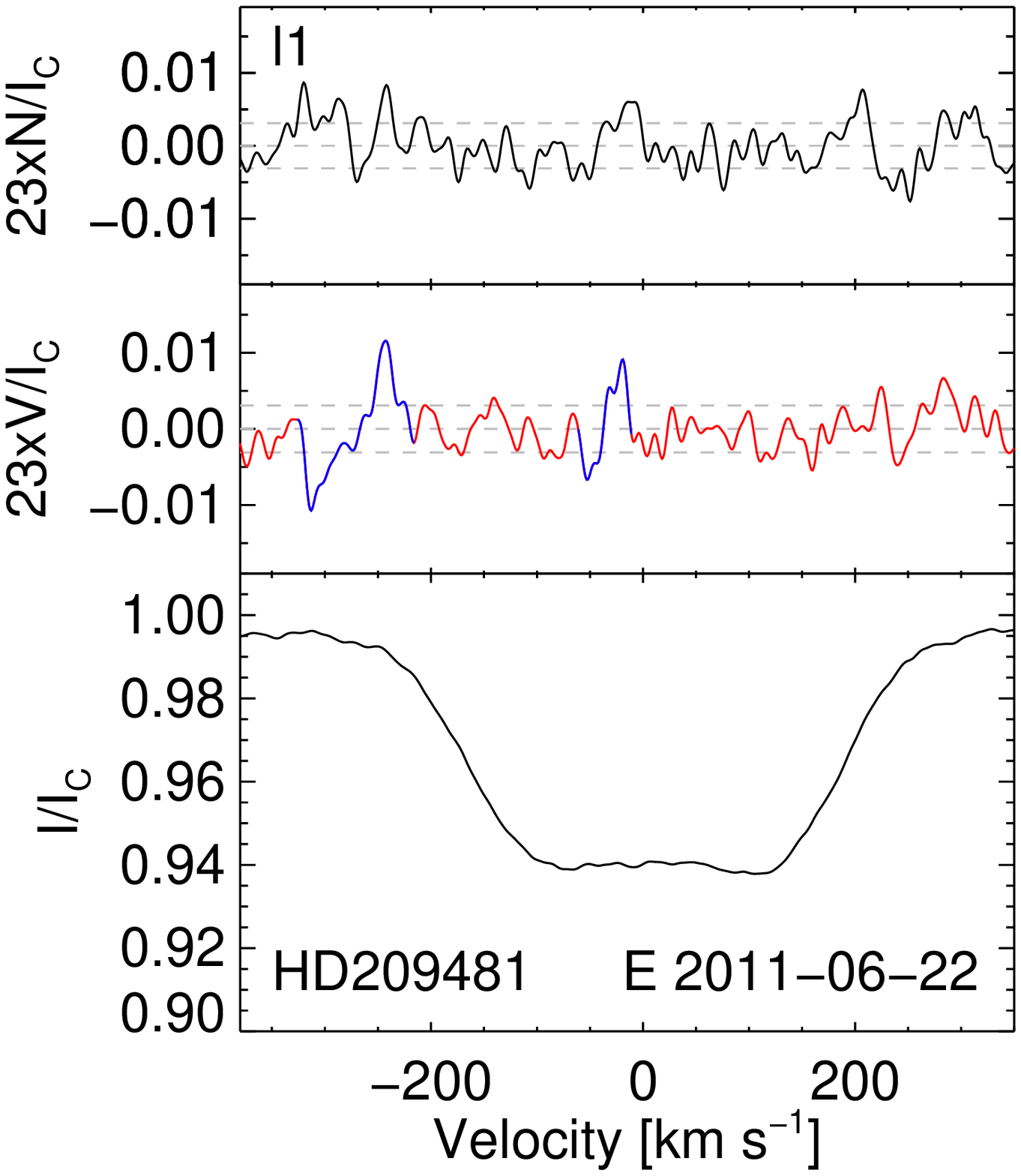}
\includegraphics[width=0.220\textwidth]{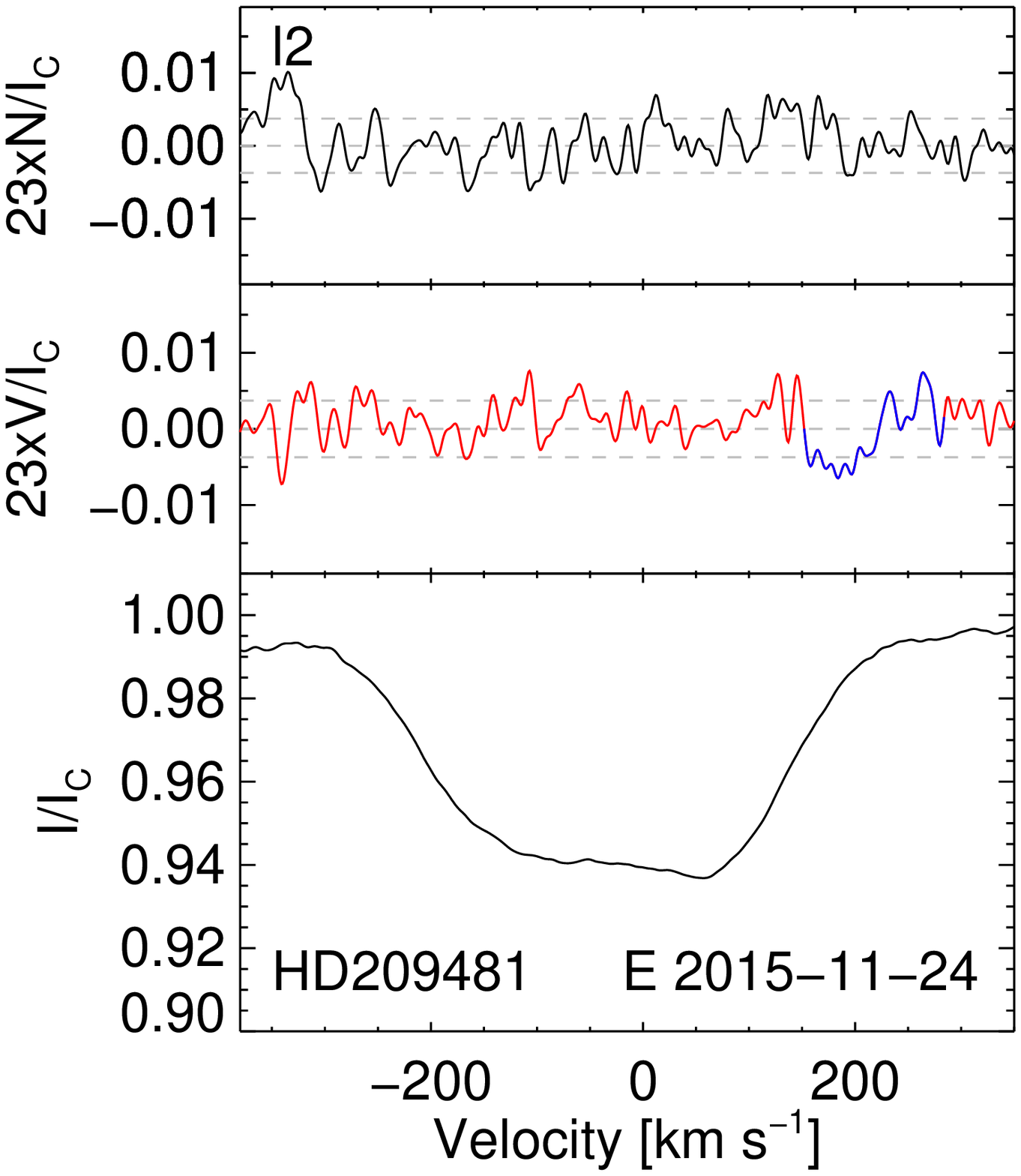}
\includegraphics[width=0.220\textwidth]{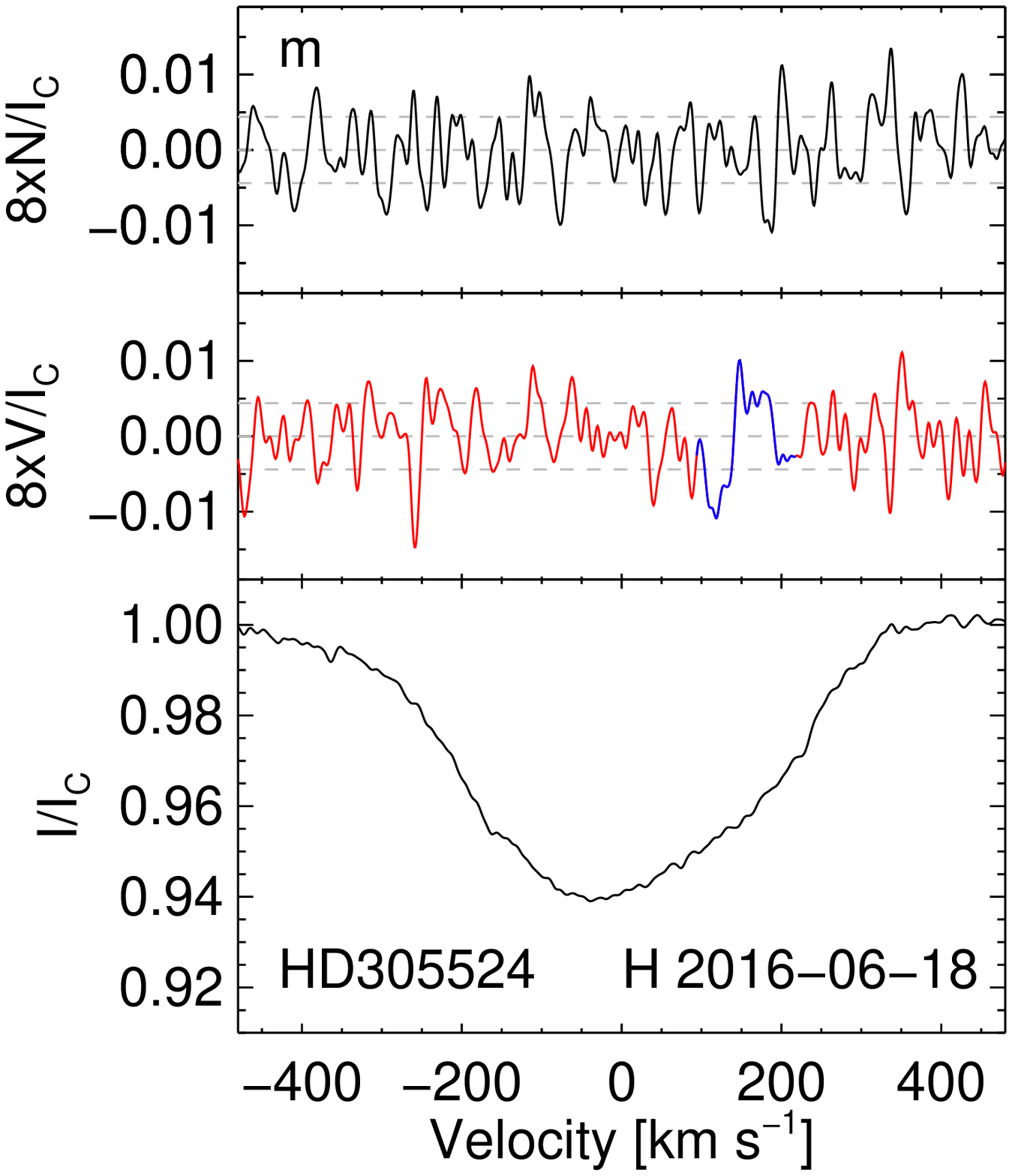}
\caption{
Continued:
k) LSD analysis results for HD\,166546 using one HARPS\-pol observation from 2013;
l1/l2) LSD analysis results for HD\,209481 using two ESPaDOnS observations acquired in 2011 and 2015;
m) LSD analysis results for HD\,305524 using one HARPS\-pol observation from 2016.
}
   \label{fig:hddef2}
\end{figure*}

\paragraph*{HD\,17505:}

In contrast to \citet{MaizApellaniz2004}, who reported for this system the spectral classification
O6.5III((f))n+O8V, in a more recent work \citet{Sota2014} characterised it as at least a triple system.
HD\,17505 is located in the centre of the
cluster IC\,1848 within the Cassiopeia OB6 association.
According to \citet{Raucq2018},
the components of the inner binary with an orbital period of 8.57\,d are both well 
inside their Roche lobe suggesting that this system has not yet experienced binary interaction, whilst
the outer orbit has a period of less than 61\,yr.

The LSD analysis of the single ESPaDOnS observation of HD\,17505
obtained in 2009 was carried out using a mask containing \ion{He}{i}, \ion{He}{ii},
and \ion{O}{iii} lines.
As presented in Table~\ref{tab:obsall} and Fig.~\ref{fig:hddef}, the detected Zeeman feature in the Stokes~$V$ spectrum is highly significant
with ${\rm FAP}<10^{-10}$. 
Given the fact that the LSD Stokes~$I$ profile consists of several components that are
heavily blended, it is unclear to which component the detected Zeeman feature belongs.

\paragraph*{HD\,35921 (LY \,Aur):}

According to \citeauthor{MaizApellaniz2019} (\citeyear{MaizApellaniz2019}, and references therein),  
this system is composed of an SB2 eclipsing system with a circular orbit with a period
of about 4\,d and a distant B0.2 IV companion located 0.6\,arcsec away and which is a SB1 system with an orbital period of 20.46\,d. 
Using Hipparcos photometry,  \citet{Lefevre2009}
reported that HD\,35921 is variable with a mean amplitude of 0.722\,mag over the orbital period of the SB2 eclipsing system.
\citet{Mayer2013} reported that the eclipsing SB2 binary 
belongs to the rare class of early-type contact systems in a phase of non-conservative mass exchange. 

The LSD analysis of the single ESPaDOnS observation of HD\,35921
obtained in 2011 was carried out using a mask containing  \ion{He}{i},
\ion{He}{ii}, \ion{O}{ii}, and \ion{O}{iii} lines.
As presented in Table~\ref{tab:obsall} and Fig.~\ref{fig:hddef}, the detected Zeeman feature in the 
Stokes~$V$ spectrum is highly significant
with ${\rm FAP}<10^{-10}$. 
According to figure~3  in \citet{Mayer2013} displaying composite Stokes~$I$ profiles in the spectra of this system,
the measured longitudinal magnetic field $\left< B_{\rm z} \right>=538\pm172$\,G corresponds to 
the secondary component.

\paragraph*{HD\,36879:}

This target is a runaway candidate from the analysis of Hipparcos proper motions
\citep{MdzinarishviliChargeishvili2005}. \citet{WalbornPanek1984}
reported that HD\,36879 shows unexplained  spectral  variability and peculiarities, among them peculiar narrow,  
variable  \ion{Si}{iv} emission lines at $\lambda$1394 and $\lambda$1403.

A magnetic field of HD\,36879 at a 3.5$\sigma$ significance level was for the first time detected in 2007 
using low-resolution FORS\,1 spectropolarimetric observations \citep{Hubrig2008}. Its presence 
was later confirmed with the moment technique in high-resolution spectropolarimetric observations acquired with 
SOFIN \citep{Hubrig2013b}.  
Our analysis of one HARPS\-pol observation acquired in 2015 and one ESPaDOnS observation from 
2009 using the LSD technique with a line mask containing \ion{He}{i}, \ion{He}{ii}, \ion{C}{iv},
\ion{N}{iv}, \ion{O}{iii},
and \ion{Si}{iv} lines, shows the presence of a clear Zeeman feature in the LSD Stokes~$V$ spectrum calculated
for the HARPS\-pol observation. As presented in Fig.~\ref{fig:hddef} and in Table~\ref{tab:obsall}, the Zeeman feature 
with  ${\rm FAP}=6\times10^{-10}$ detected 
in this observation corresponds to $\left< B_{\rm z} \right>=301\pm42$\,G.
Only a  small feature with ${\rm FAP}=4\times10^{-2}$  was detected in the LSD Stokes~$V$
spectrum calculated for the ESPaDOnS observation using the same line mask. 

Interestingly, the HARPS\-pol observation indicates that the shapes of the line profiles
for different elements are variable over a time interval of 20.5\,min, the time
difference between the beginning of subexposures. The shapes of the line profiles 
of \ion{He}{ii} and \ion{O}{iii} lines show a different character of variability compared to 
the variability of \ion{He}{i} and \ion{Si}{iv} lines.
Notably, spectrum variability based on two different observing epochs separated by 
hundreds of days was already reported in the past by \citet{Hubrig2013b}.

\paragraph*{HD\,57061 ($\tau$CMa):} 

According to \citet{Stickland1998}, this target is the brightest in NGC\,2362 and produces a bow shock in nearby gas.
NGC\,2362 is a rich young open cluster where HD\,57061
is the only target evolved away from the zero-age main sequence.
Using spatially resolved spectroscopy, \citet{MaizApellanizBarba2020} reported that HD\,57061
is a high-order multiple system. The components Aa and Ab are separated by 0.151\,arcsec and have O-type spectral classifications
with the brighter Aa component classified as an O9.2 of an uncertain luminosity classification and the
weaker Ab component classified as O9 III.
The Aa component itself is composed of an inner spectroscopic binary with a 154.92\,d period.
In addition, one of the components shows eclipses with a much shorter period of 1.282\,d
\citep{vanLeeuwenvanGenderen1997}.
Additional visual companions B, C, D, and E are listed in different literature sources.

Two components are clearly visible in
the LSD Stokes~$I$ profile calculated using the HARPS\-pol observation obtained in 2015.
We detect a clear Zeeman feature in the LSD Stokes~$V$ profile of the secondary calculated using a mask containing 
\ion{He}{i}, \ion{C}{iii}, and \ion{O}{iii} lines. As presented in Fig.~\ref{fig:hddef} and in Table~\ref{tab:obsall},
the detected Zeeman feature is highly significant with ${\rm FAP}=3\times10^{-10}$
and corresponds to a mean longitudinal magnetic field
$\left< B_{\rm z} \right>=-285\pm144$\,G.

\paragraph*{HD\,92206C:}  

According to \citet{Campillay2007}, this system with an orbital period of 2.02\,d is the second brightest member of
the open cluster NGC\,3324 after HD\,92206\,AB.
\citet{Walborn1982} reported that the spectrum of HD\,92206C displays very strong, broad hydrogen
lines, possibly similar to those in the Orion Trapezium cluster, in particular $\theta^{1}$\,Ori\,C,
and that the appearance of these lines is indicative of extreme youth.
Based on the BESO (Bochum Echelle Spectrograph for Observatorio Cerro Armazones) spectra, \citet{Mayer2017}
reported the clear presence of
a third companion visible in the \ion{He}{i} profiles at an estimated separation from the SB2 system by
less than 1\,arcsec.
Using low-resolution FORS\,2 spectra, \citet{Hubrig2013b} reported the detection of a mean longitudinal
magnetic field $\left< B_{\rm z} \right>=204\pm46$\,G using metal, helium, and
hydrogen lines.

LSD Stokes~$I$, $V$, and diagnostic null $N$ spectra have been calculated for the two available HARPS\-pol observations 
of HD\,92206C obtained in 2013 on two consecutive nights. As shown in Fig.~\ref{fig:hddef} and in
Table~\ref{tab:obsall}, using a mask containing \ion{He}{i} lines, a clear Zeeman feature in the LSD Stokes~$V$ spectrum
with ${\rm FAP}<10^{-10}$ corresponding to the third
component is detected in the observations obtained on the first night.
No feature in the LSD Stokes~$V$ spectrum using a line mask with \ion{He}{i} and \ion{He}{ii} lines is detected on
the second night, where the third component is not visible in the LSD Stokes~$I$ spectrum.

\paragraph*{HD\,101205 (V871\,Cen):}

According to \citeauthor{Sana2014} (\citeyear{Sana2014}, and references therein),
this target is a complex multiple system with three visual pairs
previously resolved at separations of 0.36\,arcsec (A,B), 1.7\,arcsec (AB, C), and 9.6\,arcsec (AB,D).
Further, the authors report that the A,B pair contains an eclipsing binary with an orbital period of about 2\,d
and a spectroscopic binary with a period of 2.8\,d.
However, it is impossible to decide which component of the
A,B pair is the eclipsing one and which is the spectroscopic one.

Our LSD analysis of the single available HARPS\-pol observation 
of HD\,101205 obtained in 2013 with a mask containing \ion{He}{i}, \ion{He}{ii}, and \ion{O}{iii} lines
reveals a Zeeman feature with ${\rm FAP}=9\times10^{-6}$.
As shown in Fig.~\ref{fig:hddef}, all components of the A,B pair are blended with each other.
Therefore, it is not clear, to which component the detected Zeeman feature should be assigned.

\paragraph*{HD\,152218A (V1294\,Sco):}

According to \citet{Rosu2022}, this system is an eclipsing eccentric binary system 
with an orbital period of 5.6\,d and rotation periods for the primary and the 
secondary  of 2.69 and 2.56\,d, respectively. It belongs to the rich
open cluster NGC\,6231 located at the core of the
Sco\,OB1 association.
From the analysis of TESS data, \citet{KolaczekSzymanski2021} reported that HD\,152218A exhibits a 
characteristic shape of brightness changes close to the periastron passage. During this passage the variable tidal potential can drive tidally 
excited oscillations (TEOs), which are usually gravity modes.
HD\,152218A also shows low-amplitude intrinsic variability in the low-frequency range, 
and can be regarded as a SPB pulsator. 

As presented in Fig.~\ref{fig:hddef},
in the first HARPS\-pol observation obtained in 2013 both components of this SB2 system
are blended with each other, whereas in the second HARPS\-pol observation obtained in 2016 they
are clearly separated. Only for the second observation we detect a clear Zeeman feature with ${\rm FAP}<10^{-10}$
corresponding to a longitudinal magnetic field $\left< B_{\rm z} \right>=-307\pm94$\,G in the secondary component. 
As shown in Table~\ref{tab:obsall}, we used for the LSD analysis of the first observation a mask containing
\ion{He}{i}, \ion{He}{ii},  \ion{C}{iv}, \ion{Si}{iii}, and 
\ion{Si}{iv} lines and for the second observation
a mask containing \ion{He}{i}, \ion{He}{ii}, \ion{C}{iii}, \ion{C}{iv}, and \ion{Si}{iv} lines.

\paragraph*{HD\,152236 ($\zeta^{01}$\,Sco):}

In the recent study by \citeauthor{Mahy2022} (\citeyear{Mahy2022}, and references therein) this system
belonging to the open cluster NGC\,6231 is considered as a candidate LBV with
a companion at a distance of about 11.5\,mas and a $\Delta$m of 6.3\,mag.
LBVs are suggested to be a brief evolutionary phase describing massive stars in transition to the Wolf-Rayet (WR)
phase.
Using linear polarization observations, \citet{Clark2012} detected position-angle
rotation through the H$\alpha$ emission line, which
can be associated either with large-scale, axisymmetric structures, or irregular wind clumps.

As shown in Fig.~\ref{fig:hddef} and in Table~\ref{tab:obsall}, we used for the first and second observations 
obtained with HARPS\-pol in 2014 a mask containing \ion{He}{i} lines, whereas for the third observation obtained in 2015 we used a mask
containing \ion{He}{i}, \ion{Si}{iii}, and \ion{Si}{iv} lines.
Distinct features in the LSD Stokes~$V$ spectra with
${\rm FAP}<10^{-10}$ and ${\rm FAP}=2\times10^{-7}$ are detected for the second and third observations, respectively.
Given the position of the definitely detected features in the Stokes~$V$ spectrum, it is possible that the 
very weak secondary component possesses a magnetic field. 

\paragraph*{HD\,152246:}

\citet{Nasseri2014}
reported that HD\,152246 belonging to the Sco\,OB1 association is at least a high-mass hierarchical triple system consisting
of an inner pair Ba/Bb with an orbital period of 6\,d and with approximately 17 and 6\,$M_{\odot}$
component mass, respectively, and an outer component A with a mass of approximately
20\,$M_{\odot}$ (see also \citealt{Mayer2017}).
The A,B pair in HD\,152246 with an orbital period of about 470\,d has an eccentricity significantly higher
than expected from the period-eccentricity diagram of other O-type systems \citep{Sana2012}.
\citet{Nasseri2014} suggested that HD\,152246 was probably born as a hierarchical quadruple system
formed by two close binaries and that the Aa,Ab initial binary may have been driven into coalescence
because of stellar evolution. 

One HARPS\-pol observation of HD\,152246 was obtained in 2013 and a second one in 2016.
As shown in Fig.~\ref{fig:hddef} and in Table~\ref{tab:obsall}, our LSD analysis using for the first observation 
a mask containing
\ion{He}{i}, \ion{He}{ii}, \ion{C}{iii}, and  \ion{C}{iv}, and a mask with \ion{He}{i}, \ion{He}{ii},
\ion{C}{iv}, \ion{Si}{iii}, and \ion{Si}{iv} lines for the second observation reveals the presence of distinct
features in the LSD Stokes~$V$ spectra with ${\rm FAP}=6\times10^{-7}$ for the first observation and ${\rm FAP}<10^{-10}$
for the second observation. Given the location of the detected features in the LSD Stokes~$V$ spectra relative to the
LSD Stokes~$I$ profiles, it is possible that they belong to the faster rotating outer component A.

\paragraph*{HD\,155775 (V1012\,Sco):} 

\citet{Lefevre2009} reported that HD\,155775 belonging to the Sco\,OB1 association is an eclipsing binary with a period of 1.515\,d 
and an amplitude of 0.061\,mag.
Our LSD analysis of the single available HARPS\-pol observation 
of HD\,155775 acquired in 2013 shows that the primary O-type component and the lower luminosity secondary component
are clearly separated in the LSD Stokes~$I$ spectrum.
As presented in Fig.~\ref{fig:hddef} and in Table~\ref{tab:obsall}, a definite Zeeman feature with ${\rm FAP}<10^{-10}$
corresponding to a mean longitudinal magnetic field $\left< B_{\rm z} \right>=-9\pm37$\,G is detected
in the primary component using a mask containing \ion{He}{i}, \ion{N}{iii}, and \ion{Si}{iii} lines.

\paragraph*{HD\,166546:}

\begin{figure}
\centering 
\includegraphics[width=0.330\textwidth]{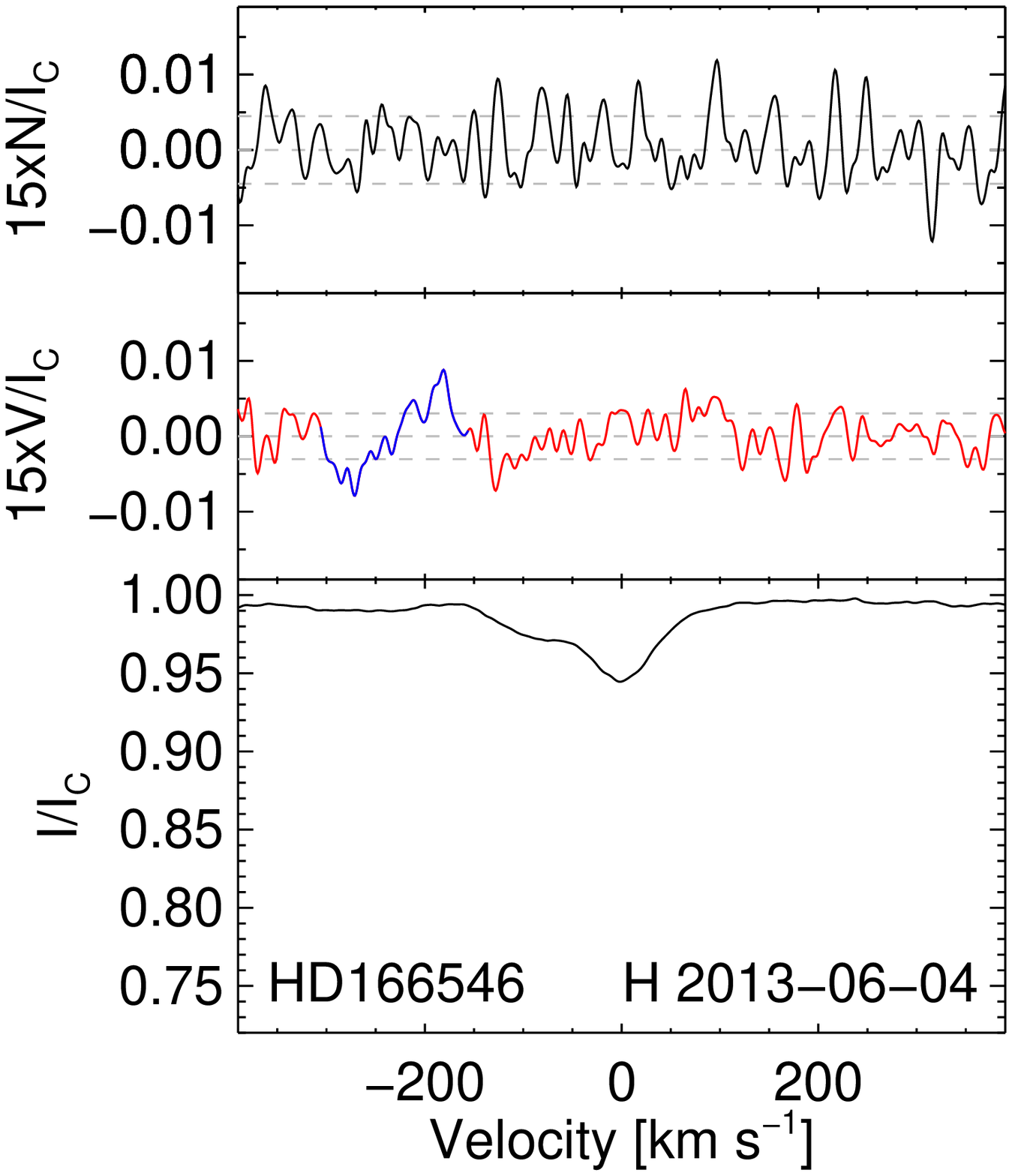}
\caption{
As Fig.~\ref{fig:collid} for the HARPS\-pol observation of
HD\,166546 acquired in 2013. In this analysis, the used line mask is populated exclusively by \ion{Si}{iii} lines. 
}
\label{fig:var_166546}
\end{figure}

Our LSD analysis of the single available
HARPS\-pol observation obtained in 2013 using a mask with exclusively \ion{Si}{iii} lines and
presented in Fig.~\ref{fig:var_166546} confirms that this target is a
binary system and could even be a triple system.
As presented in Fig.~\ref{fig:hddef} and in Table~\ref{tab:obsall}, the LSD analysis using a mask 
containing \ion{He}{i} and \ion{Si}{iii} lines reveals a definite Zeeman feature with ${\rm FAP}<10^{-10}$,
probably corresponding to the third component in this system. It can however not be excluded that the
detected feature is related to a background or foreground object. 

\paragraph*{HD\,209481 (14\,Cep):}

Using Hipparcos photometry, \citet{Lefevre2009}
reported that this semi-detached Algol eclipsing system  with an orbital
period of 3.1\,d is variable with a mean amplitude of 0.099\,mag.
\citeauthor{Mahy2011} (\citeyear{Mahy2011}, and references therein) showed that the Roche lobe filling component, either the
primary or the secondary, exhibits a strong helium and nitrogen mass fraction enhancement. The system is most likely in a
configuration where the mass transfer happened while both components were still in the hydrogen-burning core phase (e.g.
\citealt{Howarth1991}). 
Thus, this rare system composed of two evolved early-type stars is an important target for a study 
of binary interaction and its impact on massive star evolution. 

One ESPaDOnS spectrum of HD\,209481 was acquired in 2011 and another one in 2015.
As shown in Fig.~\ref{fig:hddef} and in Table~\ref{tab:obsall}, our LSD analysis using a mask 
containing \ion{He}{i} lines for the first observation in 2011 reveals the presence of
two Zeeman features in the LSD Stokes~$V$ spectrum, one with ${\rm FAP}<10^{-10}$ located close to the blue wing of the
LSD Stokes~$I$ profile and another one with ${\rm FAP}=3\times10^{-4}$ corresponding to the center of the
LSD Stokes~$I$ profile. For the second observation,
using a mask containing \ion{He}{i} and \ion{He}{ii} lines, a Zeeman feature is detected with ${\rm FAP}<10^{-10}$, corresponding to
the red wing of the LSD Stokes~$I$ profile. Given the location of the detected features relative to the observed composite 
LSD Stokes~$I$ profiles, it is not clear whether
they may belong to the additional component(s) in this system.

\paragraph*{HD\,305524:}

This target belonging to the open cluster Collinder~228 is not known to be a binary or a multiple system.
According to \citet{Levato1990}, HD\,305524 shows radial velocity
variability, probably caused by the presence of a companion.
\citet{Naze2012} used two low-resolution FORS\,2 spectropolarimetric observations of HD\,305524 to search for the presence of
a magnetic field, but no field at a significance level of 3$\sigma$ was detected.
One available HARPS\-pol observation of HD\,305524 obtained in 2016 was downloaded from the ESO archive. 
We note that the shape of the Stokes~$I$ profile calculated in our LSD analysis using a mask containing
\ion{He}{i} and \ion{He}{ii} lines indicates that this target is probably a SB2 system. As presented in
Fig.~\ref{fig:hddef} and in Table~\ref{tab:obsall}, we detect a definite Zeeman
feature with ${\rm FAP}<10^{-10}$, probably corresponding to the secondary less luminous component in this system.

\subsection{Other targets with marginal magnetic field detections}
\label{subsect:marg}

\begin{figure*}
\centering 
\includegraphics[width=0.220\textwidth]{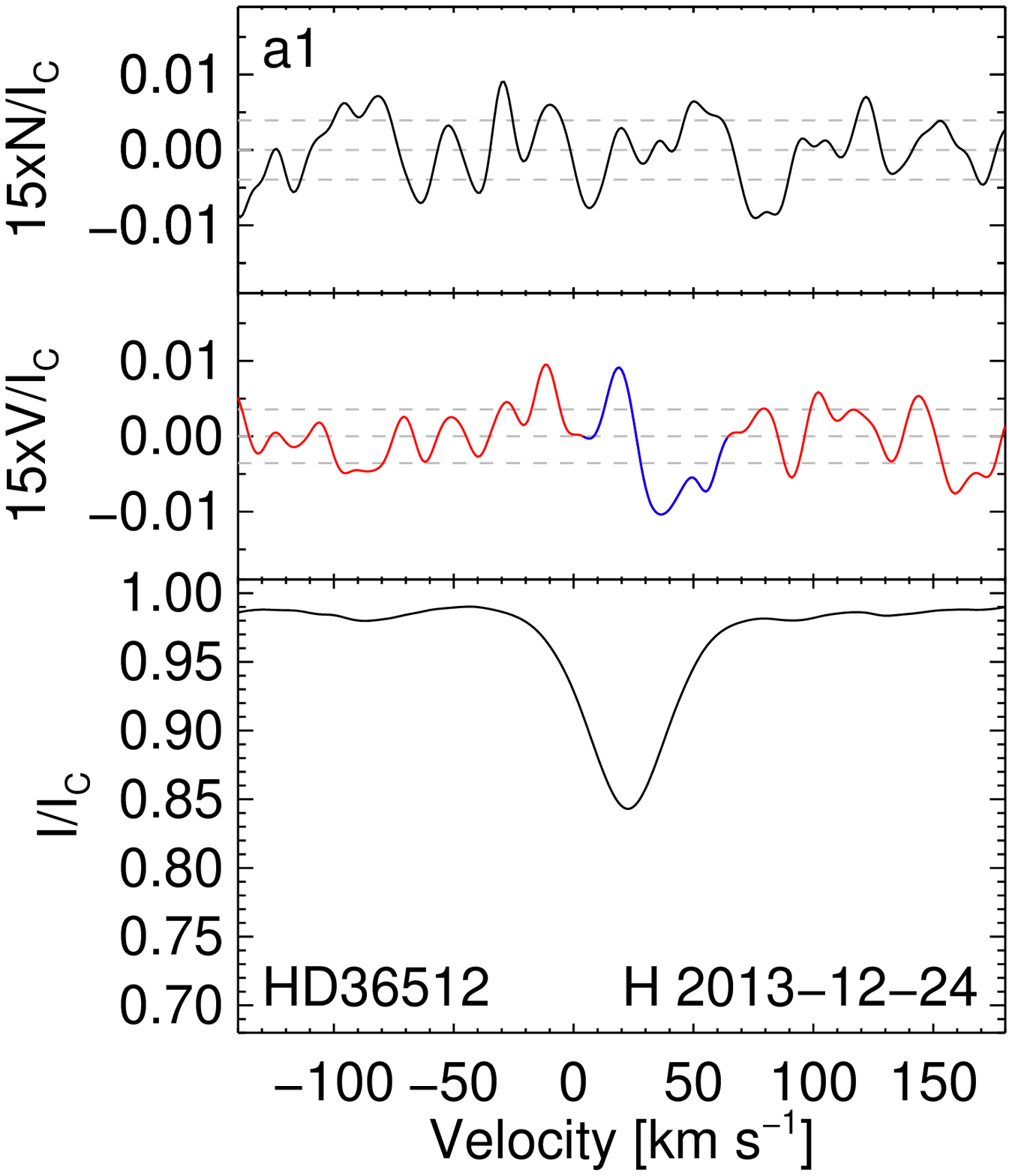}
\includegraphics[width=0.220\textwidth]{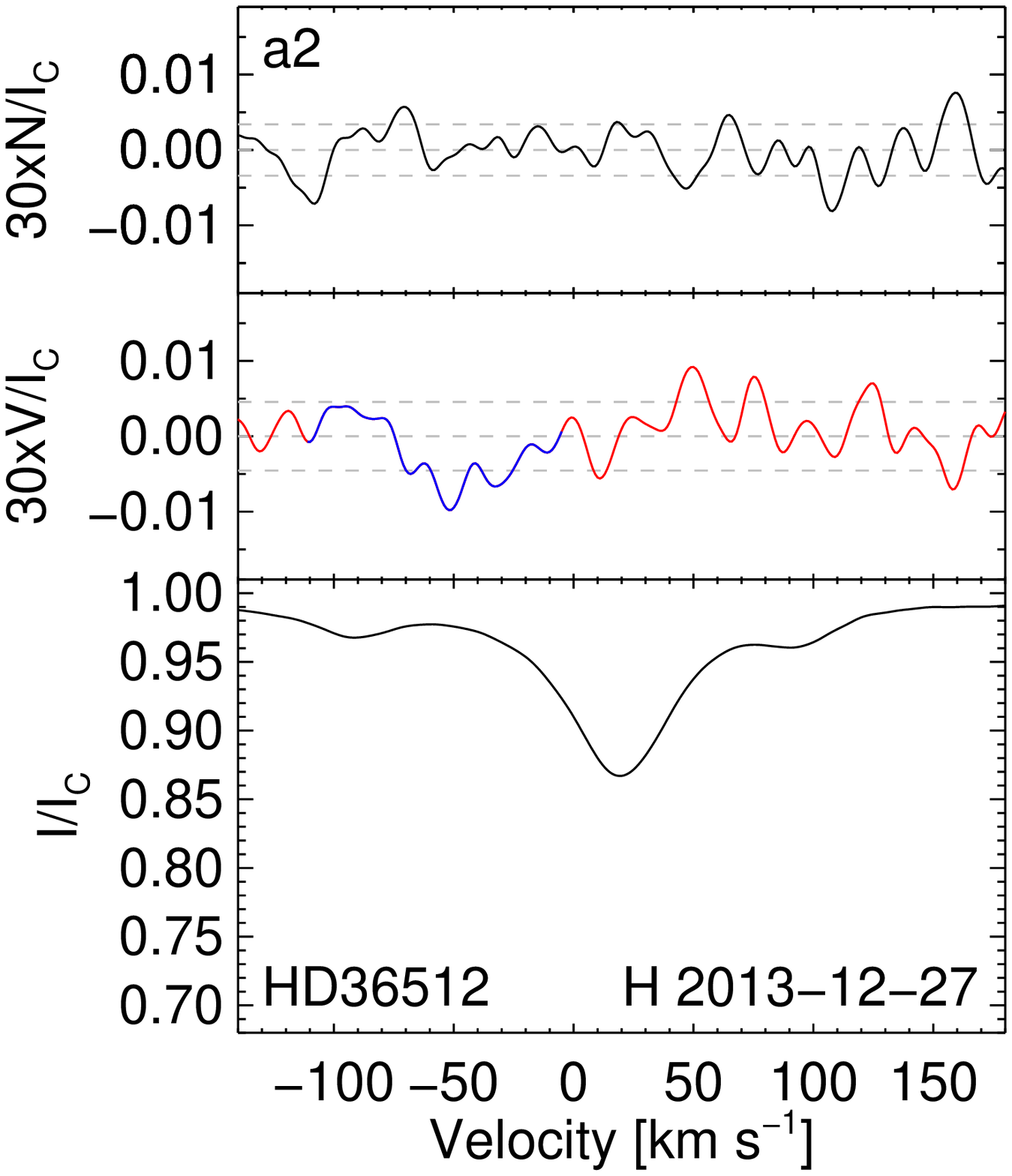}
\includegraphics[width=0.220\textwidth]{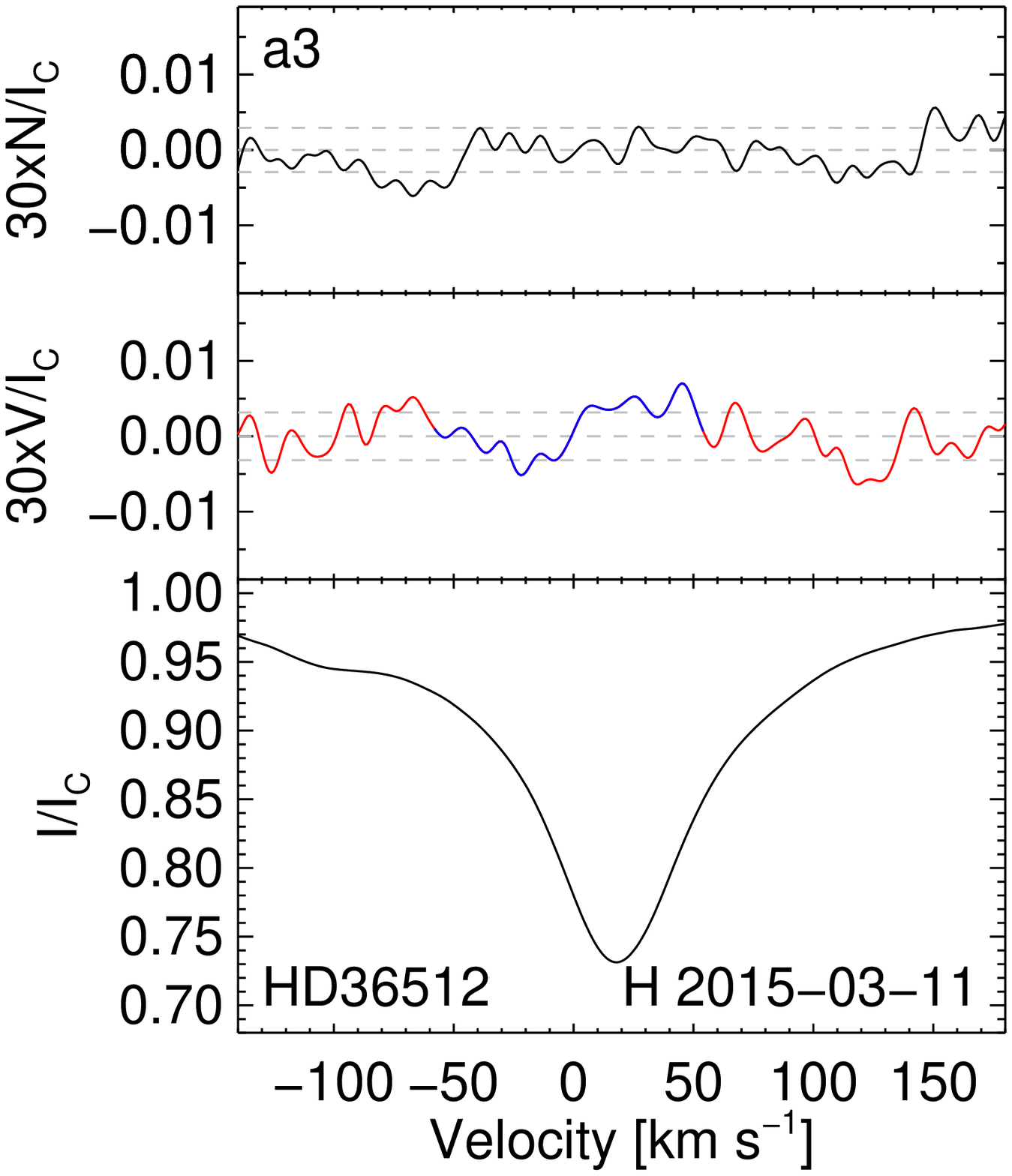}
\includegraphics[width=0.220\textwidth]{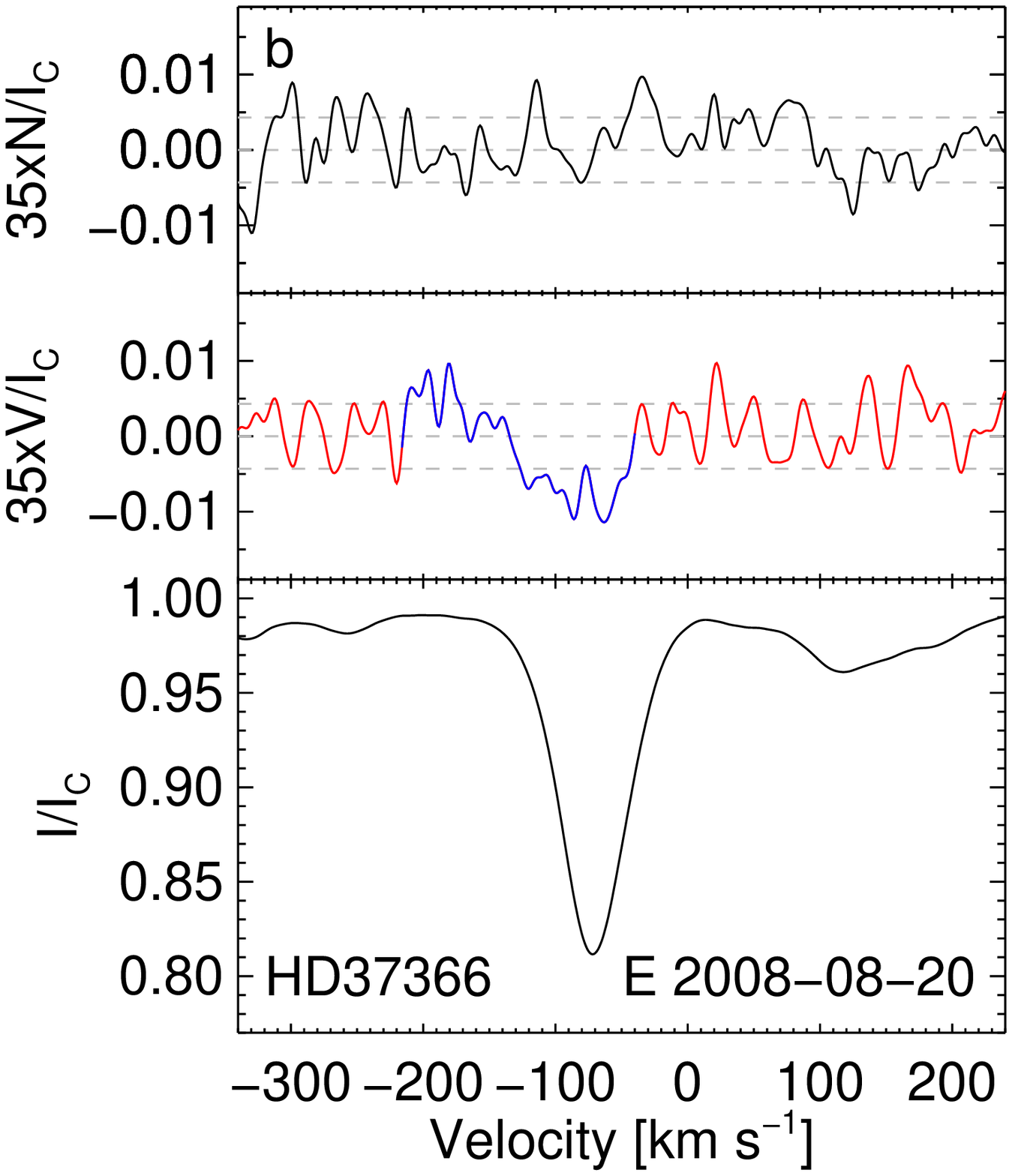}
\includegraphics[width=0.220\textwidth]{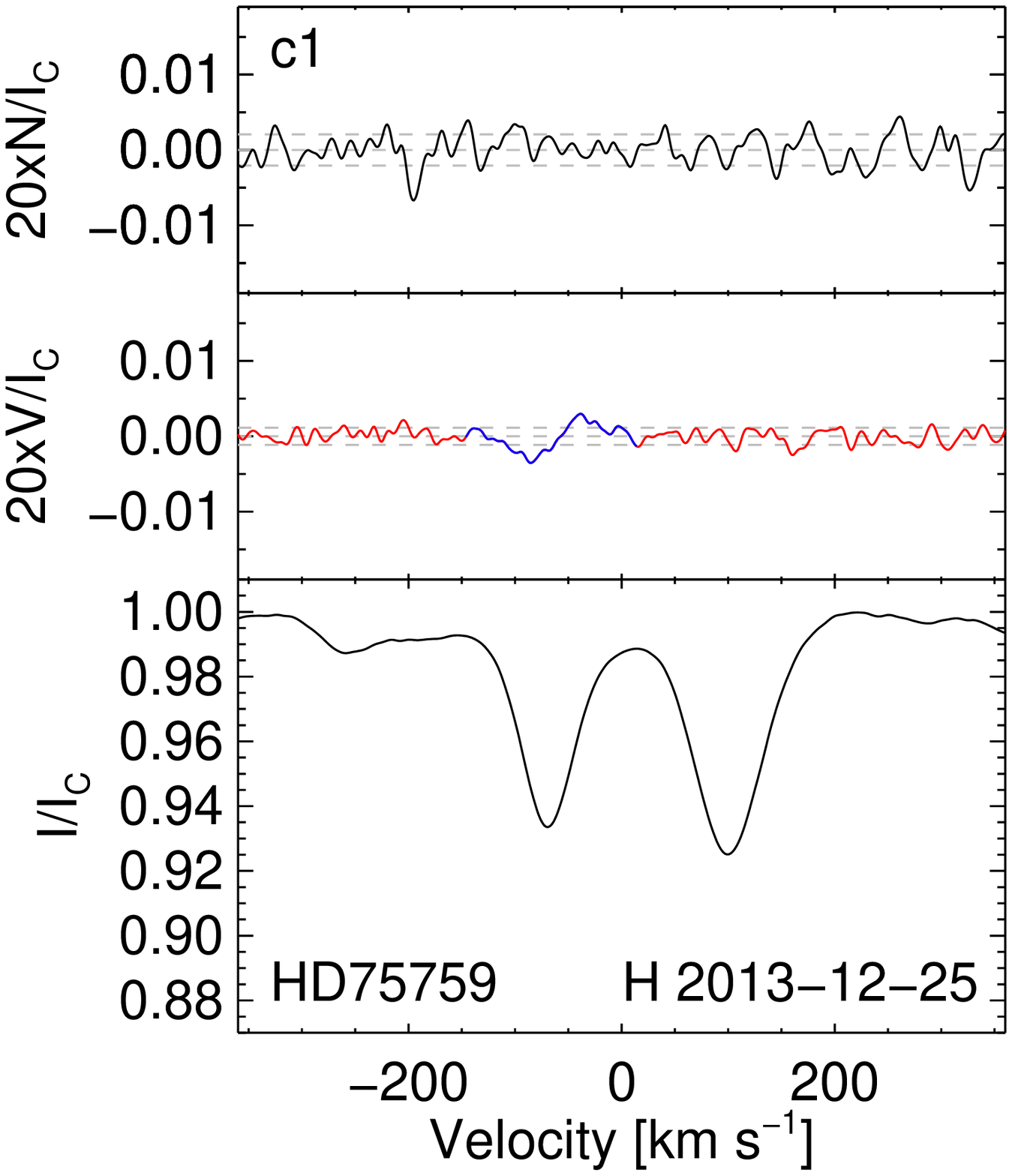}
\includegraphics[width=0.220\textwidth]{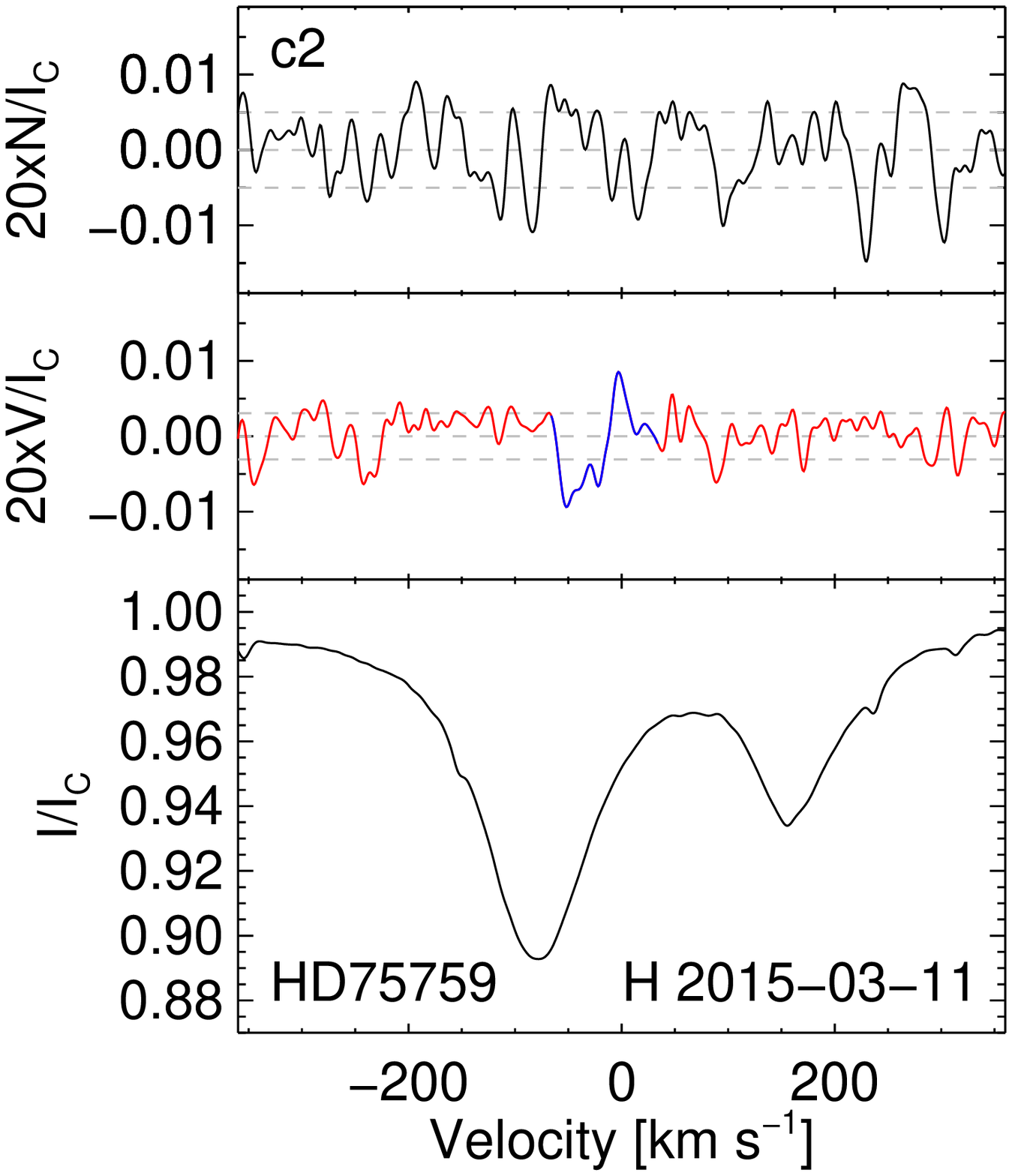}
\includegraphics[width=0.220\textwidth]{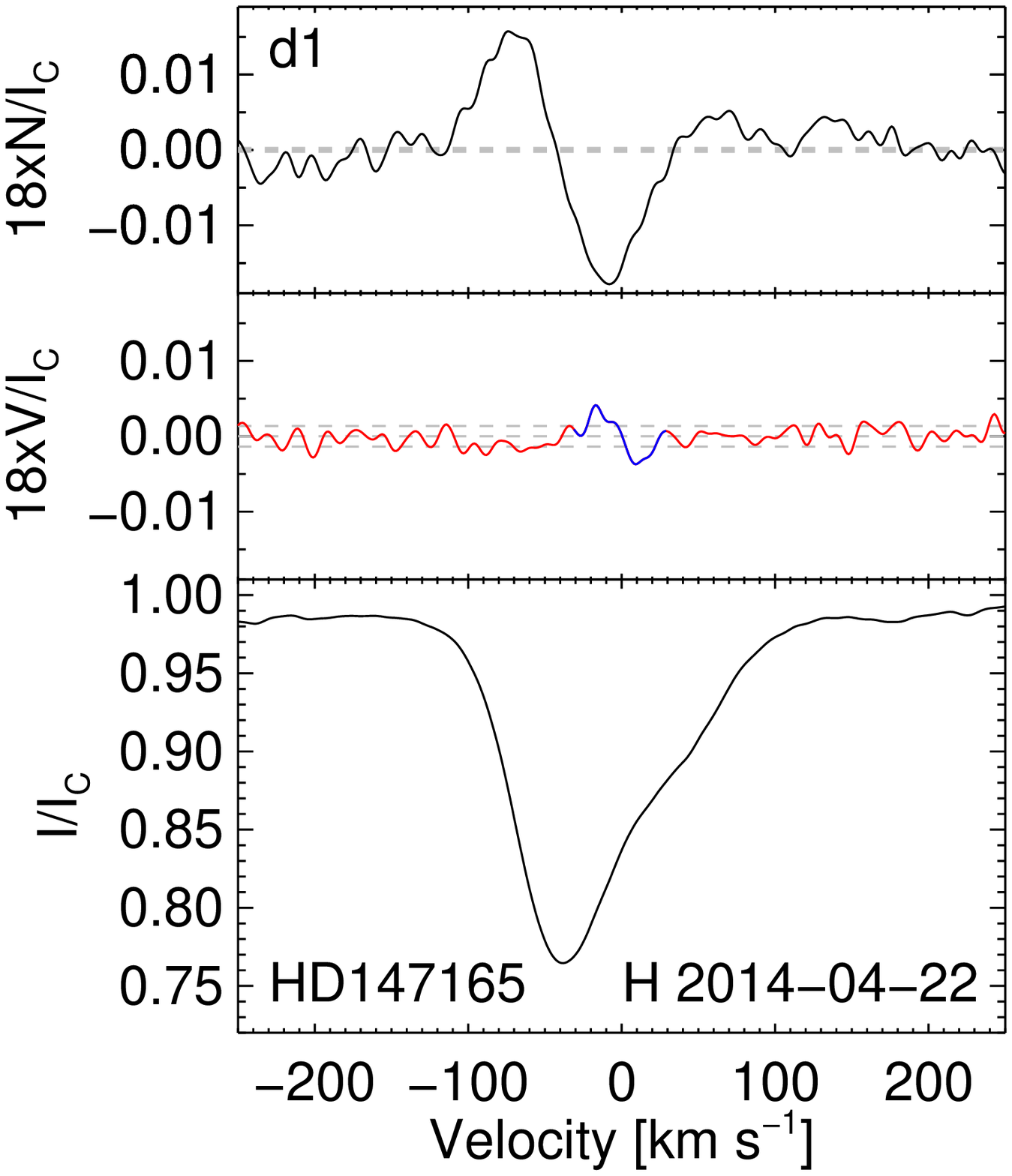}
\includegraphics[width=0.220\textwidth]{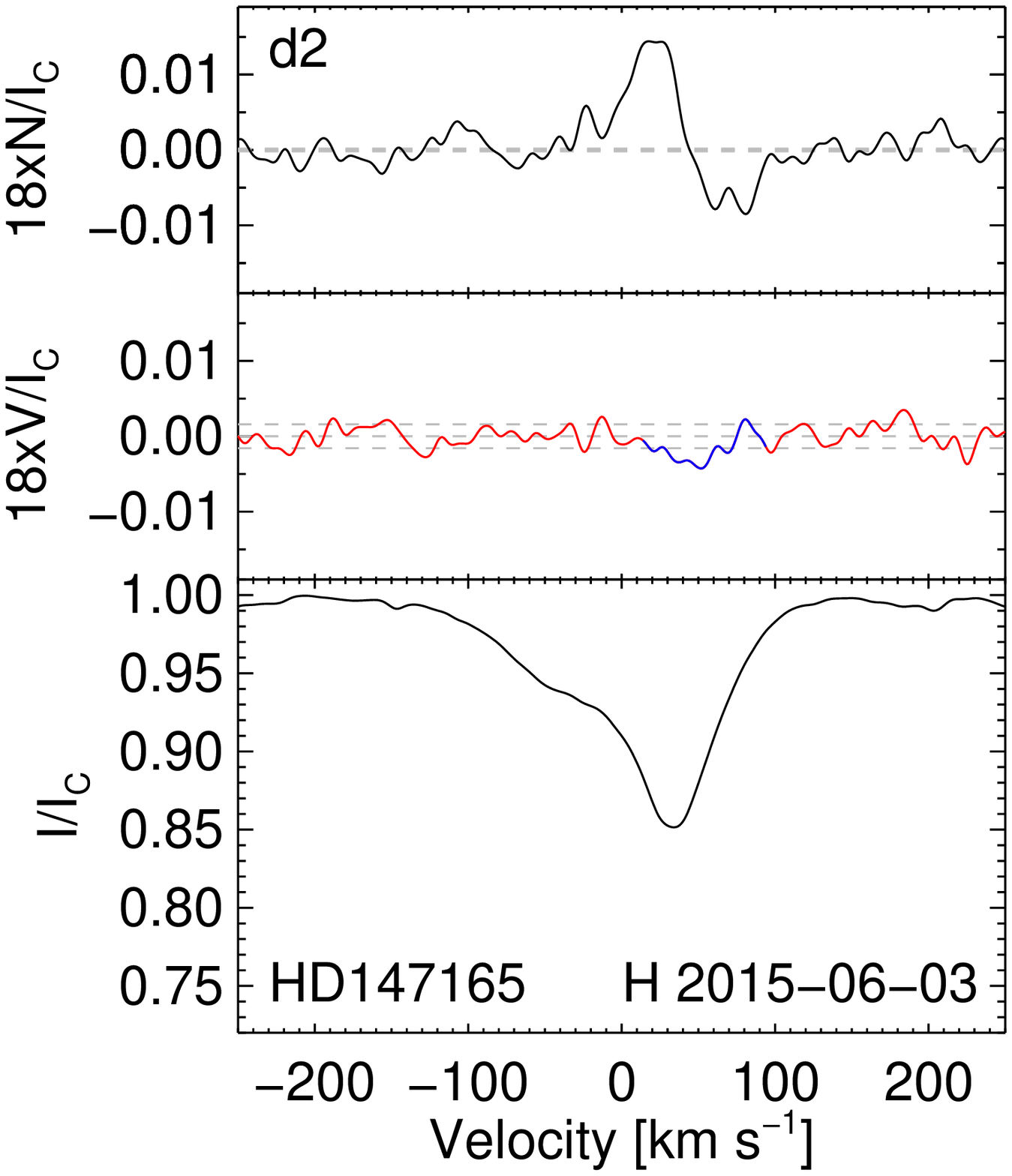}
\includegraphics[width=0.220\textwidth]{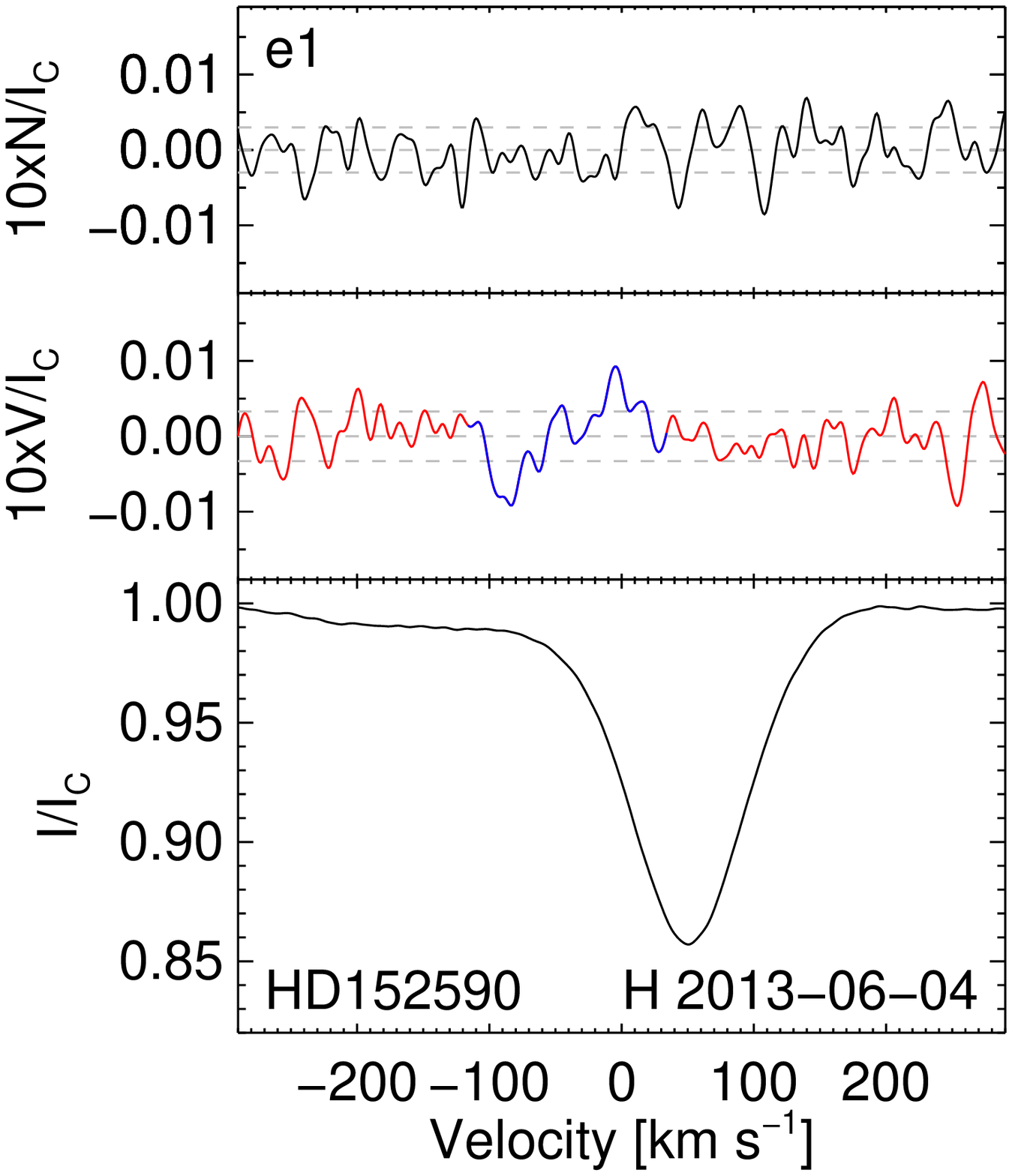}
\includegraphics[width=0.220\textwidth]{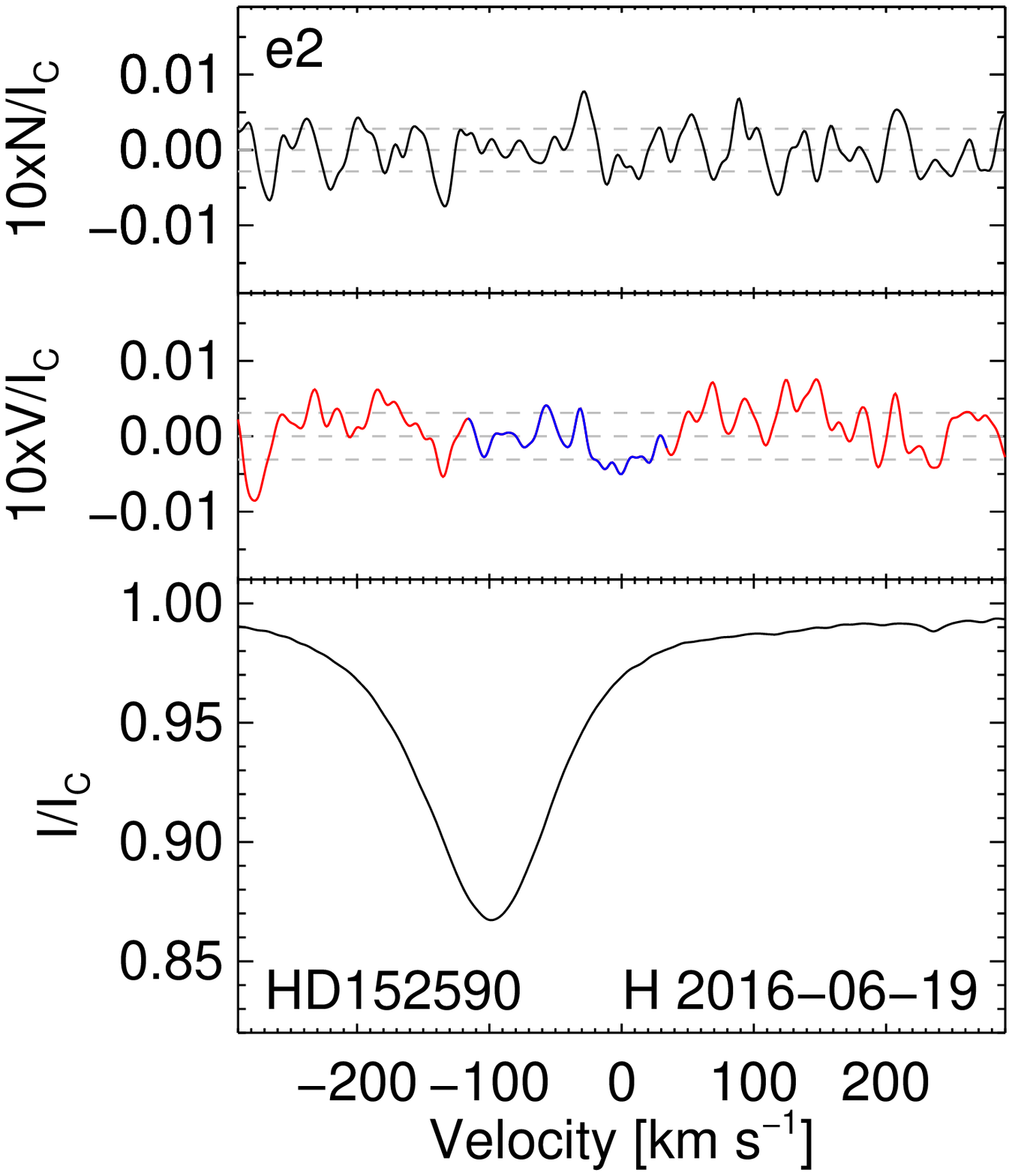}
\includegraphics[width=0.220\textwidth]{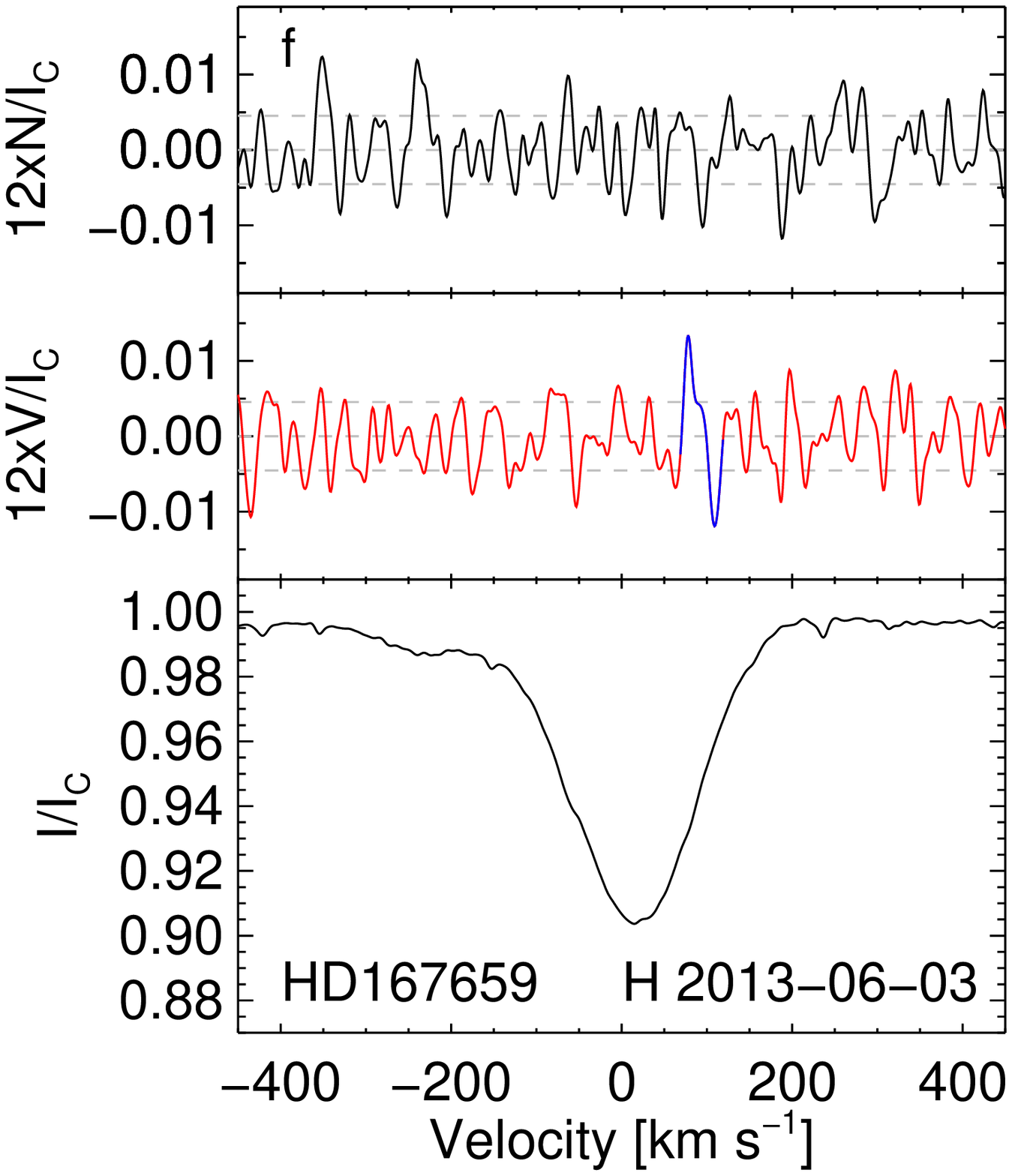}
\includegraphics[width=0.220\textwidth]{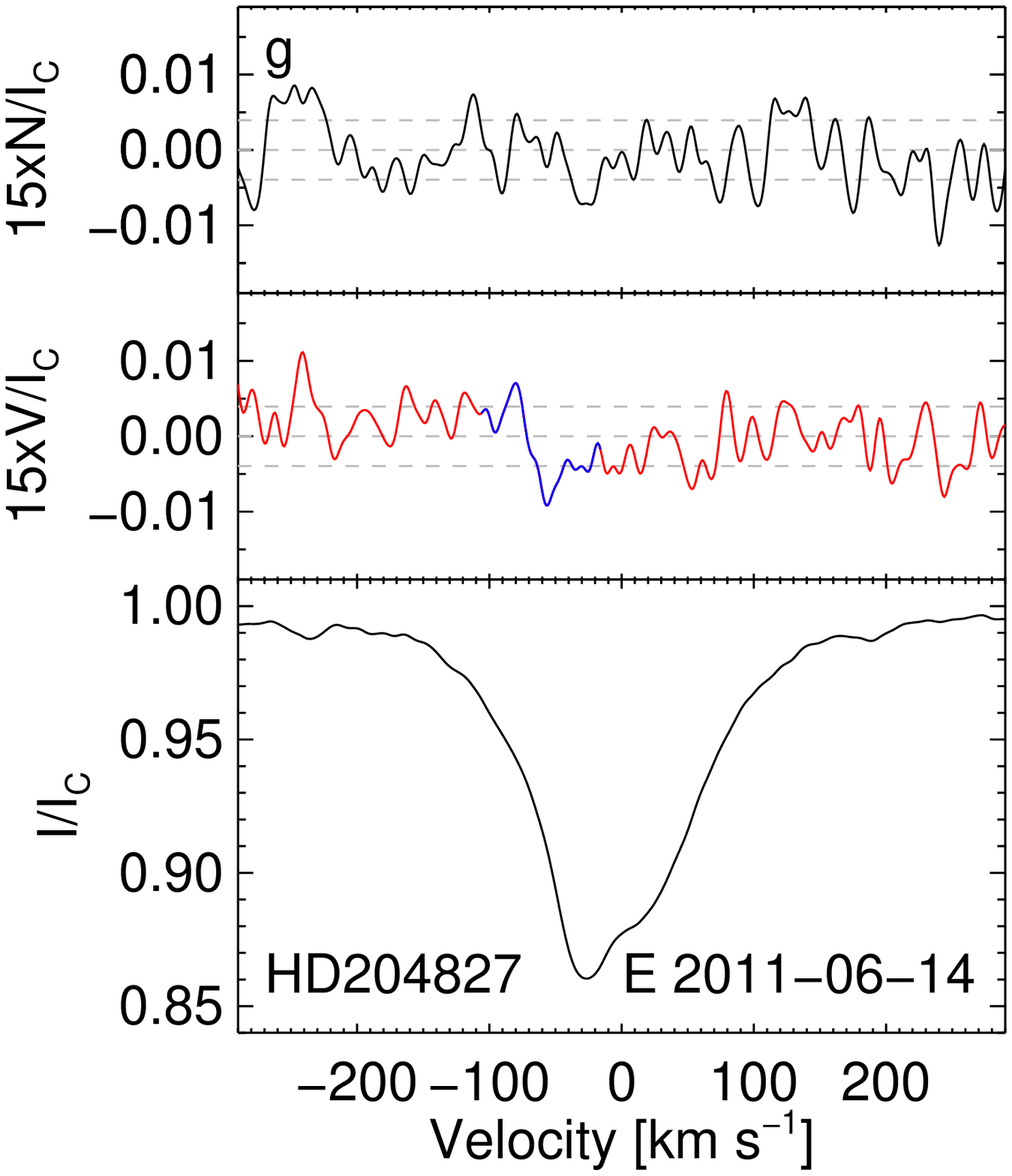}
\caption{
As Fig.~\ref{fig:collid} for the systems with marginal detections of magnetic fields:
a1-a3) LSD analysis results for HD\,36512 using three HARPS\-pol observations acquired in 2013 and 2015;
b) LSD analysis results for HD\,37366 using one ESPaDOnS observation from 2008;
c1/c2) LSD analysis results for HD\,75759 using two HARPS\-pol observations acquired in 2013 and 2015;
d1/d2) LSD analysis results for HD\,147165 with a $\beta$\,Cep primary using two HARPS\-pol observations acquired in 2014 and 2015;
e1/e2) LSD analysis results for HD\,152590 using two HARPS\-pol observations acquired in 2013 and 2016;
f) LSD analysis results for HD\,167659 using one HARPS\-pol observation from 2013; 
g) LSD analysis results for HD\,204827 using one ESPaDOnS observation from 2011.
}
\label{fig:marg}
\end{figure*}

\paragraph*{HD\,36512 ($\upsilon$\,Ori):}

According to \citet{Burssens2020}, TESS observations of this SB1 system 
suggest the presence of SLF variability, but of much lower amplitude than typical for
stars with similar spectral type. The authors also report on the
dominant frequency of 0.306\,d$^{-1}$. \citet{Smith1981} announced the presence of  non-radial pulsations. 
Although HD\,36512 is mentioned as a SB1 system in \citet{Burssens2020}, our LSD analysis using a mask containing
\ion{He}{i} lines shows that HD\,36512 is a SB2 system. 

The results of our LSD analysis of available HARPS\-pol observations, where two observations were acquired in 2013 and one in 2015, 
are presented in Fig.~\ref{fig:marg} and Table~\ref{tab:obsall}.
We achieve a marginal detection with ${\rm FAP}=3\times10^{-4}$ corresponding to $\left< B_{\rm z} \right>=123\pm10$\,G for the 
first observation using a mask
containing \ion{He}{i}, \ion{He}{ii}, \ion{C}{iii}, \ion{O}{ii}, \ion{O}{iii}, \ion{Si}{iii}, and \ion{Si}{iv} lines.
A second marginal detection is achieved for the third observation,  with ${\rm FAP}=3\times10^{-4}$ corresponding to 
$\left< B_{\rm z} \right>=-19\pm2$\,G 
using a mask containing \ion{He}{i} and \ion{He}{ii} lines. No detection was achieved for the second HARPS\-pol
observation.

\paragraph*{HD\,37366:}

According to \citeauthor{MaizApellaniz2019} (\citeyear{MaizApellaniz2019}, and references therein), this SB2 
system has a dim companion 0.6\,arcsec away.
The components of HD\,37366 are in an eccentric orbit ($P_{\rm orb}=31.9$\,d) and show different line 
broadening with projected rotational velocities of 30 and 100\,km\,s$^{-1}$ \citep{Boyajian2007}.
The secondary is a broad-lined early B-type main-sequence star. 
Our LSD analysis of the single ESPaDOnS observation of HD\,37366
obtained in 2008 using a mask containing \ion{He}{i}, \ion{He}{ii}, \ion{C}{iii}, \ion{C}{iv},
\ion{Si}{iii}, and \ion{Si}{iv} lines reveals the presence of
a Zeeman feature with ${\rm FAP}=2\times10^{-4}$ corresponding to a marginal detection of a
magnetic field in the sharp-lined O9.5IV primary component (see Fig.~\ref{fig:marg} and Table~\ref{tab:obsall}).

\paragraph*{HD\,75759 (HR\,3525):}

According to \citet{Thackeray1966}, this SB2 system which is
the central exciting source in the large \ion{H}{i} region Gum\,17, has a long orbital period of 33.3\,d 
and a large eccentricity.
\citet{Burssens2020} mention for HD\,75759 the presence of SLF  variability.

As presented in Fig.~\ref{fig:marg} and Table~\ref{tab:obsall}, our LSD analysis of the
HARPS\-pol spectrum obtained in 2013 using a mask containing  \ion{He}{i}, \ion{C}{iii}, \ion{N}{iii},
\ion{Si}{iii}, and \ion{Si}{iv} lines indicates a marginal detection in the secondary component with ${\rm FAP}=2\times10^{-4}$
corresponding to $\left< B_{\rm z} \right>=-92\pm9$\,G.
Also the LSD analysis of the second available HARPS\-pol spectrum obtained in 2015 using  a mask containing
\ion{He}{ii} and \ion{C}{iv} lines reveals a distinct feature
in the Stokes~$V$ spectrum with ${\rm FAP}=3\times10^{-4}$, but this time in the primary component.  
For this observing epoch we measure $\left< B_{\rm z} \right>=-113\pm10$\,G.

Using lower-resolution spectropolarimetric observations with the FOcal Reducer
low dispersion Spectrograph in spectropolarimetric mode (FORS\,2; \citealt{Appenzeller1998}),
\citet{Schoeller2017} reported for this system a magnetic field non-detection
with $\left< B_{\rm z} \right>=-103\pm58$\,G employing for the
measurements metal, He, and hydrogen lines.
This non-detection can probably be explained by the too low resolution of FORS\,2, only of the order of about 2000,
which does not permit to study the presence of a magnetic field in each component separately.

\paragraph*{HD\,147165 ($\sigma$\,Sco):} 

According to \citeauthor{Pigulski1992} (\citeyear{Pigulski1992}, and references therein),
this target is a quadruple system, where the primary is a high-amplitude $\beta$\,Cep-type star in a
SB2 system with an orbital period of 33\,d. The tertiary component is only about 0.4\,arcsec away whereas the fourth
component is located at a much larger distance of about 20\,arcsec.
Further, the authors reported that the primary in the SB2 system is in the shell-hydrogen burning phase presenting the
light-time effect contributing to the observed changes of the pulsation period.
\citet{Bodensteiner2018}  visually inspected WISE 22\,$\mu$m observations
and detected unaligned offset bow shocks around  HD\,147165, probably indicating binary interaction.
A spectral classification O9.5+B7 was reported by \citet{BeaversCook1980}.
A later study of this system by \citet{MaizApellaniz2021} suggests that the Aa component is a
single-lined B1\,III object and the secondary should be a B dwarf with a spectral type around B1.

One HARPS\-pol observation has been acquired in 2014 and a second one in 2015.
For the LSD analysis, we used for the first observation a mask 
containing \ion{He}{i} lines whereas for the
second observation the mask was populated using  \ion{He}{i}, \ion{He}{ii}, \ion{C}{ii},
\ion{N}{ii}, \ion{Si}{iii}, and \ion{Si}{iv} lines.
As presented in Fig.~\ref{fig:marg} and Table~\ref{tab:obsall}, a small Zeeman feature for the first
observation with ${\rm FAP}=1\times10^{-5}$, probably corresponding to the primary component, is visible in the Stokes~$V$ 
spectrum indicating a marginal detection.
No magnetic field is detected for the second observation.
The calculated diagnostic null $N$ spectra exhibit strong distinct 
features shifted to the blue in the first observation and to the red in the second observation. 
We assume that they are related to the high-amplitude pulsations in the $\beta$\,Cep primary.

\paragraph*{HD\,152590 (V1297\,Sco):}

Although \citet{Pourbaix2004} and \citet{Burssens2020} list this object as an
eclipsing SB1 system with a period of 4.5 d, HD\,152590 is clearly a SB2 system with a weak component
easily recognizable in the HARPS\-pol spectra using a LSD analysis with a mask containing \ion{He}{i} lines.

As presented in Fig.~\ref{fig:marg} and Table~\ref{tab:obsall}, using for our LSD analysis of
the two available HARPS\-pol observations acquired in 2013 and 2015 a mask 
containing \ion{He}{i}, \ion{He}{ii}, \ion{O}{iii}, and \ion{Si}{iv} lines, we achieve a marginal detection
with ${\rm FAP}=1\times10^{-4}$ in the observation acquired in 2013.

\paragraph*{HD\,167659:}

According to \citet{Sana2014}, this target in the young open cluster NGC\,6231 is a binary system
containing a faint companion with a $\Delta$m of 2.54\,mag.
The presence of a companion is also detectable in the blue wing of the LSD Stokes~$I$ profile in the single available
HARPS\-pol observation of HD\,167659 acquired in 2013. 
As shown in Fig.~\ref{fig:marg} and Table~\ref{tab:obsall}, using for our LSD analysis a mask 
containing \ion{He}{ii} lines, we detect a small Zeeman feature with ${\rm FAP}=1\times10^{-5}$ corresponding to the
red wing of the LSD Stokes~$I$ profile. It is not clear whether this feature belongs to an additional
component in this system or to a background or foreground magnetic star.

\paragraph*{HD\,204827:}

\citet{Kervella2022} show a significant detection of a companion using Gaia EDR3 data ($S/N = 3.2$),
but not using Gaia DR2 data 
($S/N = 2.47$). This star falls into the 9 per cent of their sample that has an improved $S/N$ after the release of EDR3.
\citet{Peter2012} studied massive binaries in the Cepheus OB2/3 region  and reported
that HD\,204827 has a companion at a separation of 0.09\,arcsec.

The presence of both components is clearly visible in the single available ESPaDOnS observation obtained in 2011.
As shown is Fig.~\ref{fig:marg} and Table~\ref{tab:obsall}, our LSD analysis using a mask containing
\ion{He}{i}, \ion{He}{ii}, and \ion{O}{iii} lines reveals a marginal detection of a feature with ${\rm FAP}=4\times10^{-4}$,
probably corresponding to the component exhibiting sharp lines in the LSD Stokes~$I$ spectrum.

\subsection{Other targets with no detections of magnetic fields}
\label{subsect:nodet}

\begin{figure}
\centering 
\includegraphics[width=0.220\textwidth]{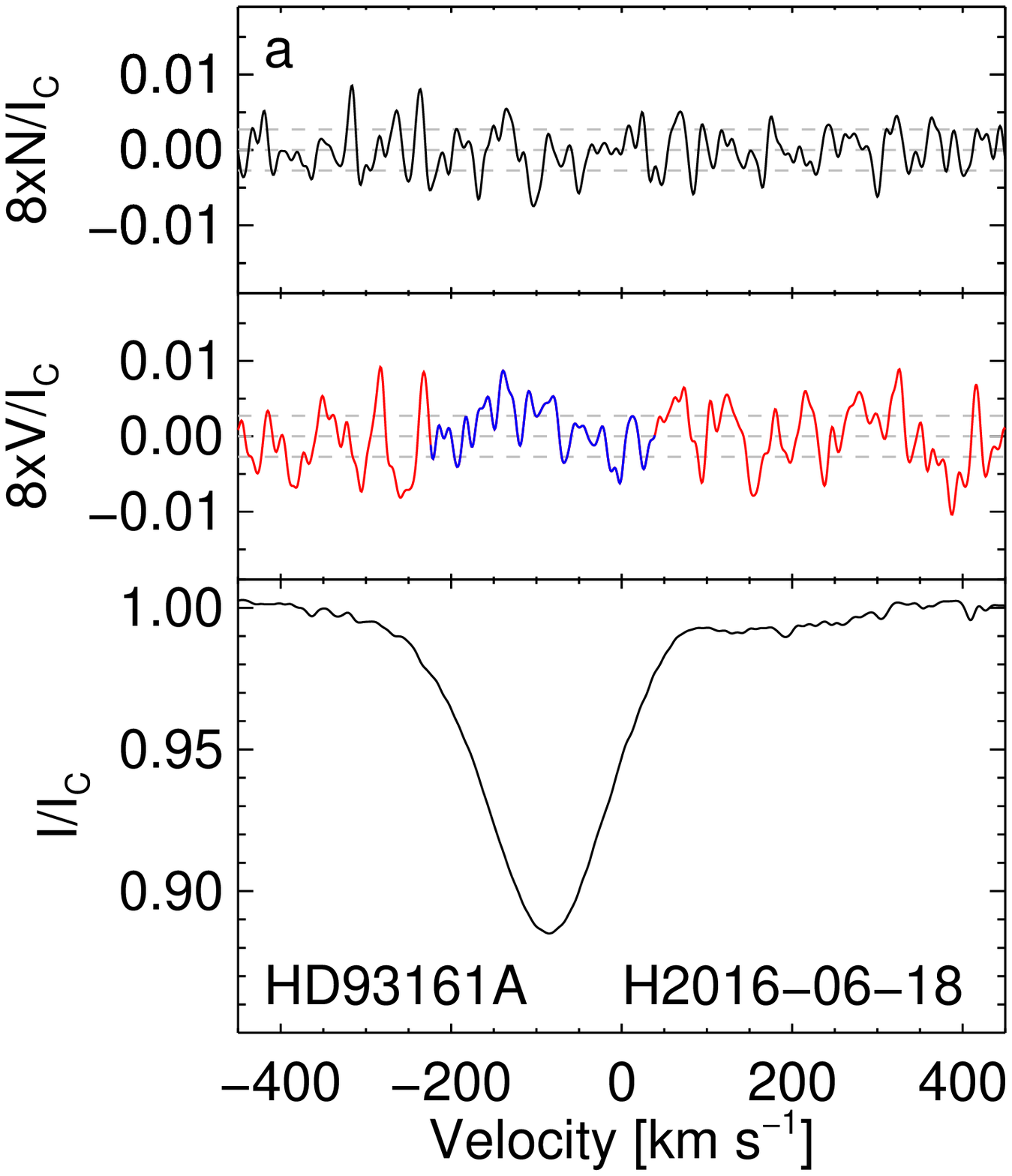}
\includegraphics[width=0.220\textwidth]{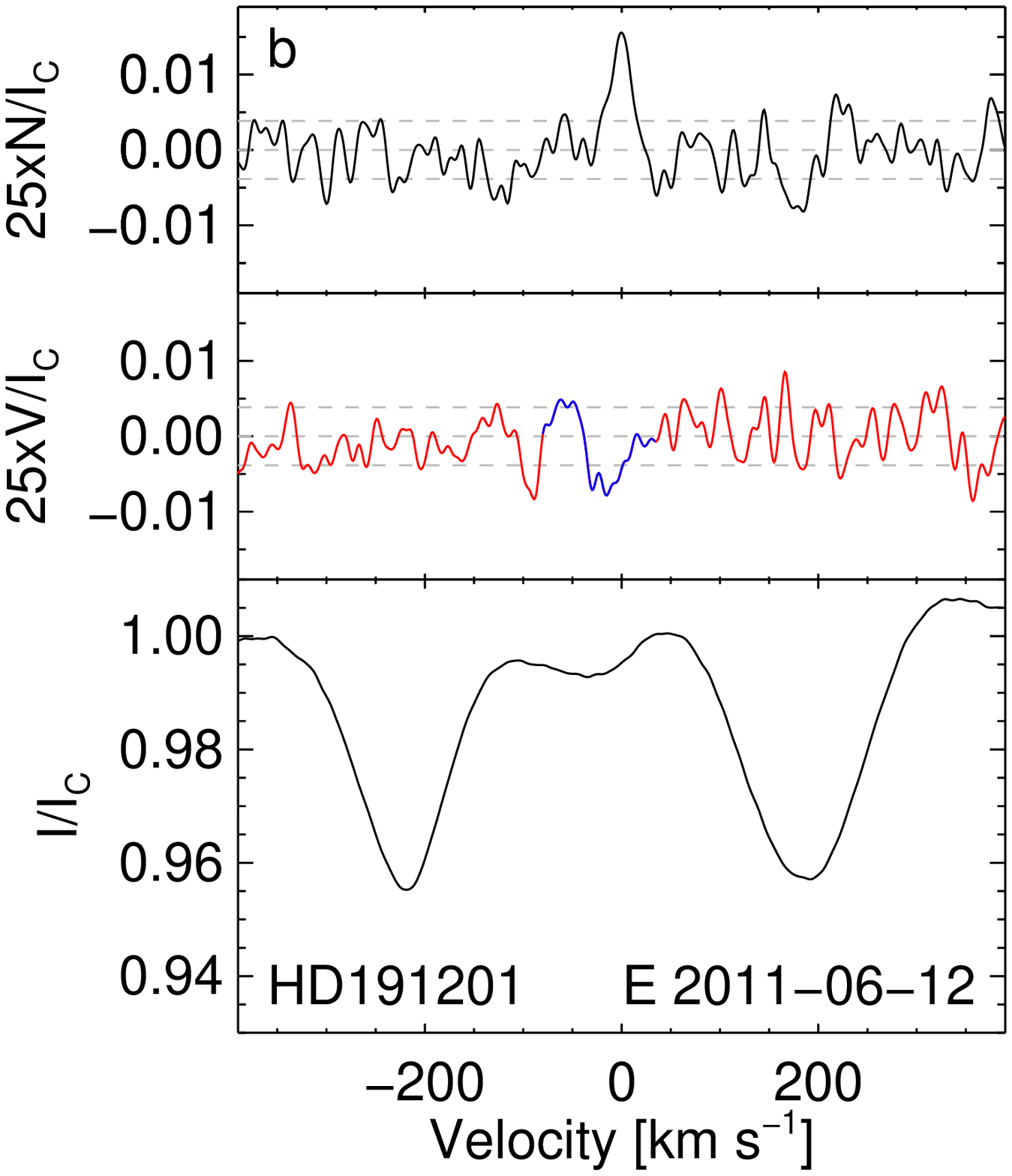}
\caption{
As Fig.~\ref{fig:collid} for the systems with non-detections:
a)  LSD analysis results for HD\,93161A using one HARPS\-pol observation from 2016;
b) LSD analysis results for HD\,191201 using one ESPaDOnS observation from 2011.
}
\label{fig:nodet}
\end{figure}

\paragraph*{HD\,93161A:}

HD\,93161 is a visual binary with a separation of 2\,arcsec in the open cluster Tr\,16.
According to \citet{Naze2005},
the primary HD\,93161A is a non-interacting SB2 system with an orbital period of 8.6\,d.
This system must be older and more evolved than main-sequence stars in the core of the Trumpler~16 cluster
\citep{Rauw2009}.

As presented in Fig.~\ref{fig:nodet} and in Table~\ref{tab:obsall}, our LSD analysis of the single HARPS\-pol
observation of HD\,93161A
obtained in 2016 using a mask containing \ion{He}{i},  \ion{He}{ii}, \ion{C}{iv}, and \ion{O}{iii} lines,
shows a non-detection, although we cannot exclude the possible presence of a very weak broad feature in the LSD Stokes
$V$ spectrum. Future high $S/N$  observations of
this system at different epochs are required to decide on the presence of a magnetic field.

\paragraph*{HD\,191201:}

This target is a visual binary with an SB2 component.
The SB2 detached system with a previously known late O-type primary and an early B-type secondary on an
orbital period of 8.3\,d is a member of
the open cluster NGC\,6871 (e.g.\ \citealt{Burkholder1997}).
According to \citet{MaizApellaniz2019},
the B component at a distance of 1\,arcsec and a $\Delta$m of 1.8\,mag 
is an O star with spectral type O9.7\,III. 

As presented in Fig.~\ref{fig:nodet} and in Table~\ref{tab:obsall}, using a mask with
\ion{He}{i}, \ion{Si}{iii}, and \ion{Si}{iv} lines in our LSD analysis of the
ESPaDOnS observation of HD\,191201
acquired in 2011, a weak feature corresponding to a 
third component is clearly visible. However, the ${\rm FAP}$ value calculated for this feature,
${\rm FAP}>4\times10^{-3}$, indicates a non-detection.  
As there is also a feature in the diagnostic null $N$ spectrum, it is possible that
the third weak component is pulsating. Additional observations are necessary
to better characterise the third component and to search for the presence of a magnetic field.

\section{Discussion}
\label{sec:meas}

We present for the first time a search for magnetic fields in a
sample of binary and multiple systems with O-type primaries, utilizing a special procedure not previously 
applied in spectropolarimetric studies of massive multiple systems.
The difficulty in detecting magnetic fields in such systems is related to a number of 
obstacles in the treatment of  their polarimetric spectra, among them a low number of useful spectral lines and 
severe blending between the spectral lines of the individual components.
In the presented study, only eleven systems show in 
their spectra sufficient component separations to allow us to study the presence of a magnetic field in the individual components.
Moreover, since nearly all massive systems belong to open clusters that usually form the densest parts in OB associations,
the polarimetric spectra can be contaminated by background or foreground stars and this contamination can lead to 
a misinterpretation of the detected Zeeman features in the Stokes~$V$ spectra.

Our spectropolarimetric survey of 36 systems reveals that 22 systems exhibit in their LSD Stokes~$V$ profiles
definitely detected Zeeman features with ${\rm FAP}\leq 10^{-5}$. Among them, for fourteen systems the detected 
features are most likely associated with an O-type component whereas for three systems we suggest an association with 
a B-type component. 
For five systems with definite Zeeman feature detections, the quadruple system HD\,101205, the PACWB systems
HD\,152248 and HD\,164794, the presumably triple system HD\,166546, and the Algol system HD\,209481,  
it is not clear whether the origin of the Zeeman features is related to 
one of the already known companions, or to additional companions, background, or foreground stars. 
Since all five stars belong to open clusters or associations, the presence of background or foreground stars cannot be excluded.

\begin{figure}
\centering 
\includegraphics[width=0.50\textwidth]{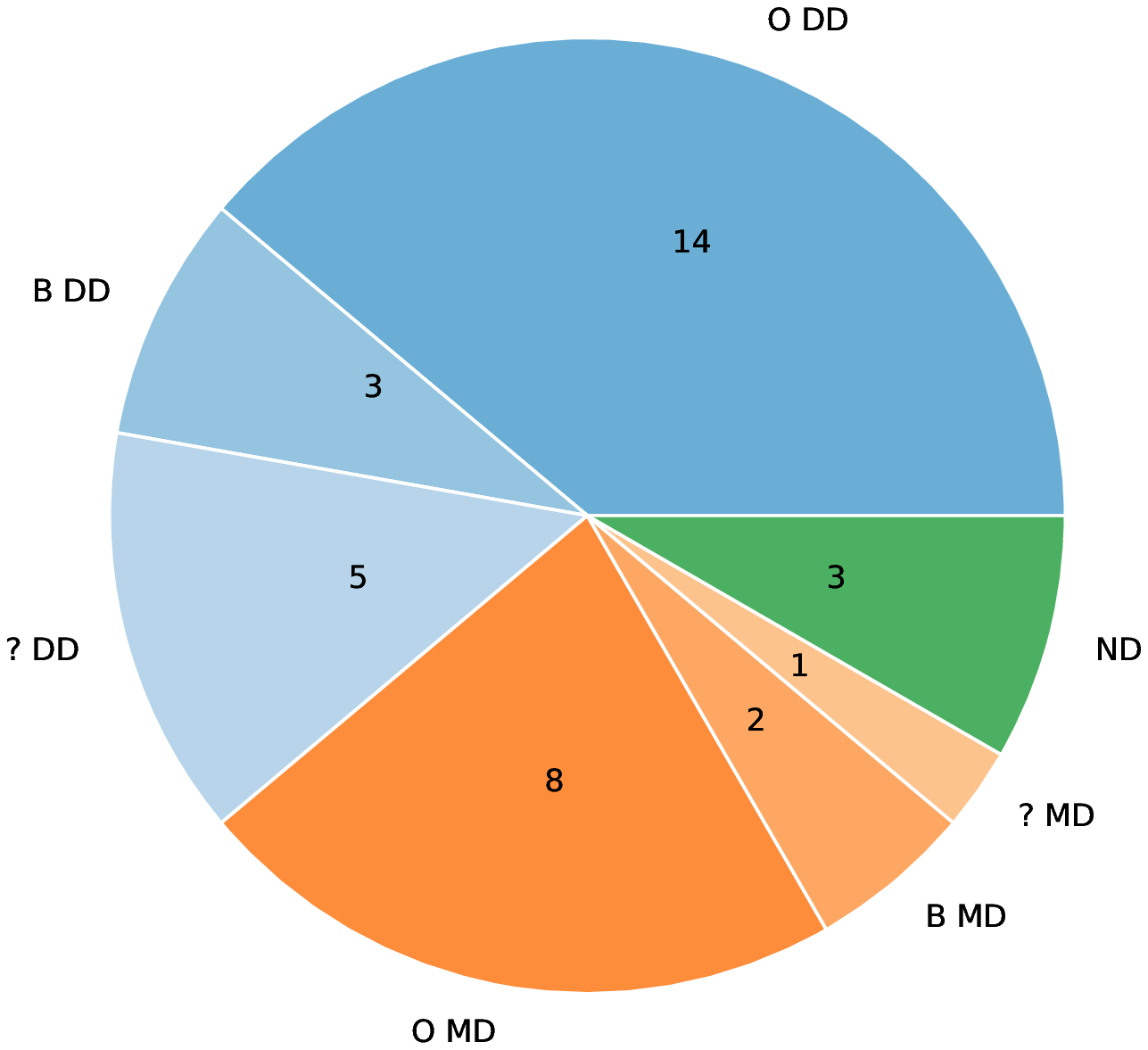}
\caption{
Schematic representation of the results of our LSD analysis of 36 systems with O-type primaries.
The labels O\,DD and B\,DD stand for definite detections in O and B components and O\,MD and B\,MD correspond to 
marginal  detections in O and B components.
?\,DD and ?\,MD indicate the number of systems with magnetic field detections
not obviously associated with already known system components. 
The label ND indicates the number of systems with non-detections.
}
\label{fig:pie}
\end{figure}

For eight systems, HD\,36512, HD\,37366, HD\,75759, HD\,147165, HD\,204827,
the PACWBs HD\,36486 and HD\,151804, and the CWB system HD\,149404,  
only marginal evidence for the detection of a Zeeman feature in the O-type components was found.
Marginal field detections have been achieved for two systems with early B-type components, HD\,152590 and the PACWB HD\,37468.
The correspondance to the known system components is not clear for the system HD\,167659 with a marginal detection, 
belonging to the young open cluster NGC\,6231.
Detections were not achieved in three systems, in the visual binary HD\,93161A,
the HMXB system HD\,153919, and in the triple system HD\,191201.
A schematic representation of the results of our LSD analysis is shown in the pie chart of Fig.~\ref{fig:pie}.

Definite detections of Zeeman features were achieved in two systems classified in the literature as Wolf-Rayet  stars,
HD\,152408 (=WR\,79a) and HD\,190918 (=WR\,133) with $\left< B_{\rm z} \right>=-355\pm44$\,G.
Reports on the detection of magnetic fields
in WR stars, which are descendants of massive O stars, are still very scarce.
Due to the strong line broadening, predominantly low-resolution spectropolarimetric data obtained with FORS\,2 were used in
the past (e.g.\ \citealt{Hubrig2016}; \citealt{Hubrig2020}). We also report on definite 
detections of a Zeeman feature in two observations of the system containing the candidate LBV HD\,152236.
These two features can probably be
associated with a very weak secondary component in this system.

Interestingly, out of the 36 systems, six binary and multiple systems, HD\,75759, HD\,92206C, HD\,152218A, HD\,152246, HD\,152590,
and HD\,164794,  were previously observed using low-resolution ($R\approx2000$) spectropolarimetry 
with FORS\,2, but, in contrast to our study, no detection was achieved for any of these systems 
(e.g.\ \citealt{Schoeller2017}). These results imply that
exclusively high-resolution spectropolarimetric observations should be used to study magnetic fields in massive binary and multiple
systems.

In this work, we  present the first observational evidence that PACWBs host magnetic stars that are possibly
responsible for the generation of their synchrotron radio emission. The LSD analysis of these systems presented
in Fig.~\ref{fig:collid} and Table~\ref{tab:obsall} shows that among the ten PACWBs, definite detections of 
Zeeman features are achieved in seven systems. Marginal detections were achieved for 
three systems, HD\,36486, HD\,37468, and HD\,151804. For the systems HD\,36486, HD\,167971, and
HD\,190918, it was possible to estimate their mean longitudinal magnetic field strengths, which are in the range from 
a few hundred Gauss up to the highest value of 1.3\,kG measured in one of the three components of the blue supergiant 
system HD\,167971.
This triple system is of extraordinary interest for future studies of magnetism in O stars because for all 
three components either definite or marginal detections are achieved. Furthermore, this system
with two colliding-wind regions \citep{DeBecker2018} contains interacting components: 
the third component is orbiting the short period ($P_{\rm orb}=3.3$\,d) eclipsing overcontact pair with both components filling their 
corresponding Roche lobes.
Obviously, the confirmation of the presence of such a strong field of 1.3\,kG in this system using
additional polarimetric observations obtained at higher $S/N$ is highly desirable. 
Definite detections of Zeeman features have also been achieved in the two interacting X-rayCWB systems 
HD\,1337 and HD\,48099 having short orbital periods of 3.5\,d and 3.1\,d, respectively. 

Especially studies of magnetic fields in systems
with periods short enough to allow for interaction between the components are important because they can permit to get a deeper 
insight into the role of the magnetic field not only in binary and multiple system evolution, but also in the context
of supernova progenitors. According to \citet{Sana2012}, massive stars in close binaries with primaries of mass
$M\gtrsim15\,M_{\odot}$ will interact with a stellar companion before they explode as core-collapse supernovae with neutron stars 
or black holes as end products.
A significant fraction of the targets in our sample have rather short orbital periods, below 10\,d.
This is probably due to the fact that 
systems with massive components usually belong to open clusters:
according to \citet{vanLeeuwenvanGenderen1997}, massive binary and multiple  systems
play an important role for the dynamical evolution of star clusters because they can serve as an energy exchange
mechanism, leading to the creation of very tight binaries and
escaping cluster members, but also to temporary and usually very unstable three to four star configurations.
Nine targets in our sample,  HD\,1337, HD\,35921, HD\,36486, HD\,48099, HD\,149404, HD\,152246, HD\,167971, HD\,190916, and HD\,209481
are mentioned in the literature as interacting systems where the components, depending on 
their evolutionary state, undergo or in the past have already undergone episodes
of conservative or non-conservative mass transfer or even coalescence.

The role of magnetic fields in binary interactions as well as magnetic field geometries in components of binary systems
and multiple systems are currently 
almost unexplored: among O-type systems, only Plaskett's star with $P_{\rm orb}=14.4$\,d
is known to harbour a broad-line secondary with a longitudinal magnetic field of a semi-amplitude of about 500\,G \citep{Grunhut2013}. 
However, significant difficulties in the interpretation of the observed Stokes~$V$ profiles were recently 
reported by \citet{Grunhut2022}.
As for systems with B-type stars, \citet{Hubrig2014a} reported on a study of the field geometry of a 
magnetic Ap secondary in the close SB2 system HD\,161701 
with a B-type HgMn primary: the obtained measurements indicated that the field on the surface of the Ap star permanently facing the
primary component is positive and that it is negative on the far side.
The authors suggested that the alignment of the magnetic axis with the orbital radius vector may indicate that the
generation of the magnetic field was a dynamic process during tidal synchronization. 
Obviously, binary systems and multiple systems with more than one component possessing a magnetic field are of considerable interest 
for observations of the configuration and strength of their fields.
In the presented study we identified only two such systems, 
the SB2 system HD\,75759 with $P_{\rm orb}=33.3$\,d, for which marginal detections 
were achieved in both components, and the triple system HD\,167971 with $P_{\rm orb}=3.3$\,d for the inner eclipsing binary
and a $P_{\rm orb}$ of about 22\,yr for the outer system. As already mentioned above, for this triple system 
we achieved definite and marginal detections in all three spectroscopic components.

Notably, previously reported spectropolarimetric observations of magnetic Bp and O stars 
in double and multiple systems
seem to indicate that for the majority of the systems only one of the components is magnetic.
No magnetic field was detected in the primary component of Plaskett's star \citep{Grunhut2022}.
\citet{Hubrig2014b}
presented low-resolution FORS\,2 spectropolarimetric observations of three components in the multiple 
system ADS\,10991 in the Trifid 
nebula with an age of just a few 0.1\,Myr: the O7.5 star HD\,164492A, the B1 triple system HD\,164492C \citep{Gonzalez2017}, 
and the Herbig Be star HD\,164492D. 
No magnetic field at a significance level of 3$\sigma$ was detected in HD\,164492A and HD\,164492D, whereas a longitudinal
magnetic field of the order of 600\,G is present in the primary of the system HD\,164492C.
Among the SB2 systems with early B-type primaries studied by \citet{Shultz2018}, most systems 
contain only one magnetic component.   

Our survey of high-resolution ESO HARPS\-pol and CFHT 
ESPaDOnS archival spectropolarimetric observations of O-type 
binary and multiple systems revealed a sizable sample of such systems with 
potential magnetic components, pointing out that the incidence rate of magnetic fields 
in massive binaries has probably been greatly underestimated. 
This is a puzzling and rather unexpected result, indicating that multiplicity probably plays a role in the generation of 
magnetic fields in massive stars. Assuming that interactions between the components in massive binary and multiple systems
constitutes a dominant process for the generation of magnetic fields, we should indeed expect that a majority of 
these systems will harbour a magnetic component. 

Evidently, to answer the principal question of the possible origin of magnetic fields in massive stars, 
further work based on dedicated spectropolarimetric monitoring of the studied systems is needed.
As most of the targets have been observed only once or twice,
our LSD analysis solely targeted the detection of their magnetic fields using the presence of 
conspicuous features in their LSD Stokes~$V$ spectra.
Clearly, to establish the magnetic nature of the surveyed targets with more confidence,
additional spectropolarimetric observations are needed to investigate the temporal variability of
the detected Zeeman features.
Furthermore, to better understand the physical processes playing a role in massive stars, it is also important to investigate the 
three-dimensional structure of these systems based on monitoring of 
the strength of the magnetic field, the intensity variability of different spectral lines, and of the radial velocity shifts.
Spectropolarimetric monitoring will allow the disentangling of the individual stellar spectra
and will provide crucial information on the magnetic field geometries, 
rotation periods, and the geometry of the orbits as well as on the peculiar properties 
and the evolutionary history of the components from the study of their radii, abundances, and rotation rates.
We also plan to use TESS and Kepler data to look more closely
at the O-type systems studied in this paper
in order to further constrain their properties,
as well as search for low-frequency power excess
associated with massive stars
(e.g.\ \citealt{Bowman2019}).

\section*{Acknowledgements}

We thank the referee, Gautier Mathys, for insightful comments.
Our work is based on observations made with ESO telescopes at the La Silla
and Paranal observatories under programmes 187.D-0917(A) and 191.D-0255
and observations collected at the Canada-France-Hawaii Telescope (CFHT), which is operated by the National
Research Council Canada, the Institut National des Sciences de l'Univers of the Centre National de la
Recherche Scientifique of France, and the University of Hawaii.

\section*{Data Availability}

The HARPS and ESPaDOnS spectropolarimetric observations underlying this article can be obtained
from the ESO and CFHT archives.

\bsp	
\label{lastpage}

\begin{thebibliography}{99}

\bibitem[\protect\citeauthoryear{Abbott et al.}{2020}]{Abbott2020}
Abbott R., et al.,
2020, ApJ, 896, L44

\bibitem[\protect\citeauthoryear{Alecian et al.}{2015}]{Alecian2015}
Alecian E., et al.,
2015, in Meynet G., Georgy C., Groh J., Stee P., eds,
New Windows on Massive Stars, IAUS, 307, 330

\bibitem[\protect\citeauthoryear{Ankay et al.}{2001}]{Ankay2001}
Ankay A., Kaper L., de Bruijne J.~H.~J., Dewi J., Hoogerwerf R., Savonije G.~J.,
2001, A\&A, 370, 170

\bibitem[\protect\citeauthoryear{Appenzeller et al.}{1998}]{Appenzeller1998}
Appenzeller I., et al.,
1998, The Messenger, 94, 1

\bibitem[\protect\citeauthoryear{Beavers \& Cook}{1980}]{BeaversCook1980}
Beavers W.~I., Cook D.~B.,
1980, ApJS, 44, 489

\bibitem[\protect\citeauthoryear{Benaglia \& Koribalski}{2004}]{BenagliaKoribalski2004}
Benaglia P., Koribalski B.,
2004, A\&A, 416, 171

\bibitem[\protect\citeauthoryear{Berdyugin et al.}{2016}]{Berdyugin2016}
Berdyugin A., et al.,
2016, A\&A, 591, A92

\bibitem[\protect\citeauthoryear{Bodensteiner et al.}{2018}]{Bodensteiner2018}
Bodensteiner J., Baade D., Greiner J., Langer N.,
2018, A\&A, 618, A110

\bibitem[\protect\citeauthoryear{Boroson et al.}{2003}]{Boroson2003}
Boroson B., Vrtilek S.~D., Kallman T., Corcoran M.,
2003, ApJ, 592, 516

\bibitem[\protect\citeauthoryear{Bowman et al.}{2019}]{Bowman2019}
Bowman D.~M., et al.,
2019, {\aap}, {621}, {A135}

\bibitem[\protect\citeauthoryear{Boyajian et al.}{2007}]{Boyajian2007}
Boyajian T.~S., et al.,
2007, ApJ, 664, 1121

\bibitem[\protect\citeauthoryear{Brinkmann}{1981}]{Brinkmann1981}
Brinkmann W.,
1981, A\&A, 94, 323

\bibitem[\protect\citeauthoryear{Burkholder, Massey \& Morrell}{1997}]{Burkholder1997}
Burkholder V., Massey P., Morrell N.,
1997, ApJ, 490, 328

\bibitem[\protect\citeauthoryear{Burssens et al.}{2020}]{Burssens2020}
Burssens S., et al.,
2020, A\&A, 639, A81

\bibitem[\protect\citeauthoryear{Buscombe}{1969}]{Buscombe1969}
Buscombe W.,
1969, MNRAS, 144, 1

\bibitem[\protect\citeauthoryear{Campillay et al.}{2007}]{Campillay2007}
Campillay A., et al.,
2007, VI Reunion Anual SociedadChilena de Astronomia (SOCHIAS), 63

\bibitem[\protect\citeauthoryear{Clark et al.}{2002}]{Clark2002}
Clark J.~S., Goodwin S.~P., Crowther P.~A., Kaper L., Fairbairn M., Langer N., Brocksopp C.,
2002, A\&A, 392, 909

\bibitem[\protect\citeauthoryear{Clark et al.}{2012}]{Clark2012}
Clark J.~S., Najarro F., Negueruela I., Ritchie B.~W., Urbaneja M.~A., Howarth I.~D.,
2012, A\&A, 541, A145

\bibitem[\protect\citeauthoryear{De Becker \& Raucq}{2013}]{DeBeckerRaucq2013}
De Becker M., Raucq F.,
2013, A\&A, 558, A28

\bibitem[\protect\citeauthoryear{De Becker}{2018}]{DeBecker2018}
De Becker M.,
2018, A\&A, 620, A144

\bibitem[\protect\citeauthoryear{Donati, Semel \& Rees}{1992}]{Donati1992}
Donati J.-F., Semel M., Rees D.~E.,
1992, \aap, 265, 669

\bibitem[\protect\citeauthoryear{Donati et al.}{1997}]{Donati1997} 
Donati J.-F., Semel M., Carter B.~D., Rees D.~E., Collier Cameron A.,
1997, MNRAS, 291, 658

\bibitem[\protect\citeauthoryear{Dsilva et al.}{2022}]{Dsilva2022}
Dsilva K., Shenar T., Sana H., Marchant P.,
2022, A\&A, {\sl accepted}, arXiv:2212.06927

\bibitem[\protect\citeauthoryear{Fabry et al.}{2021}]{Fabry2021}
Fabry M., Hawcroft C., Frost A.~J., Mahy L., Marchant P., Le Bouquin J.-B., Sana H.,
2021, A\&A, 651, A119 

\bibitem[\protect\citeauthoryear{Falanga et al.}{2015}]{Falanga2015}
Falanga M., Bozzo E., Lutovinov A., Bonnet-Bidaud J.~M., Fetisova Y., Puls J.,
2015, A\&A, 577, A130

\bibitem[\protect\citeauthoryear{Ferrario et al.}{2009}]{Ferrario2009}
Ferrario L., Pringle J.~E., Tout C.~A., Wickramasinghe D.~T.,
2009, MNRAS, 400, L71

\bibitem[\protect\citeauthoryear{Frost et al.}{2021}]{Frost2021}
Frost A.~J., et al.,
2021, OBA Stars: Variability and Magnetic Fields, 19

\bibitem[\protect\citeauthoryear{Garmany, Conti \& Massey}{1980}]{Garmany1980}
Garmany C.~D., Conti P.~S., Massey P.,
1980, ApJ, 242, 1063

\bibitem[\protect\citeauthoryear{Gonz\'alez et al.}{2017}]{Gonzalez2017}
Gonz\'alez J.~F., et al.,
2017, MNRAS, 467, 437

\bibitem[\protect\citeauthoryear{Grunhut et al.}{2013}]{Grunhut2013}
Grunhut J.~H., et al.,
2013, MNRAS, 428, 1686

\bibitem[\protect\citeauthoryear{Grunhut et al.}{2017}]{Grunhut2017}
Grunhut J.~H., et al.,
2017, MNRAS, 465, 2432

\bibitem[\protect\citeauthoryear{Grunhut et al.}{2022}]{Grunhut2022}
Grunhut J.~H., et al.,
2022, MNRAS, 512, 1944

\bibitem[\protect\citeauthoryear{Harvin et al.}{2002}]{Harvin2002}
Harvin J.~A., Gies D.~R., Bagnuolo W.~G.~Jr., Penny L.~R., Thaller M.~L.,
2002, ApJ, 565, 1216

\bibitem[\protect\citeauthoryear{Hill, Crawford \& Barnes}{1974}]{Hill1974}
Hill G., Crawford D.~L., Barnes J.~V.,
1974, AJ, 79, 1271

\bibitem[\protect\citeauthoryear{Holgado et al.}{2020}]{Holgado2020}
Holgado G., et al.,
2020, A\&A, 638, A157

\bibitem[\protect\citeauthoryear{Holgado et al.}{2022}]{Holgado2022}
Holgado G., Sim{\'o}n-D{\'\i}az S., Herrero A., Barb{\'a} R.~H.,
2022, A\&A, 665, A150

\bibitem[\protect\citeauthoryear{Howarth et al.}{1991}]{Howarth1991}
Howarth I.~D., Stickland D.~J., Prinija R.~K., Koch R.~H., Pfeiffer R.~J.,
1991, The Observatory, 111, 167 

\bibitem[\protect\citeauthoryear{Hubrig et al.}{2008}]{Hubrig2008}
Hubrig S., Sch{\"o}ller M., Schnerr R.~S., Gonz{\'a}lez J.~F., Ignace R., Henrichs H.~F.,
2008, A\&A, 490, 793

\bibitem[\protect\citeauthoryear{Hubrig et al.}{2011}]{Hubrig2011}
Hubrig S., et al.,
2011, A\&A, 528, A151

\bibitem[\protect\citeauthoryear{Hubrig et al.}{2013a}]{Hubrig2013}
Hubrig S., Ilyin I., Sch\"oller M., Lo Curto G.,
2013a, Astron.\ Nachr., 334, 1093

\bibitem[\protect\citeauthoryear{Hubrig et al.}{2013b}]{Hubrig2013b}
Hubrig S., et al.,
2013b, A\&A, 551, A33

\bibitem[\protect\citeauthoryear{Hubrig et al.}{2014a}]{Hubrig2014a}
Hubrig S., et al.,
2014a, MNRAS, 440, L6

\bibitem[\protect\citeauthoryear{Hubrig et al.}{2014b}]{Hubrig2014b}
Hubrig S., et al.,
2014b, A\&A, 564, L10

\bibitem[\protect\citeauthoryear{Hubrig et al.}{2016}]{Hubrig2016}
Hubrig S., Scholz K., Hamann W.-R., Sch{\"o}ller M., Ignace R., Ilyin I., Gayley K.~G., Oskinova L.~M.,
2016, MNRAS, 458, 3381

\bibitem[\protect\citeauthoryear{Hubrig et al.}{2018}]{Hubrig2018}
Hubrig S., J\"arvinen S., Madej J., Bychkov V., Ilyin I., Sch\"oller M., Bychkova L.,
2018, MNRAS, 477, 3791

\bibitem[\protect\citeauthoryear{Hubrig et al.}{2020}]{Hubrig2020}
Hubrig S., Sch{\"o}ller M., Cikota A., J{\"a}rvinen S.~P.,
2020, MNRAS, 499, L116

\bibitem[\protect\citeauthoryear{Ibanoglu, {\c{C}}ak{\i}rl{\i} \& Sipahi}{2013}]{Ibanoglu2013}
Ibanoglu C., {\c{C}}ak{\i}rl{\i} {\"O}., Sipahi E.,
2013, MNRAS, 436, 750

\bibitem[\protect\citeauthoryear{J\"arvinen et al.}{2020}]{Jarvinen2020}
J\"arvinen S.~P., Hubrig S., Mathys G., Khalack V., Ilyin I., Adigozalzade H.,
2020, MNRAS, 499, 2734 

\bibitem[\protect\citeauthoryear{J{\"a}rvinen et al.}{2021}]{Jarvinen2021}
J{\"a}rvinen S.~P., Hubrig S., Sch{\"o}ller M., K{\"u}ker M., Ilyin I., Chojnowski S.~D.,
2021, MNRAS, 501, 4534

\bibitem[\protect\citeauthoryear{J\"arvinen et al.}{2022}]{Jarvinen2022}
J{\"a}rvinen S.~P., Hubrig S., Sch{\"o}ller M., Cikota A., Ilyin I., Hummel C.~A., K{\"u}ker M.,
2022, MNRAS, 510, 4405

\bibitem[\protect\citeauthoryear{Kaper et al.}{1996}]{Kaper1996}
Kaper L., Henrichs H.~F., Nichols J.~S., Snoek L.~C., Volten H., Zwarthoed G.~A.~A.,
1996, A\&AS, 116, 257

\bibitem[\protect\citeauthoryear{Kervella et al.}{2022}]{Kervella2022}
Kervella P., Arenou F., Th{\'e}venin F.,
2022, A\&A, 657, A7

\bibitem[\protect\citeauthoryear{Kolaczek-Szymanski et al.}{2021}]{KolaczekSzymanski2021}
Ko{\l}aczek-Szyma{\'n}ski P.~A., Pigulski A., Michalska G., Mo{\'z}dzierski D., R{\'o}{\.z}a{\'n}ski T.,
2021, A\&A, 647, A12

\bibitem[\protect\citeauthoryear{Kupka, Dubernet \& the VAMDC Collaboration}{2011}]{Kupka2011}
Kupka F., Dubernet M.-L., VAMDC Collaboration,
2011, Baltic Astronomy, 20, 503

\bibitem[\protect\citeauthoryear{Lef\'evre et al.}{2009}]{Lefevre2009}
Lef{\`e}vre L., Marchenko S.~V., Moffat A.~F.~J., Acker A.,
2009, A\&A, 507, 1141

\bibitem[\protect\citeauthoryear{Levato et al.}{1990}]{Levato1990}
Levato H., Malaroda S., Garcia B., Morrell N., Solivella G.,
1990, ApJS, 71, 323

\bibitem[\protect\citeauthoryear{Linder}{2008}]{Linder2008}
Linder N.,
2008, PhD thesis, Univ.\ of Li\'ege

\bibitem[\protect\citeauthoryear{Mahy et al.}{2010}]{Mahy2010}
Mahy L., Rauw G., Martins F., Naz{\'e} Y., Gosset E., De Becker M., Sana H., Eenens P.,
2010, ApJ, 708, 1537

\bibitem[\protect\citeauthoryear{Mahy et al.}{2011}]{Mahy2011}
Mahy L., Martins F., Machado C., Donati J.-F., Bouret J.-C.,
2011, A\&A, 533, A9

\bibitem[\protect\citeauthoryear{Mahy et al.}{2018}]{Mahy2018}
Mahy L., et al.,
2018, A\&A, 616, A75 

\bibitem[\protect\citeauthoryear{Mahy et al.}{2022}]{Mahy2022}
Mahy L., et al.,
2022, A\&A, 657, A4

\bibitem[\protect\citeauthoryear{Ma\'iz-Apell\'aniz et al.}{2004}]{MaizApellaniz2004}
Ma{\'\i}z-Apell{\'a}niz J., Walborn N.~R., Galu{\'e} H.~{\'A}., Wei L.~H.,
2004, ApJS, 151, 103

\bibitem[\protect\citeauthoryear{Ma\'iz Apell\'aniz et al.}{2018}]{MaizApellaniz2018}
Ma{\'\i}z Apell{\'a}niz J., Barb{\'a} R.~H., Sim{\'o}n-D{\'\i}az S., Sota A., Trigueros P{\'a}ez E., Caballero J.~A., Alfaro E.~J.,
2018, A\&A, 615, A161

\bibitem[\protect\citeauthoryear{Ma\'iz-Apell\'aniz et al.}{2019}]{MaizApellaniz2019}
Ma\'iz-Apell\'aniz J., et al.,
2019, A\&A, 626, A20

\bibitem[\protect\citeauthoryear{Ma\'iz-Apell\'aniz \& Barb\'a}{2020}]{MaizApellanizBarba2020}
Ma{\'\i}z Apell{\'a}niz J., Barb{\'a} R.~H.,
2020, A\&A, 36, A28

\bibitem[\protect\citeauthoryear{Ma\'iz-Apell\'aniz et al.}{2021}]{MaizApellaniz2021}
Ma{\'\i}z Apell{\'a}niz J., Barb{\'a} R.~H., Fari{\~n}a C., Sota A., Pantaleoni Gonz{\'a}lez M., Holgado G., Negueruela I., Sim{\'o}n-D{\'\i}az S.,
2021, A\&A, 646, A11

\bibitem[\protect\citeauthoryear{Mason et al.}{1998}]{Mason1998}
Mason B.~D., Gies D.~R., Hartkopf W.~I., Bagnuolo W.~G.~Jr., ten Brummelaar T., McAlister H.~A.,
1998, AJ, 115, 821

\bibitem[\protect\citeauthoryear{Mathys}{1995}]{Mathys1995}
Mathys G.,
1995, A\&A, 293, 733

\bibitem[\protect\citeauthoryear{Mayer et al.}{2013}]{Mayer2013}
Mayer P., Drechsel H., Harmanec P., Yang S., {\v{S}}lechta M.,
2013, A\&A, 559, A22

\bibitem[\protect\citeauthoryear{Mayer et al.}{2017}]{Mayer2017}
Mayer P., et al.,
2017, A\&A, 600, A33

\bibitem[\protect\citeauthoryear{Mdzinarishvili \& Chargeishvili}{2005}]{MdzinarishviliChargeishvili2005}
Mdzinarishvili T.~G., Chargeishvili K.~B.,
2005, A\&A, 431, L1

\bibitem[\protect\citeauthoryear{Nasseri et al.}{2014}]{Nasseri2014}
Nasseri A., et al.,
2014, A\&A, 568, A94

\bibitem[\protect\citeauthoryear{Naz\'e et al.}{2005}]{Naze2005}
Naz{\'e} Y., Antokhin I.~I., Sana H., Gosset E., Rauw G.,
2005, MNRAS, 359, 688

\bibitem[\protect\citeauthoryear{Naz\'e et al.}{2012}]{Naze2012}
Naz{\'e} Y., Bagnulo S., Petit V., Rivinius T., Wade G., Rauw G., Gagn{\'e} M.,
2012, MNRAS, 423, 3413

\bibitem[\protect\citeauthoryear{Neiner et al.}{2015}]{Neiner2015}
Neiner C., Grunhut J., Leroy B., De Becker M., Rauw G.,
2015, A\&A, 575, A66

\bibitem[\protect\citeauthoryear{Offner et al.}{2022}]{Offner2022}
Offner S.~S.~R., Moe M., Kratter K.~M., Sadavoy S.~I., Jensen E.~L.~N., Tobin J.~J.,
2022, in Inutsuka S., Aikawa Y., Muto T., Tomida K., Tamura M., eds,
Protostars and Planets VII, arXiv:2203.10066

\bibitem[\protect\citeauthoryear{Peter et al.}{2012}]{Peter2012}
Peter D., Feldt M., Henning Th., Hormuth F.,
2012, A\&A, 538, A74

\bibitem[\protect\citeauthoryear{Pigulski}{1992}]{Pigulski1992}
Pigulski A.,
1992, A\&A, 261, 203

\bibitem[\protect\citeauthoryear{Portegies Zwart, Pooley \& Lewin }{2002}]{PortegiesZwart2002}
Portegies Zwart S.~F., Pooley D., Lewin W.~H.~G.,
2002, ApJ, 574, 762

\bibitem[\protect\citeauthoryear{Pourbaix et al.}{2004}]{Pourbaix2004}
Pourbaix D., et al.,
2004, A\&A, 424, 727

\bibitem[\protect\citeauthoryear{Raucq et al.}{2018}]{Raucq2018}
Raucq F., Rauw G., Mahy L., Sim{\'o}n-D{\'\i}az S.,
2018, A\&A, 614, A60

\bibitem[\protect\citeauthoryear{Rauw et al.}{2001}]{Rauw2001}
Rauw G., Naz{\'e} Y., Carrier F., Burki G., Gosset E., Vreux J.-M.,
2001, A\&A, 368, 212

\bibitem[\protect\citeauthoryear{Rauw et al.}{2009}]{Rauw2009}
Rauw G., Naz{\'e} Y., Fern{\'a}ndez Laj{\'u}s E., Lanotte A.~A., Solivella G.~R., Sana H., Gosset E.,
2009, MNRAS, 398, 1582

\bibitem[\protect\citeauthoryear{Rauw et al.}{2012}]{Rauw2012}
Rauw G., Sana H., Spano M., Gosset E., Mahy L., De Becker M., Eenens P.,
2012, A\&A, 542, A95 

\bibitem[\protect\citeauthoryear{Rauw et al.}{2016}]{Rauw2016}
Rauw G., et al.,
2016, A\&A, 589, A121

\bibitem[\protect\citeauthoryear{Richardson et al.}{2021}]{Richardson2021}
Richardson N.~D., et al.,
2021, ApJ, 908, L3

\bibitem[\protect\citeauthoryear{Rosu et al.}{2022}]{Rosu2022}
Rosu S., Rauw G., Naz{\'e} Y., Gosset E., Sterken C.,
2022, A\&A, 664, A98

\bibitem[\protect\citeauthoryear{Sana, Rauw \& Gosset}{2001}]{Sana2001}
Sana H., Rauw G., Gosset E.,
2001, A\&A, 370, 121

\bibitem[\protect\citeauthoryear{Sana et al.}{2012}]{Sana2012}
Sana H., et al.,
2012, Sci., 337, 444

\bibitem[\protect\citeauthoryear{Sana et al.}{2014}]{Sana2014}
Sana H., et al.,
2014, ApJS, 215, 15

\bibitem[\protect\citeauthoryear{Schneider et al.}{2016}]{Schneider2016}
Schneider F.~R.~N., Podsiadlowski Ph., Langer N., Castro N., Fossati L.,
2016, MNRAS, 457, 2355

\bibitem[\protect\citeauthoryear{Sch\"oller et al.}{2017}]{Schoeller2017}
Sch\"oller M., et al.,
2017, A\&A, 599, A66

\bibitem[\protect\citeauthoryear{Shultz et al.}{2018}]{Shultz2018}
Shultz M.~E., et al.,
2018, MNRAS, 475, 5144

\bibitem[\protect\citeauthoryear{Skinner et al.}{2010}]{Skinner2010}
Skinner S.~L., Zhekov S.~A., G{\"u}del M., Schmutz W., Sokal K.~R.,
2010, AJ, 139, 825

\bibitem[\protect\citeauthoryear{Smith}{1981}]{Smith1981}
Smith M.~A.,
1981, ApJ, 248, 214

\bibitem[\protect\citeauthoryear{Snik et al.}{2008}]{Harps}
Snik F., Jeffers S., Keller C., Piskunov N., Kochukhov O., Valenti J., Johns-Krull C.,
2008, in McLean I.~S., Casali M.~M., eds,
Proc.\ SPIE Conf.\ Ser.\ Vol.~7014,
Ground-based and Airborne Instrumentation for Astronomy~II.
SPIE, Bellingham, p.~E22

\bibitem[\protect\citeauthoryear{Sota et al.}{2014}]{Sota2014}
Sota A., Ma{\'\i}z Apell{\'a}niz J., Morrell N.~I., Barb{\'a} R.~H., Walborn N.~R., Gamen R.~C., Arias J.~I., Alfaro E.~J.,
2014, ApJS, 211, 10

\bibitem[\protect\citeauthoryear{Stickland, Lloyd \& Sweet}{1998}]{Stickland1998}
Stickland D.~J., Lloyd C., Sweet I.,
1998, The Observatory, 118, 7

\bibitem[\protect\citeauthoryear{Takahashi \& Langer}{2021}]{TakahashiLanger2021}
Takahashi K., Langer N.,
2021, A\&A, 646, A19

\bibitem[\protect\citeauthoryear{Thackeray}{1966}]{Thackeray1966}
Thackeray A.~D.,
1966, MNRAS, 134, 97

\bibitem[\protect\citeauthoryear{Thaller}{1998}]{Thaller1998}
Thaller M.~L.,
1998, in I.~D.~Howarth, eds,
Proc.\ Boulder-Munich II.\ Worksh., Properties of Hot, Luminous Stars,
ASP Conf.\ Ser., 131, 417

\bibitem[\protect\citeauthoryear{Tout et al.}{2008}]{Tout2008}
Tout C.~A., Wickramasinghe D.~T., Liebert J., Ferrario L., Pringle J.~E.,
2008, MNRAS, 387, 897

\bibitem[\protect\citeauthoryear{ud-Doula, Owocki \& Townsend}{2008}]{udDoula2008}
ud-Doula A., Owocki S.~P., Townsend R.~H.~D.,
2008, MNRAS, 385, 97

\bibitem[\protect\citeauthoryear{van der Hucht}{2001}]{vanderHucht2001}
van der Hucht K.~A.,
2001, NewAR, 45, 135

\bibitem[\protect\citeauthoryear{van der Meij et al.}{2021}]{vanderMeij2021}
van der Meij V., Guo D., Kaper L., Renzo M.,
2021, A\&A, 655, A31

\bibitem[\protect\citeauthoryear{van Leeuwen \& van Genderen}{1997}]{vanLeeuwenvanGenderen1997}
van Leeuwen F., van Genderen A.~M.,
1997, A\&A, 327, 1070

\bibitem[\protect\citeauthoryear{Walborn}{1973}]{Walborn1973}
Walborn N.~R.,
1973, ApJ, 179, 517

\bibitem[\protect\citeauthoryear{Walborn}{1982}]{Walborn1982}
Walborn N.~R.,
1982, AJ, 87, 1300

\bibitem[\protect\citeauthoryear{Walborn \& Panek}{1984}]{WalbornPanek1984}
Walborn N.~R., Panek R.~J.,
1984, ApJ, 286, 718

\bibitem[\protect\citeauthoryear{Walborn}{2006}]{Walborn2006}
Walborn N.~R.,
2006, The Ultraviolet Universe: Stars from Birth to Death,
26th meeting of the IAU, Joint Discussion~4, 16-17 August 2006, Prague, Czech Republic, JD04, 19

\bibitem[\protect\citeauthoryear{Wang, Zhu \& Yue}{2011}]{Wang2022}
Wang Z.~H., Zhu L.~Y., Yue Y.~F.,
2022, MNRAS, 511, 488

\bibitem[\protect\citeauthoryear{Wickramasinghe, Tout \& Ferrario}{2014}]{Wickramasinghe2014}
Wickramasinghe D.~T., Tout C.~A., Ferrario L.,
2014, MNRAS, 437, 675

\bibitem[\protect\citeauthoryear{Wurster, Bate \& Price}{2019}]{Wurster2019}
Wurster J., Bate M.~R., Price D.~J.,
2019, MNRAS, 489, 1719

\end{thebibliography}
\end{document}